\documentclass[12pt]{article}
\usepackage[latin9]{inputenc}
\usepackage{geometry}
\usepackage[active]{srcltx}
\usepackage{float}
\usepackage{amsmath}
\usepackage{amsthm}
\usepackage{amssymb}
\usepackage{graphicx}
\usepackage{rotfloat}
\usepackage{setspace}
\usepackage{esint}
\usepackage[authoryear]{natbib}
\usepackage[font=normalsize,labelfont=bf]{caption}
\usepackage{comment}
\usepackage{xr-hyper}
\usepackage{hyperref}
\usepackage{nicefrac}
\usepackage{mathtools}
\usepackage[caption=false]{subfig}
\usepackage{algorithm}
\usepackage{algpseudocode}
\makeatletter
\providecommand{\tabularnewline}{\\}
\floatstyle{ruled}
\newfloat{algorithm}{tbp}{loa}
\providecommand{\algorithmname}{Algorithm}
\floatname{algorithm}{\protect\algorithmname}

\usepackage{latexsym}
\usepackage{amsthm}\usepackage{amsfonts}\usepackage{graphicx}\usepackage{epsfig}
\expandafter\def\csname bm\endcsname{\boldmath}

\newcommand*{\patchAmsMathEnvironmentForLineno}[1]{%
      \expandafter\let\csname old#1\expandafter\endcsname\csname #1\endcsname
      \expandafter\let\csname oldend#1\expandafter\endcsname\csname end#1\endcsname
      \renewenvironment{#1}%
         {\linenomath\csname old#1\endcsname}%
         {\csname oldend#1\endcsname\endlinenomath}}%
    \newcommand*{\patchBothAmsMathEnvironmentsForLineno}[1]{%
      \patchAmsMathEnvironmentForLineno{#1}%
      \patchAmsMathEnvironmentForLineno{#1*}}%
    \AtBeginDocument{%
    \patchBothAmsMathEnvironmentsForLineno{equation}%
    \patchBothAmsMathEnvironmentsForLineno{align}%
    \patchBothAmsMathEnvironmentsForLineno{flalign}%
    \patchBothAmsMathEnvironmentsForLineno{alignat}%
    \patchBothAmsMathEnvironmentsForLineno{gather}%
    \patchBothAmsMathEnvironmentsForLineno{multline}%
    }

\usepackage{setspace}
\def\dispmuskip{\thinmuskip= 3mu plus 0mu minus 2mu \medmuskip=  4mu plus 2mu minus 2mu \thickmuskip=5mu plus 5mu minus 2mu}
\def\textmuskip{\thinmuskip= 0mu                    \medmuskip=  1mu plus 1mu minus 1mu \thickmuskip=2mu plus 3mu minus 1mu}
\def\beq{\dispmuskip\begin{equation}}    \def\eeq{\end{equation}\textmuskip}
\def\beqn{\dispmuskip\begin{displaymath}}\def\eeqn{\end{displaymath}\textmuskip}
\def\bea{\dispmuskip\begin{eqnarray}}    \def\eea{\end{eqnarray}\textmuskip}
\def\bean{\dispmuskip\begin{eqnarray*}}  \def\eean{\end{eqnarray*}\textmuskip}

\usepackage{bm}  

\usepackage[mathlines]{lineno} 

\makeatletter
\newcommand*{\addFileDependency}[1]{
  \typeout{(#1)}
  \@addtofilelist{#1}
  \IfFileExists{#1}{}{\typeout{No file #1.}}
}
\makeatother

\usepackage{subfig}
\usepackage[export]{adjustbox}

\usepackage{comment}

\let\oldref\ref
\renewcommand{\ref}[1]{(\oldref{#1})}

\usepackage{amsthm}
\theoremstyle{definition}
\usepackage{authblk}

\usepackage{bbm}
\usepackage{xr}
\usepackage{tikz}
\usetikzlibrary{positioning, decorations.pathreplacing, shapes.misc, fit}
\usetikzlibrary{arrows.meta}

\begin{document}

\title{\bf Flexible Bayesian Models for Time-Varying Income Distributions}
\author{David Gunawan}
\affil{School of Mathematics and Physics, University of Wollongong, Wollongong, New South Wales, Australia}
\maketitle

\begin{abstract}
Survey data are widely used to study how income inequality, poverty, and welfare evolve over time. A common practice is to estimate the income distribution separately for each year, treating annual observations as independent cross-sections. For population subgroups with relatively small sample sizes, however, this approach can produce unstable parameter estimates, imprecise inference for inequality and poverty measures, and potentially misleading posterior probabilities of Lorenz and stochastic dominance. This paper develops flexible Bayesian models for time-varying income distributions that borrow strength across adjacent years by allowing the parameters of income distributions to evolve dynamically. We consider a random walk specification and an extended model with shrinkage priors. The proposed framework yields coherent inference for the full income distributions over time, as well as for associated inequality measures, poverty indices, and dominance probabilities. Simulation studies show that, relative to independent year-by-year models, the proposed approach produces substantially more precise and stable inference, while avoiding spurious variation in welfare comparisons. An application to the Aboriginal and residents of the Australian Capital Territory (ACT) population subgroups in the Household, Income and Labour Dynamics in Australia survey shows that the dynamic models deliver improved inference for income distributions and related welfare measures, and can change conclusions about distributional dominance over time.
\end{abstract}


\section{Introduction}
The estimation of income distributions plays a central role in the measurement of inequality and poverty and, more broadly, in welfare comparisons across time and across populations. For useful overviews of the extensive literature on income distribution modelling, including alternative specifications, their properties, and estimation methods, see the monograph by \citet{KleiberKotz2003}, the edited volume by \citet{Chotikapanich2008}, and the articles by \citet{BandourianMcDonaldTurley2003} and \citet{McDonaldXu1995}. The present paper focuses on inference for income distributions of population subgroups for which only a small number of observations are available in each year.


The availability of detailed survey data has transformed empirical research on inequality and poverty. These data provide a rich basis for tracking distributional change, evaluating policy
reforms, and quantifying the persistence of economic disadvantage. Prominent examples
include the Household Income and Labour Dynamics in Australia (HILDA) survey \citep{WatsonWooden2012HILDA} and the
Panel Study of Income Dynamics (PSID) in the United States \citep{McGonagle2012PSID}. In Australia, HILDA has become a key resource for  analysing
how welfare, inequality and poverty evolve for different demographic and labour-market groups.
As such datasets continue to mature and expand, they increasingly support inherently dynamic questions: whether inequality and poverty are rising or falling, and how future distributions might evolve under plausible trajectories.

Policy discussions typically focus on inequality measures, poverty indices, and welfare comparisons that depend on the full income distribution. These include widely used measures of inequality, such as the Gini coefficient and generalised entropy measures, including the Theil indices \citep{Cowell2011}, as well as poverty measures such as the headcount ratio, poverty gap, and related indices \citep{FosterGreerThorbecke1984}. In addition, distributional comparisons based on Lorenz and stochastic dominance provide partial orderings that are often more informative than any single summary measure \citep{barrett2003consistent,barrett2014consistent}.


A common strategy in applied work is to estimate a parametric income distribution separately
for each year, treating each year as an independent cross-section, for example Dagum \citep{Dagum1977}, Singh-Maddala \citep{SinghMaddala1976}, and Generalised Beta 2 (GB2) distributions \citep{McDonaldXu1995}. Under this approach, the distributional parameters are estimated using only within-year observations. The posterior densities of inequality and poverty measures, together with the posterior probabilities of dominance, are then computed as functions of parameter draws obtained from Markov chain Monte Carlo (MCMC) algorithms \citep{gunawan2021posterior}. While simple,
this year-by-year estimation ignores an important empirical feature: income distributions
typically evolve smoothly over time, interrupted by occasional shifts associated with macroeconomic
shocks, labour-market changes, or policy reforms. When adjacent years are in fact similar,
estimating each year independently can lead to noisy parameter paths and unstable year-to-year
movements in derived welfare, inequality and poverty measures, particularly for small sample sizes. As a result, the independent cross-sectional approach may produce spurious volatility in estimates of inequality and poverty, and may also yield misleading posterior probabilities of dominance.


This paper makes several contributions. First, this paper develops a flexible dynamic statistical model for income distributions that
explicitly links information across time. The key idea is to treat the distributional parameters
as latent, time-varying states that evolve according to a stochastic process. 
Second, it shows that dynamic modelling improves inference for income distributions, especially for population subgroups with small annual sample sizes. By linking adjacent years, the proposed approach yields smoother and more stable estimates of income distribution parameters and substantially more precise inference for derived welfare summaries. Third, the paper shows that the benefits are not limited to narrower credible intervals. Since poverty, inequality, and dominance comparisons are sensitive to noise in the fitted distributions, independent year-by-year models can lead to substantively different and potentially misleading conclusions about welfare, poverty, and inequality changes over time. In contrast, the proposed dynamic models provide posterior inference that is more stable, and better aligned with the underlying temporal structure in the data.

We illustrate the approach using both a simulation study and an application to income data from HILDA, with particular emphasis on the Aboriginal population subgroup and the residents of the Australian Capital Territory (ACT) subgroup. The simulation results show that, relative to independent year-by-year models, the proposed dynamic specifications recover the underlying time-varying income distribution more accurately and provide tighter uncertainty quantification for poverty and inequality measures. In the empirical application, the dynamic models yield smoother parameter trajectories and more stable welfare summaries, and they can change posterior inferences about Lorenz and stochastic dominance over time.




The remainder of the paper is organised as follows. Section~\ref{sec:model} presents the proposed dynamic income models and discusses the prior distributions of the model parameters. Section \ref{sec:posteriorinference} discusses Bayesian inference for the proposed models using Markov chain Monte Carlo.  Section \ref{sec:inequalitymeasures} discusses poverty and inequality measures. Section \ref{sec:stochasticdominance} discusses Lorenz and stochastic dominance conditions. Section \ref{sec:simulationstudy} discusses a simulation study. Section \ref{sec:empiricalapplications} discusses the real data application using HILDA survey. Section \ref{sec:conclusions} concludes. The paper also has an online supplement with additional technical details and examples.

\section{Model\label{sec:model}}

In this section, we develop dynamic parametric models for income distributions observed over time. Rather than estimating each year's income distribution independently, which can lead to noisy parameter trajectories and unstable estimates of inequality and poverty measures, we treat each annual observed income as a realisation from an underlying distribution that evolves over time. The key idea is to allow the parameters of a flexible income distribution to vary by year, while borrowing strength across adjacent years to regularise estimation and improve robustness. This is particularly important for subgroup analysis, where some annual samples are small and year-by-year estimation is highly variable.

Let $\mathbf y_t=(y_{t,1},\ldots,y_{t,n_t})^\top$ denote the vector of observed incomes in year $t$, where $n_t$ is the number of individuals or households observed in that year. Collecting all years,
$\mathbf y = \big(\mathbf y_1^\top,\ldots,\mathbf y_T^\top\big)^\top.$
We model the evolution of the income distribution over time through a set of time-varying latent state parameters.

\subsection{Dynamic models for income distributions\label{subsec:dynamic_income_model}}
This section discusses the proposed dynamic models for income distributions.
Let $p_{Y_t}(\cdot \mid \boldsymbol\phi_t)$ denote the density of a parametric income distribution at time $t$ with the parameter vector $\boldsymbol\phi_t$. Examples include the Dagum, Singh--Maddala, and GB2 distributions discussed in Section \ref{sec:incomedistributions} of the online supplement. Conditional on $\boldsymbol\phi_t$, the incomes observed in year $t$ are assumed independent and identically distributed:
\[
y_{t,i}\mid \boldsymbol\phi_t \stackrel{\mathrm{iid}}{\sim} p_{Y_t}(\,\cdot \mid \boldsymbol\phi_t),
\qquad i=1,\ldots,n_t,\quad t=1,\ldots,T.
\]
Hence the likelihood factorises as
\[
p(\mathbf y\mid \boldsymbol\phi_{1:T})
=
\prod_{t=1}^T \prod_{i=1}^{n_t}
p_{Y_t}(y_{t,i}\mid \boldsymbol\phi_t),
\]
where $\boldsymbol\phi_{1:T}=(\boldsymbol\phi_1^{\top},\ldots,\boldsymbol\phi_T^{\top})^{\top}$. 
The income distribution parameters are often constrained; for example, scale and shape parameters are typically positive. To work on an unconstrained space, we introduce a latent state vector $\boldsymbol\theta_t = (\theta_{1,t},\ldots,\theta_{d,t})^\top \in \mathbb R^d$
and a smooth one-to-one transformation $g$, such that
\[
\boldsymbol\phi_t = g(\boldsymbol\theta_t), \qquad t=1,\ldots,T.
\]
For instance, positivity constraints can be enforced using exponential transformations. Under this reparameterisation, the observation model becomes
\[
y_{t,i}\mid \boldsymbol\theta_t \stackrel{\mathrm{iid}}{\sim}
p_{Y_t}\!\left(\,\cdot \mid g(\boldsymbol\theta_t)\right),
\qquad i=1,\ldots,n_t,\quad t=1,\ldots,T.
\]

\noindent To model gradual changes in the income distribution over time, we place a dynamic prior on the latent states $\{\boldsymbol\theta_t\}_{t=1}^T$. A natural starting point is a random walk (RW) model.  
Assume that ${\phi_{1,1},\ldots,\phi_{d,1}>0}$ and define
${\boldsymbol{\theta}_1
=
\bigl(\log(\phi_{1,1}),\ldots,\log(\phi_{d,1})\bigr)^\top,}
$
with
${
\phi_{1,1},\ldots,\phi_{d,1}
\stackrel{\mathrm{ind}}{\sim}
\mathrm{Half\text{-}Cauchy}(0,1).
}$
Under this log-transformation, the density of the initial state
$\boldsymbol{\theta}_1$ is given by
\begin{equation}
p(\boldsymbol{\theta}_1)
=
\prod_{k=1}^d
\frac{2\exp(\theta_{k,1})}{\pi\bigl(1+\exp(2\theta_{k,1})\bigr)},
\label{eq:initialstate}
\end{equation}
where $\phi_{k,1}=\exp(\theta_{k,1})$, for $k=1,\ldots,d$.
For $t=2,\ldots,T$, the state evolution is given by
\[
\boldsymbol\theta_t \mid \boldsymbol\theta_{t-1},\boldsymbol\sigma
\sim
\mathcal N\!\left(\boldsymbol\theta_{t-1},\,\mathbf Q(\boldsymbol\sigma)\right),
\]
where
$\mathbf Q(\boldsymbol\sigma)=\mathrm{diag}(\sigma_1^2,\ldots,\sigma_d^2)$
and $\boldsymbol\sigma=(\sigma_1,\ldots,\sigma_d)^\top$ controls the magnitude of year-to-year change in each latent state. Equivalently, componentwise,
\[
\theta_{k,t} = \theta_{k,t-1} + \sigma_k \varepsilon_{k,t},
\qquad
\varepsilon_{k,t}\stackrel{\mathrm{iid}}{\sim}\mathcal N(0,1),
\qquad
k=1,\ldots,d,\quad t=2,\ldots,T.
\]
This formulation allows different parameters of the income distribution to evolve at different rates. The innovation scales determine how strongly the model borrows information across adjacent years. Small values imply smoother trajectories, whereas larger values allow more pronounced temporal variation. We use a weakly informative half-Cauchy prior:
${
\sigma_k \sim \mathrm{Half\mbox{-}Cauchy}(0,1),
}$
for $k=1,...,d$. Because the half-Cauchy prior is concentrated near zero but has heavy tails, it encourages small value of $\sigma_k$, while still allowing for large value of $\sigma_k$ when warranted by the data. Combining the observation and state equations yields the joint posterior density
\[
p(\boldsymbol\theta_{1:T},\boldsymbol\sigma\mid \mathbf y)
\propto
\left[
\prod_{t=1}^T \prod_{i=1}^{n_t}
p_{Y_t}\!\left(y_{t,i}\mid g(\boldsymbol\theta_t)\right)
\right]
p(\boldsymbol\theta_1)
\left[
\prod_{t=2}^T
p(\boldsymbol\theta_t\mid \boldsymbol\theta_{t-1},\boldsymbol\sigma)
\right]
p(\boldsymbol\sigma),
\]
where $\boldsymbol\theta_{1:T}=(\boldsymbol\theta_1^\top,\ldots,\boldsymbol\theta_T^\top)^\top$. This makes clear how the dynamic model differs from year-by-year estimation: inference for $\boldsymbol\theta_t$ is informed not only by the data observed in year $t$, but also indirectly by neighbouring years through the latent trajectory. As a result, the model yields smoother and typically more stable estimates of time-varying income distributions and of the inequality and poverty measures derived from them.





\subsection{Random walk income model with horseshoe shrinkage priors\label{subsec:rw_horseshoe}}

The random walk model can be extended by considering a horseshoe shrinkage prior \citep{carvalho2010horseshoe} on the innovation scales. This yields a {random walk income model with horseshoe shrinkage priors} (RW-HS). The main idea is to replace the innovation scales by a global-local shrinkage structure: a global parameter controls the overall amount of temporal smoothing, while local parameters allow individual components of the latent state to deviate from this common level of smoothness when supported by the data.


Under the horseshoe specification, the latent states evolve according to \eqref{eq:initialstate} for $t=1$
and, for $t=2,\ldots,T$,
\[
\boldsymbol\theta_t \mid \boldsymbol\theta_{t-1},\tau^2,\boldsymbol\Lambda
\sim
\mathcal N\!\left(\boldsymbol\theta_{t-1},\,\tau^2\mathbf\Lambda\right),
\]
where
$\mathbf\Lambda=\mathrm{diag}(\lambda_1^2,\ldots,\lambda_d^2).
$
Equivalently, componentwise,
\[
\theta_{k,t}
=
\theta_{k,t-1} + \tau\lambda_k \varepsilon_{k,t},
\qquad
\varepsilon_{k,t}\stackrel{\mathrm{iid}}{\sim}\mathcal N(0,1),
\qquad
k=1,\ldots,d,\quad t=2,\ldots,T.
\]
Here, $\tau>0$ is a global shrinkage parameter and $\lambda_k>0$ is a local shrinkage parameter for the $k$th latent state. When $\tau$ is small, the latent trajectories are strongly smoothed across time. When a particular component requires more flexibility, a large value of $\lambda_k$ allows that component to have a larger jump. We assign half-Cauchy priors to both the global and local scales: 
${\lambda_k \sim \mathrm{Half\mbox{-}Cauchy}(0,1),
 k=1,\ldots,d,}
$
and
$
\tau \sim \mathrm{Half\mbox{-}Cauchy}(0,1).
$

\section{Posterior inference\label{sec:posteriorinference}}

This section describes the Metropolis-within-Gibbs (MwG) samplers \citep{RobertCasella2004} for the proposed dynamic models for income distributions. 



\subsection{Metropolis-within-Gibbs sampler for the random walk  model for income distributions \label{subsec:mcmc_rw}}

We first consider posterior inference for the random walk model for income distributions. 
Algorithm~\ref{alg:mwg_rw_income_model} gives the full MwG sampler. Each MwG iteration alternates between two steps:  
(i) updating the latent states $\boldsymbol\theta_{1:T}$ one time point at a time using random-walk Metropolis-within-Gibbs updates, and  
(ii) updating the innovation variances $\boldsymbol\sigma^2$ using Metropolis-within-Gibbs steps with inverse-gamma proposals. 

We define the time-$t$ log-likelihood contribution as
$
\ell_t(\boldsymbol\theta_t)
=
\sum_{i=1}^{n_t}
\log p_{Y_t}\!\left(y_{t,i}\mid g(\boldsymbol\theta_t)\right).
$
At each iteration, we update $\boldsymbol\theta_t$ conditionally on its neighbours using the Gaussian proposal
$
\boldsymbol\theta_t^\star
\sim
\mathcal N\!\left(\boldsymbol\theta_t,\kappa_t\boldsymbol\Sigma_t\right),
$
where $\boldsymbol\Sigma_t$ is a positive definite proposal covariance matrix and $\kappa_t>0$ is a tuning constant. In practice, $\boldsymbol\Sigma_t$ is obtained from the empirical covariance matrix  of previous MCMC draws, and $\kappa_t$ is tuned to target an acceptance probability around $0.2$ using the algorithm proposed by \citet{garthwaite2016adaptive}.

Because the latent process is first-order Markov, $\boldsymbol\theta_t$ interacts with the rest of the trajectory only through its immediate neighbours $\boldsymbol\theta_{t-1}$ and $\boldsymbol\theta_{t+1}$. Hence, when proposing $\boldsymbol\theta_t^\star$, the acceptance probability depends only on $\ell_t(\boldsymbol\theta_t)$ and the transition densities linking $\boldsymbol\theta_t$ to adjacent states. Define
\[
\log \pi_t^{\mathrm{RW}}(\boldsymbol\theta_t)=
\begin{cases}
\ell_1(\boldsymbol\theta_1)
+\log  p(\boldsymbol\theta_1)
+\log \mathcal N(\boldsymbol\theta_2\mid \boldsymbol\theta_1,\mathbf Q),
& t=1,
\\[0.6em]
\ell_t(\boldsymbol\theta_t)
+\log \mathcal N(\boldsymbol\theta_t\mid \boldsymbol\theta_{t-1},\mathbf Q)
+\log \mathcal N(\boldsymbol\theta_{t+1}\mid \boldsymbol\theta_t,\mathbf Q),
& 1<t<T,
\\[0.6em]
\ell_T(\boldsymbol\theta_T)
+\log \mathcal N(\boldsymbol\theta_T\mid \boldsymbol\theta_{T-1},\mathbf Q),
& t=T.
\end{cases}
\]
Since the proposal is symmetric, the acceptance probability is
${
\alpha_t
=
\min\Bigl\{1,
\exp\bigl(
\log \pi_t^{\mathrm{RW}}(\boldsymbol\theta_t^\star)
-
\log \pi_t^{\mathrm{RW}}(\boldsymbol\theta_t)
\bigr)
\Bigr\}.
}$
We then set $\boldsymbol\theta_t= \boldsymbol\theta_t^\star$ with probability $\alpha_t$. Detailed updates for the innovative variances are given in Section \ref{sec:additionaldetailrandomwalkincome} of the online supplement. 

\begin{algorithm}[h]
\caption{Metropolis-within-Gibbs sampler for the random walk model for income distributions}
\label{alg:mwg_rw_income_model}
\begin{algorithmic}[1]
\Require Data $\{\mathbf y_t\}_{t=1}^T$, initial values $(\boldsymbol\theta_{1:T}^{(0)},\boldsymbol\sigma^{2(0)})$
\For{$m=1,\ldots,M$}
    \State \textbf{(A) Update latent states $\boldsymbol\theta_{1:T}$}
    \For{$t=1,\ldots,T$}
        \State Propose $\boldsymbol\theta_t^\star \sim \mathcal N\!\left(\boldsymbol\theta_t^{(m-1)},\,\kappa_t\boldsymbol\Sigma_t\right)$
        \State Compute $\log \pi_t^{\mathrm{RW}}(\boldsymbol\theta_t^\star)$ and $\log \pi_t^{\mathrm{RW}}(\boldsymbol\theta_t^{(m-1)})$
        \State Set
        \[
        \alpha_t=
        \min\left\{
        1,\exp\!\left(
        \log \pi_t^{\mathrm{RW}}(\boldsymbol\theta_t^\star)
        -
        \log \pi_t^{\mathrm{RW}}(\boldsymbol\theta_t^{(m-1)})
        \right)
        \right\}
        \]
        \State With probability $\alpha_t$, set $\boldsymbol\theta_t^{(m)}=\boldsymbol\theta_t^\star$; otherwise set $\boldsymbol\theta_t^{(m)}=\boldsymbol\theta_t^{(m-1)}$
    \EndFor

    \State \textbf{(B) Update innovation variances $\boldsymbol\sigma^2$}
    \For{$k=1,\ldots,d$}
        \State Compute $S_k=\sum_{t=2}^T (\theta_{k,t}^{(m)}-\theta_{k,t-1}^{(m)})^2$
        \State Set $\alpha_k=(T-2)/2$ and $\beta_k=S_k/2$
        \State Propose $\sigma_k^{2\star}\sim \mathrm{IG}(\alpha_k,\beta_k)$
        \State Set
        \[
        \alpha_{\sigma_k^2}
        =
        \min\left\{
        1,\frac{1+\sigma_k^{2(m-1)}}{1+\sigma_k^{2\star}}
        \right\}
        \]
        \State With probability $\alpha_{\sigma_k^2}$, set $\sigma_k^{2(m)}=\sigma_k^{2\star}$; otherwise set $\sigma_k^{2(m)}=\sigma_k^{2(m-1)}$
    \EndFor

    \State Store $\bigl(\boldsymbol\theta_{1:T}^{(m)},\boldsymbol\sigma^{2(m)}\bigr)$
\EndFor
\end{algorithmic}
\end{algorithm}

\subsection{Metropolis-within-Gibbs sampler for the random walk model with horseshoe shrinkage priors for income distributions\label{subsec:mcmc_rw_hs}}

We now extend the MwG sampler to the random walk income model with horseshoe shrinkage priors \citep{carvalho2010horseshoe}. Under this specification, the state innovations satisfy
\[
\Delta\theta_{k,t}
=
\theta_{k,t}-\theta_{k,t-1},
\qquad t=2,\ldots,T,\quad k=1,\ldots,d,
\]
with
\[
\Delta\theta_{k,t}\mid \tau^2,\lambda_k^2
\stackrel{\mathrm{ind}}{\sim}
\mathcal N(0,\tau^2\lambda_k^2).
\]

A convenient augmented representation of horseshoe priors is obtained through inverse-gamma mixtures \citep{makalic2015simple}. Specifically,
\[
\lambda_k^2 \mid \nu_k \sim \mathrm{IG}\!\left(\frac{1}{2},\frac{1}{\nu_k}\right),
\qquad
\nu_k \sim \mathrm{IG}\!\left(\frac{1}{2},1\right),
\qquad k=1,\ldots,d,
\]
and
\[
\tau^2 \mid \xi \sim \mathrm{IG}\!\left(\frac{1}{2},\frac{1}{\xi}\right),
\qquad
\xi \sim \mathrm{IG}\!\left(\frac{1}{2},1\right),
\]
where $\mathrm{IG}(a,b)$ denotes the inverse-gamma distribution with density proportional to
$
x^{-(a+1)}\exp\!\left(-\frac{b}{x}\right), x>0.
$
Let $\boldsymbol{\nu}=(\nu_1,\ldots,\nu_d)^\top$. The joint posterior distribution of the latent states $\boldsymbol\theta_{1:T}$ and shrinkage parameters is given by
\begin{align*}
&p(\boldsymbol\theta_{1:T},\tau^2,\boldsymbol\lambda^2,\xi,\boldsymbol\nu \mid \mathbf y) \\
&\quad\propto
\left\{
\prod_{t=1}^T \prod_{i=1}^{n_t}
p_{Y_t}\!\left(y_{t,i}\mid g(\boldsymbol\theta_t)\right)
\right\}
p(\boldsymbol\theta_1)
\left\{
\prod_{t=2}^T
p\!\left(\boldsymbol\theta_t\mid \boldsymbol\theta_{t-1},\tau^2,\boldsymbol\lambda^2\right)
\right\} \\
&\qquad\times
p(\tau^2\mid \xi)\,p(\xi)
\prod_{k=1}^d p(\lambda_k^2\mid \nu_k)\,p(\nu_k).
\end{align*}

Algorithm~\ref{alg:mwg_rw_hs_income_model} in Section \ref{sec:additionaldetailshorseshoe} of the online supplement summarises the MwG sampler for the random walk income model with horseshoe shrinkage priors. The two dynamic income models differ only in how the temporal innovation variances are modelled. In the RW model, each latent state has its own innovation variance $\sigma_k^2$. In the RW-HS specification, these variance terms are replaced by the product 
$\tau^2\lambda_k^2$, where $\tau^2$ controls the overall degree of smoothing 
and $\lambda_k^2$ allows each latent state component to depart from this 
global level. 


\section{Inequality and poverty measures\label{sec:inequalitymeasures}}

This section summarises the inequality and poverty indices used in our empirical analysis.
Given a parametric income model with density $p_{Y_t}(\cdot\mid\boldsymbol{\phi_t})$ and distribution
function ${F_{Y_t}(\cdot\mid\boldsymbol{\phi_t})}$ at time $t$, we evaluate each index as a functional of
$\boldsymbol{\phi}_t$ and a poverty line $z>0$. Let
$\mu_t$ denotes the mean income, assumed finite, and 
$L_t(u;\boldsymbol{\phi}_t)$ denotes the Lorenz curve at time $t$, defined as the cumulative share of total
income held by the poorest $u\in[0,1]$ proportion of the population. 


The most widely used inequality measure is the Gini coefficient, $G_t$, defined as twice the
area between the Lorenz curve and the line of equality. It ranges from $0$ 
to $1$, and is invariant to scale, in the sense that multiplying
all incomes by a positive constant does not change $G_t$. The Gini index at time $t$ is given by
\begin{equation}
G_t
=
1-2\int_{0}^{1} L_t(u;\boldsymbol{\phi}_t)\,du.
\label{eq:gini_lorenz}
\end{equation}
An equivalent expression, convenient for parametric modelling, is
\begin{equation}
G_t
=
-1+\frac{2}{\mu_t}\int_{0}^{\infty} y\,F_{Y_t}(y\mid\boldsymbol{\phi})\,p_{Y_t}(y\mid\boldsymbol{\phi})\,dy,
\label{eq:gini_cdf_pdf}
\end{equation}
see, for example, \citet{Gastwirth1971}. In practice, \eqref{eq:gini_cdf_pdf} is useful because
closed-form expressions for $F_{Y_t}(\cdot\mid\boldsymbol{\phi})$ and for moment distribution
functions often yield analytic or numerically stable evaluations of $G_t$ under parametric
models.

We now briefly describe well-known poverty measures.
Let $z>0$ denote a fixed poverty line, for example an absolute poverty threshold or a
fraction of median income. We consider poverty measures that quantify complementary
aspects of poverty: {incidence} (how many are poor), and {depth} (how far below the
poverty line the poor are on average). These measures are standard in the poverty
measurement literature; see, for example, \citet{FosterGreerThorbecke1984}.
The headcount ratio ($\mathrm{HC}_t$) is simply the proportion of the population with income below $z$ at time $t$:
\begin{equation}
\mathrm{HC}_t
=
F_{Y_t}(z\mid\boldsymbol{\phi}_t).
\label{eq:hc}
\end{equation}
The headcount ratio is easy to interpret but ignores the depth of poverty: it does not change
when incomes of the poor fall further below $z$, provided they remain below the poverty line.
The poverty gap ($\mathrm{PG}_t$) measures the average proportional shortfall from the poverty line at time $t$:
\begin{equation}
\mathrm{PG}_t
=
\int_{0}^{z}\left(\frac{z-y}{z}\right)p_{Y_t}(y\mid\boldsymbol{\phi}_t)\,dy.
\label{eq:pg}
\end{equation}
Unlike \(\mathrm{HC}_t\), the poverty gap accounts for how far incomes fall below the poverty line and can be interpreted as the minimum share of total income needed to raise all poor individuals to \(z\).
More generally, the Foster--Greer--Thorbecke (FGT) class at time $t$ is defined by
\begin{equation}
\mathrm{FGT}\alpha_t
=
\int_{0}^{z}\left(\frac{z-y}{z}\right)^{\alpha} p_{Y_t}(y\mid\boldsymbol{\phi_t})\,dy,
\qquad \alpha\ge 0.
\label{eq:fgt_general}
\end{equation}
Thus, $\mathrm{FGT}0_t=\mathrm{HC}_t$ and $\mathrm{FGT}1_t=\mathrm{PG}_t$. In this paper, we estimate the indices by Monte Carlo simulation:
We draw a large sample from the fitted income distribution and compute the corresponding sample-based estimates.

\section{Lorenz and stochastic dominance \label{sec:stochasticdominance}}

This section discusses Lorenz dominance, generalised Lorenz dominance, and first-order stochastic dominance. Consider two income distributions at time $t$ indexed by parameter vectors
$\boldsymbol{\phi}_{A,t}$ and $\boldsymbol{\phi}_{B,t}$, with corresponding cumulative distribution
functions $F_{Y_{A,t}}(\cdot\mid\boldsymbol{\phi}_{A,t})$ and $F_{Y_{B,t}}(\cdot\mid\boldsymbol{\phi}_{B,t})$, means
$\mu_{A,t}$ and $\mu_{B,t}$, and Lorenz curves
$L_{A,t}(\cdot;\boldsymbol{\phi}_{A,t})$ and $L_{B,t}(\cdot;\boldsymbol{\phi}_{B,t})$. We state the dominance
conditions using the population share $u\in[0,1]$. Distribution $A$ Lorenz (LD) dominates  distribution $B$ at time $t$ if
\begin{equation}
\begin{aligned}
L_{A,t}(u;\boldsymbol{\phi}_{A,t})
&\ge L_{B,t}(u;\boldsymbol{\phi}_{B,t})
\quad \text{for all } u\in[0,1], \\
L_{A,t}(u;\boldsymbol{\phi}_{A,t})
&> L_{B,t}(u;\boldsymbol{\phi}_{B,t})
\quad \text{for some } u\in(0,1).
\end{aligned}
\label{eq:lorenz_dom_phi}
\end{equation}
Distribution $A$ generalised Lorenz (GLD) dominates  distribution $B$ at time $t$ if
\begin{equation}
\begin{aligned}
\mu_{A,t}\,L_{A,t}(u;\boldsymbol{\phi}_{A,t})
&\ge
\mu_{B,t}\,L_{B,t}(u;\boldsymbol{\phi}_{B,t})
\quad \text{for all } u\in[0,1],\\
\mu_{A,t}\,L_{A,t}(u;\boldsymbol{\phi}_{A,t})
&>
\mu_{B,t}\,L_{B,t}(u;\boldsymbol{\phi}_{B,t})
\quad \text{for some } u\in(0,1).
\end{aligned}
\label{eq:ssd_lorenz_phi}
\end{equation}
Distribution $A$ first-order (FSD) stochastically dominates distribution $B$ at time $t$ if
\begin{equation}
\begin{aligned}
F_{Y_{A,t}}^{-1}(u\mid\boldsymbol{\phi}_{A,t})\ge F_{Y_{B,t}}^{-1}(u\mid\boldsymbol{\phi}_{B,t})
&\quad \text{for all } u\in[0,1],\\
\text{and}\qquad
F_{Y_{A,t}}^{-1}(u\mid\boldsymbol{\phi}_{A,t})> F_{Y_{B,t}}^{-1}(u\mid\boldsymbol{\phi}_{B,t})
&\quad \text{for some } u\in(0,1).
\end{aligned}
\label{eq:fsd_quantile_phi}
\end{equation}

We follow \citet{lander2020bayesian} and \citet{gunawan2021posterior}  to compute posterior probabilities of Lorenz and stochastic dominance. The posterior probabilities of Lorenz and stochastic dominance are estimated by counting the proportion of MCMC draws for which the estimated distributional functions satisfy the relevant dominance conditions. Since income is a continuous variable, when calculating the proportion of MCMC draws that satisfy the relevant dominance conditions for all $u\in [0,1]$, the best we can do is to check the conditions for a finite grid set of points. Accordingly, the resulting posterior probabilities should be interpreted relative to the adopted grid resolution and the total number of MCMC draws used in the analysis. In this paper, we see Section \ref{sec:additionaldetailLorenzstochastic} for further details. In practice, this approach is analogous to widely used frequentist procedures, such as the tests of \citet{barrett2003consistent} for first-order stochastic and generalized Lorenz dominance and \citet{barrett2014consistent} for Lorenz dominance, which approximate continuum-based conditions by evaluating them over a discrete set of support points. 

We also plot probability curves, which give the posterior probability of pointwise dominance at
each population share $u$. Over any range of $u$, the posterior probability of dominance on that
range can be no larger than the minimum value of the probability curve within the range. This makes
the probability curve a useful device for identifying which parts of the distribution, such as the tails
or the middle, drive the overall dominance probability. In particular, if dominance is largely
determined by tail behaviour, one can assess sensitivity by omitting extreme values of $u$.
Likewise, if interest focuses on a particular segment of the population, such as the poor,
one can examine how the dominance probability changes when attention is restricted to the
corresponding range of $u$.

\section{Simulation study\label{sec:simulationstudy}}

This section presents a simulation study designed to assess how borrowing strength over time improves inference for income distributions modelled by the Dagum distribution \citep{Dagum1977}. We generate annual incomes for $T=25$ years with $n=250$ observations per year, and we compare three models: an {independent} approach (ind) that fits a Dagum distribution separately for each year, and a {random-walk} (RW) approach in which transformed Dagum parameters evolve over time through a latent process following random walk model, and a random walk approach with horseshoe priors (RW-HS). All models are estimated in a fully Bayesian framework using the Markov chain Monte Carlo (MCMC) sampler discussed in Section \ref{sec:posteriorinference}.


Our evaluation assesses the ability of different models to estimate the true time-varying Dagum parameters, $(a_t,b_t,p_t)$, as well as a range of key distributional functionals. In particular, we examine posterior estimates of mean income $\mu_t$, the Gini coefficient $G_t$, and the $\mathrm{FGT}{0_t}$ and $\mathrm{FGT}{1_t}$ poverty indices, for $t=1,...,T$. We also compare the estimated densities, cumulative distribution functions, generalised Lorenz curves, and Lorenz curves over time obtained from posterior draws, with posterior credible bands reported for each curve.


We now discuss the data-generating process. We simulate data from the random-walk model specification. The log of the Dagum parameters follow a Gaussian random walk, producing gradual year-to-year changes in the income distribution. For $t=1,\ldots,T$ and $i=1,\ldots,n$, we assume
$
y_{t,i}\mid (a_t,b_t,p_t) \overset{\text{iid}}{\sim} \mathrm{Dagum}(a_t,b_t,p_t).
$
The initial values are set to
$
a_1=3.54, b_1=329.58, p_1=0.61,
$
and the corresponding latent log-parameters are defined as
$
\tilde a_t=\log a_t, 
\tilde b_t=\log b_t,
\tilde p_t=\log p_t.
$
For $t=2,\ldots,T$, these latent log-parameters evolve according to
$
\tilde a_t=\tilde a_{t-1}+\sigma_a \varepsilon_{a,t},
\tilde b_t=\tilde b_{t-1}+\sigma_b \varepsilon_{b,t},
\tilde p_t=\tilde p_{t-1}+\sigma_p \varepsilon_{p,t},
$
where
$
\varepsilon_{a,t},\varepsilon_{b,t},\varepsilon_{p,t}
\overset{\mathrm{iid}}{\sim}\mathcal N(0,1),
$
and $  
\sigma_a=\sigma_b=\sigma_p=0.02.$

\begin{figure}[h]
    \centering
    \includegraphics[width=0.8\linewidth]{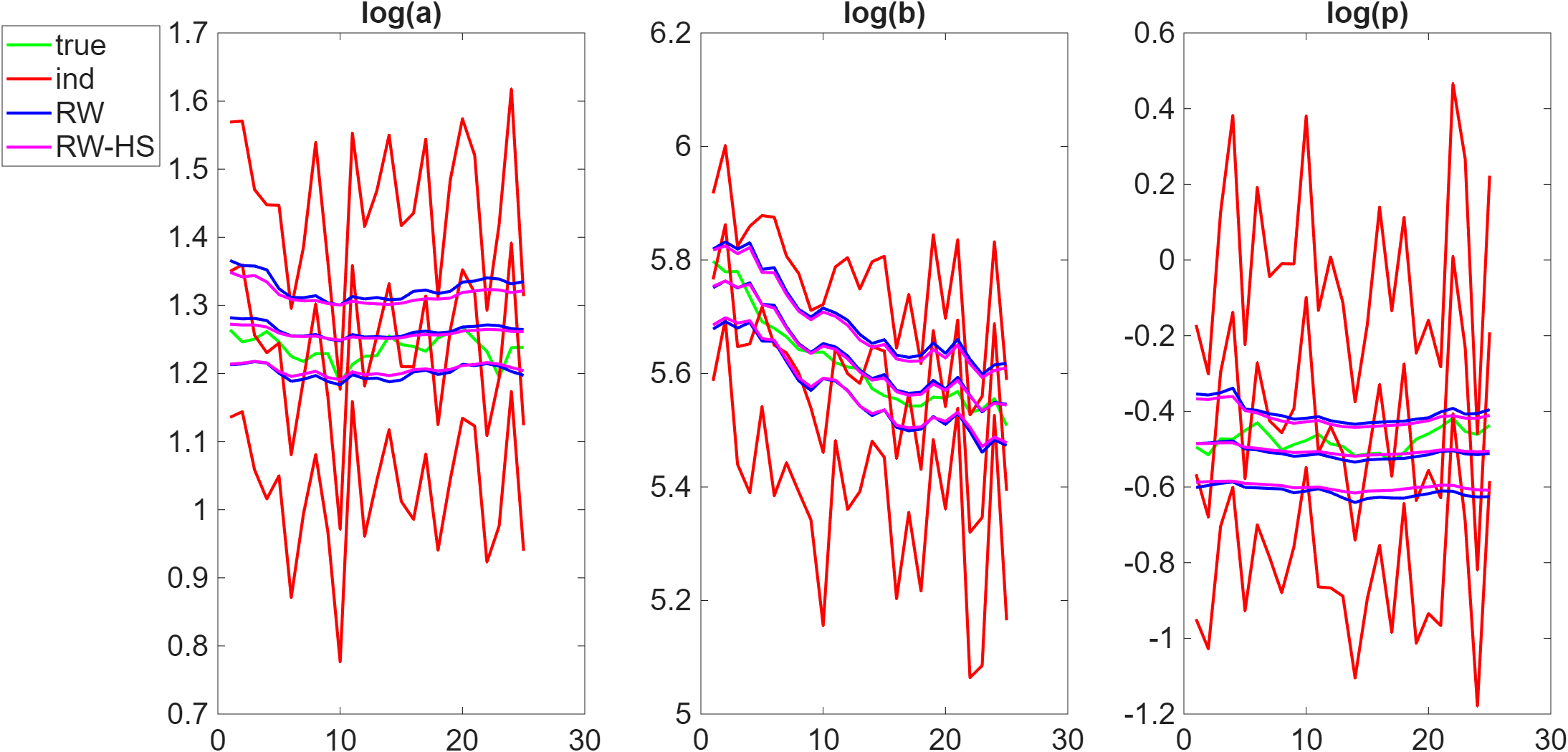}
    \caption{The posterior means (with 95\% credible intervals) of model parameters over time obtained from the independent Dagum income model (ind), the random walk Dagum income model (RW), and the random walk Dagum income model with horseshoe priors (RW-HS) for the simulated dataset. The true parameter values are also plotted.  
}
    \label{fig:Figure_param_dagum_sim1}
\end{figure}

Figure~\ref{fig:Figure_param_dagum_sim1} compares the log of the true parameters of the Dagum distribution with MCMC-based posterior summaries under the independent year-by-year model (ind), the random walk (RW) model, and the random walk model with horseshoe priors (RW-HS). For each parameter, the center curve represents the posterior mean, and the outer curves give the 95\% credible intervals. The RW model tracks the true trajectories closely: it reproduces the gradual decline in $\log(b_t)$ and the mild fluctuations in $\log(a_t)$ and $\log(p_t)$, while delivering markedly tighter credible bands. In contrast, the independent model yields highly volatile year to year estimates and substantially wider uncertainty bands, most notably for $\log(p_t)$. The RW-HS model produces slightly smoother parameter trajectories than the RW model. 
Overall, the figure highlights the efficiency gains from modelling temporal dependence in the parameters.

\begin{figure}[H]
    \centering
    \includegraphics[width=0.8\linewidth]{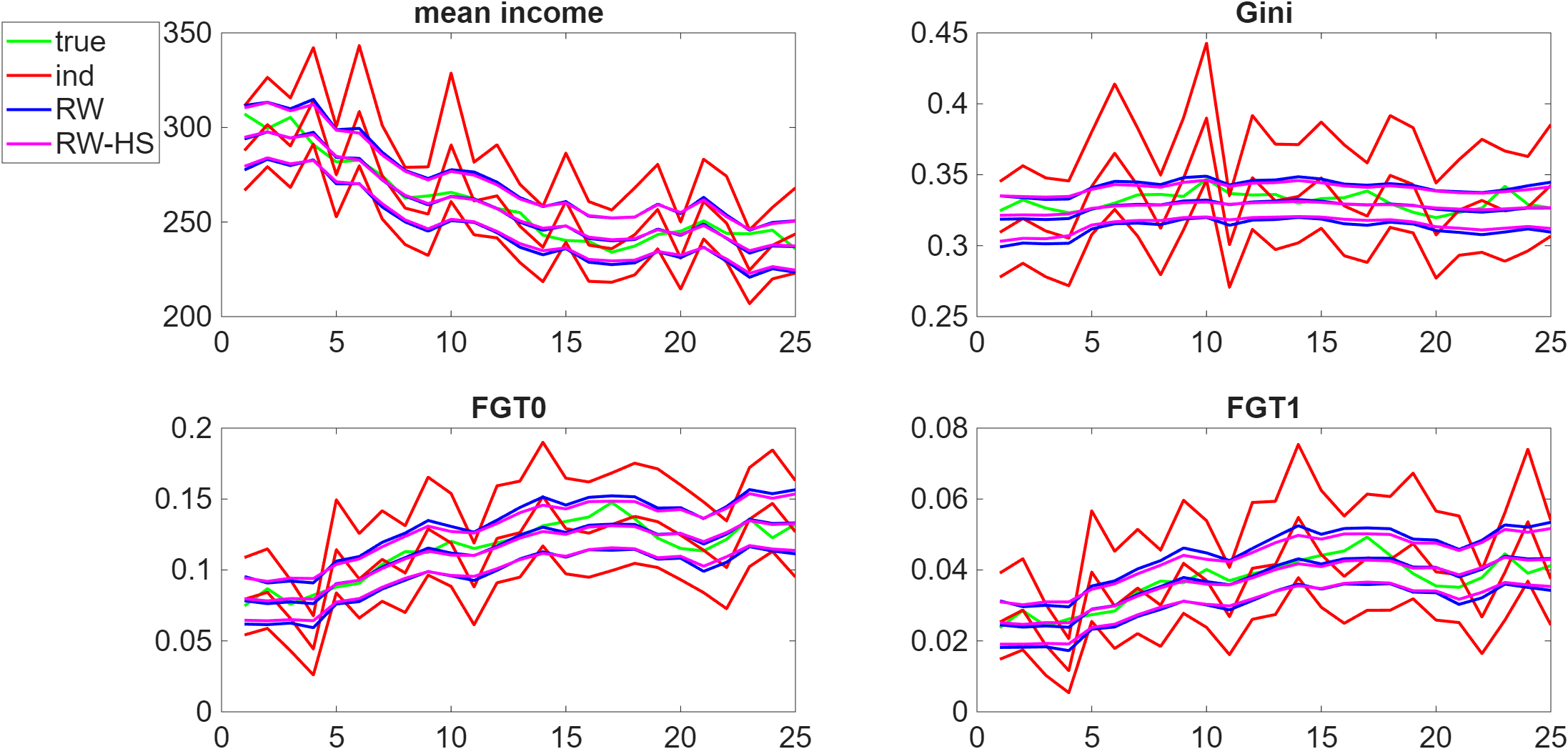}
    \caption{The posterior means (with 95\% credible intervals) of the mean income, Gini coefficient, FGT0, and FGT1 indices over time obtained from the independent Dagum income model (ind), the random walk Dagum income model (RW), and the random walk Dagum income model with horseshoe priors (RW-HS) for the simulated dataset. The true parameter values are also plotted. 
}
    \label{fig:Figure_dagum_mean_trajectory_sim1}
\end{figure}

Figure \ref{fig:Figure_dagum_mean_trajectory_sim1} shows the true and estimated mean income, Gini coefficients, FGT0, and FGT1 indices over time obtained using the three models for the simulated dataset. 
The true mean income exhibits a clear downward trajectory over time, with only modest local fluctuations. The RW and RW-HS models track this gradual decline closely: their posterior means follow the true trajectory, and their pointwise credible bands are tight, with the RW-HS model producing slightly narrower intervals. In contrast, the independent model produces much noisier mean estimates and substantially wider credible bands, with several large year-to-year swings that are not present in the true trajectory. A similar pattern is observed for the headcount ratio (FGT0). The true FGT0 increases from the early years to the middle of the sample and then fluctuates only mildly thereafter. The RW and RW-HS models capture this smooth evolution with narrow credible intervals, whereas the independent model exhibits erratic movements and inflated uncertainty, leading to substantial overestimation or underestimation of the true values in some years.

For inequality, the true Gini coefficient trajectory remains relatively stable over the sample, with only small changes around its long-run level. The RW and RW-HS models again provide a close match to the true series, with smooth posterior means and narrow credible bands. By comparison, the independent model yields highly volatile Gini estimates, including pronounced spikes and dips (with wide bands). 


The PDF and CDF plots in Figures \ref{fig:Figure_PDF_sim1} and \ref{fig:Figure_CDF_sim1} in Section \ref{sec:additionalfiguresimulationstudy} of the online supplement (with 99\% credible intervals) show that all models capture the shape of the true density and CDF. The independent model produces wide bands and large variations in peak height and location, indicating that year-by-year estimation with $n=250$ can translate sampling noise into spurious changes in the density and CDF. The RW and RW-HS models yield narrower credible bands and posterior means that are closer to the true density curve. The RW-HS model provides slightly tighter credible intervals than the RW model.

Figures~\ref{fig:Figure_GLC_sim1_t25_099} and~\ref{fig:Figure_LC_sim1_t25_099} in Section~\ref{sec:additionalfiguresimulationstudy} of the online supplement show the GLCs and LCs for time period $t=25$, obtained under the independent, RW, and RW-HS models. For the GLCs, uncertainty is smallest at low population shares and increases toward the upper tail; accordingly, the independent model exhibits a wider 99\% credible band near the top of the distribution. The RW and RW-HS models deliver markedly tighter bands over the entire curve, especially for the upper deciles, implying more precise inference. The same conclusion holds for the LCs at $t=25$: the independent fit implies much greater uncertainty about income shares, particularly beyond the median and into the top quantiles, whereas the RW and RW-HS models produce narrower credible regions and smoother implied inequality profiles. 


Tables~\ref{tab:tableprobsim1} and~\ref{tab:tableprobsim1_poorest} show that the estimated dominance probabilities can differ markedly across the independent model, the random walk model, and the random walk model with horseshoe shrinkage priors, with the two dynamic specifications generally producing much more similar results to each other than to the independent model. For the full population, the RW and RW-HS models tend to assign higher probabilities than the independent model for Lorenz dominance (LD), especially in the 2005--2001, 2015--2010, and 2025--2015 comparisons. For example, the LD probability for 2025 relative to 2015 increases from $0.2730$ under the independent model to $0.5667$ under RW and $0.5574$ under RW-HS, indicating substantially stronger evidence of inequality improvement once temporal smoothing is imposed. A similar pattern appears for 2005 relative to 2001 and 2015 relative to 2010, where the dynamic models roughly double the LD probabilities relative to the independent specification. By contrast, for first-order stochastic dominance (FSD) and generalised Lorenz dominance (GLD), the most striking discrepancy occurs for the 2010--2005 comparison. Here the independent model suggests moderate evidence of welfare improvement, with probabilities $0.1211$ for FSD and $0.1974$ for GLD, whereas the RW and RW-HS models reduce these probabilities to values close to zero. This indicates that the independent model may overstate welfare improvement. For the remaining comparisons, FSD and GLD probabilities are generally low under all three models, although the RW and RW-HS specifications often produce slightly larger probabilities than the independent model for 2005--2001 and 2025--2015. 



The differences across models are even more pronounced when attention is restricted to the poorest 10\% of the population. In this case, the independent model often yields substantially larger probabilities of FSD and GLD than the dynamic models, particularly for the 2010--2005 and 2025--2015 comparisons. For example, for 2010 relative to 2005, the independent model reports probabilities of $0.4511$ for FSD and $0.6004$ for GLD, whereas the corresponding probabilities under RW and RW-HS fall dramatically to around $0.02$ or lower. Likewise, for 2025 relative to 2015, the independent model gives much stronger evidence of dominance at the lower end of the distribution than either dynamic specification.  In contrast, the RW and RW-HS give substantially smaller dominance probabilities. However, the dynamic models give larger probabilities of Lorenz dominance relative to the independent model in several cases, such as 2005--2001 and 2015--2010. The RW and RW-HS results are very close throughout.


\begin{table}[h]
\caption{Estimated posterior probabilities of Lorenz dominance, generalised Lorenz dominance, and first-order stochastic dominance for the simulated dataset, based on the independent Dagum income model (ind), the random walk Dagum income model (RW), and the random walk Dagum income model with horseshoe priors (RW-HS).\label{tab:tableprobsim1}}

\centering{}%
\begin{tabular}{cccc}
\hline 
 & ind & RW & RW-HS\tabularnewline
\hline 
2005 FSD 2001 & 0.0088 & 0.0351 & 0.0362\tabularnewline
2005 GLD 2001 & 0.0148 & 0.0499 & 0.0526\tabularnewline
2005 LD 2001 & 0.0178 & 0.1458 & 0.1845\tabularnewline
\hline 
2010 FSD 2005 & 0.1211 & 0.0047 & 0.0034\tabularnewline
2010 GLD 2005 & 0.1974 & 0.0071 & 0.0045\tabularnewline
2010 LD 2005 & 0.0143 & 0.1497 & 0.1654\tabularnewline
\hline 
2015 FSD 2010 & 0.0064 & 0.0095 & 0.0087\tabularnewline
2015 GLD 2010 & 0.0234 & 0.0226 & 0.0169\tabularnewline
2015 LD 2010 & 0.1981 & 0.3818 & 0.3742\tabularnewline
\hline 
2025 FSD 2015 & 0.0181 & 0.0360 & 0.0386\tabularnewline
2025 GLD 2015 & 0.0603 & 0.0870 & 0.0884\tabularnewline
2025 LD 2015 & 0.2730 & 0.5667 & 0.5574\tabularnewline
\hline 
\end{tabular}
\end{table}

\begin{table}[h]
\caption{Estimated posterior probabilities of Lorenz dominance, generalised Lorenz dominance, and first-order stochastic dominance over the poorest 10\% of the population for the simulated dataset, based on the independent Dagum income model (ind), the random walk Dagum income model (RW), and the random walk Dagum income model with horseshoe priors (RW-HS).\label{tab:tableprobsim1_poorest}}

\centering{}%
\begin{tabular}{cccc}
\hline 
 & ind & RW & RW-HS\tabularnewline
\hline 
2005 FSD 2001 & 0.0470 & 0.0858 & 0.0798\tabularnewline
2005 GLD 2001 & 0.0716 & 0.1024 & 0.0936\tabularnewline
2005 LD 2001 & 0.0993 & 0.2025 & 0.2450\tabularnewline
\hline 
2010 FSD 2005 & 0.4511 & 0.0216 & 0.0134\tabularnewline
2010 GLD 2005 & 0.6004 & 0.0338 & 0.0206\tabularnewline
2010 LD 2005 & 0.4550 & 0.2015 & 0.2174\tabularnewline
\hline 
2015 FSD 2010 & 0.1507 & 0.1152 & 0.0706\tabularnewline
2015 GLD 2010 & 0.1638 & 0.1490 & 0.0971\tabularnewline
2015 LD 2010 & 0.2196 & 0.4449 & 0.4388\tabularnewline
\hline 
2025 FSD 2015 & 0.6197 & 0.3274 & 0.2760\tabularnewline
2025 GLD 2015 & 0.7637 & 0.4057 & 0.3340\tabularnewline
2025 LD 2015 & 0.8808 & 0.6803 & 0.6615\tabularnewline
\hline 
\end{tabular}
\end{table}

\begin{figure}[h]
    \centering
    \includegraphics[width=0.8\linewidth]{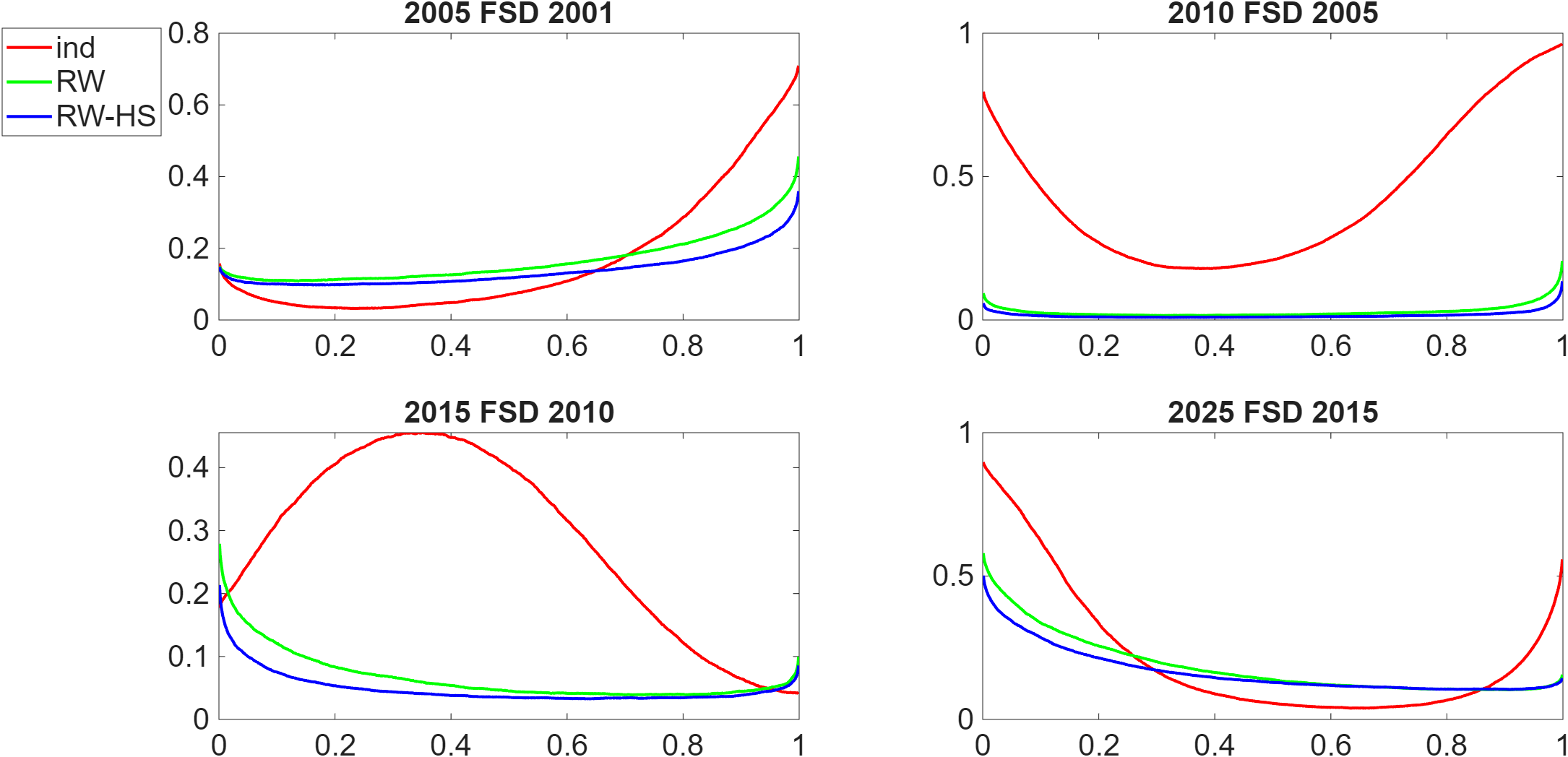}
    \caption{Estimated probability curves for first order stochastic dominance obtained from the independent Dagum income model (ind), the random walk Dagum income model (RW), and the random walk Dagum income model with horseshoe priors (RW-HS) for the simulated dataset.  
}
    \label{fig:yFSDx_prob_sim1}
\end{figure}

\begin{figure}[h]
    \centering
    \includegraphics[width=0.8\linewidth]{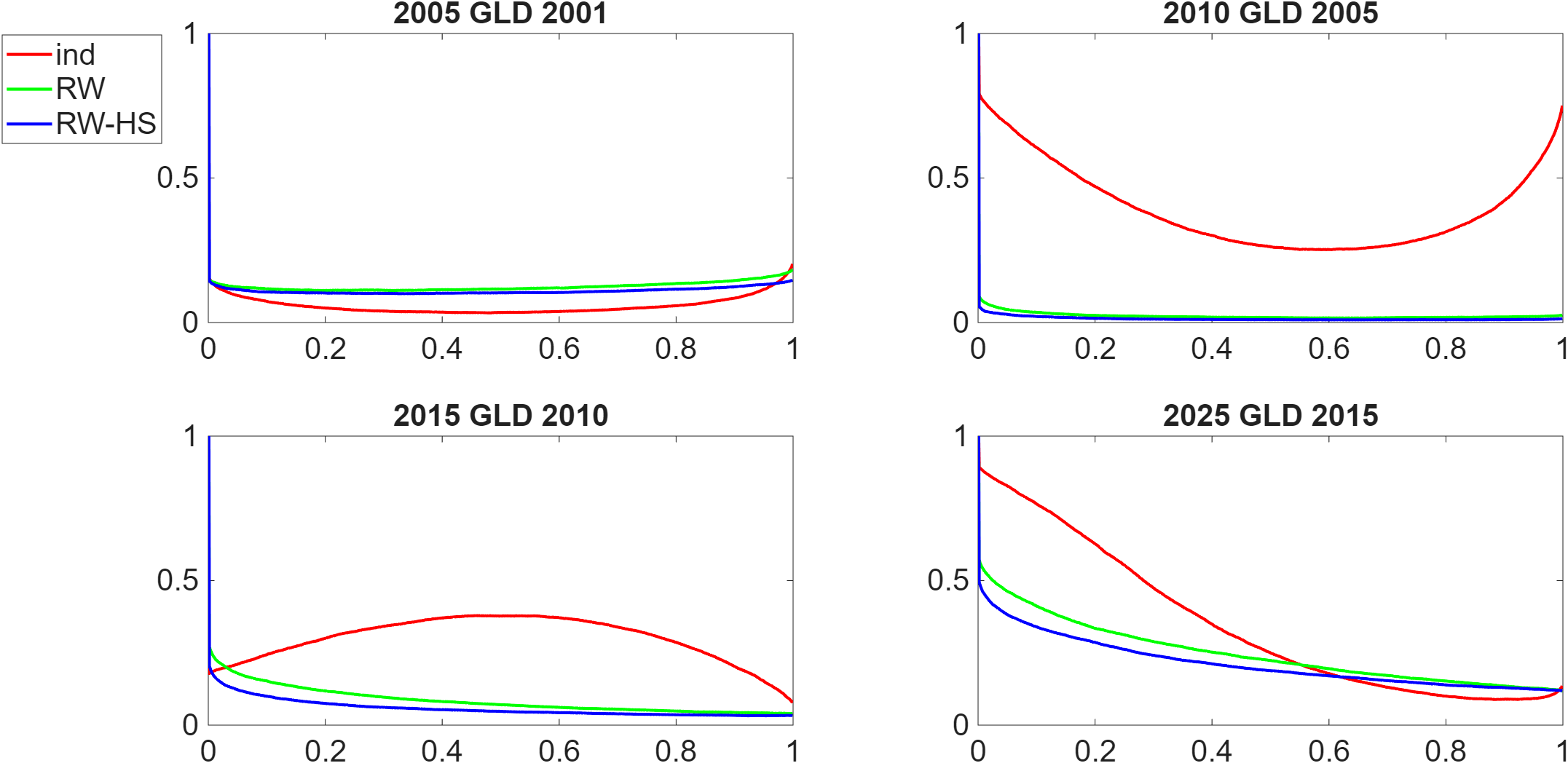}
    \caption{Estimated probability curves for generalised Lorenz dominance obtained from the independent Dagum income model (ind), the random walk Dagum income model (RW), and the random walk Dagum income model with horseshoe priors (RW-HS) for the simulated dataset.  
}
    \label{fig:yGLDx_prob_sim1}
\end{figure}

\begin{figure}[h]
    \centering
    \includegraphics[width=0.8\linewidth]{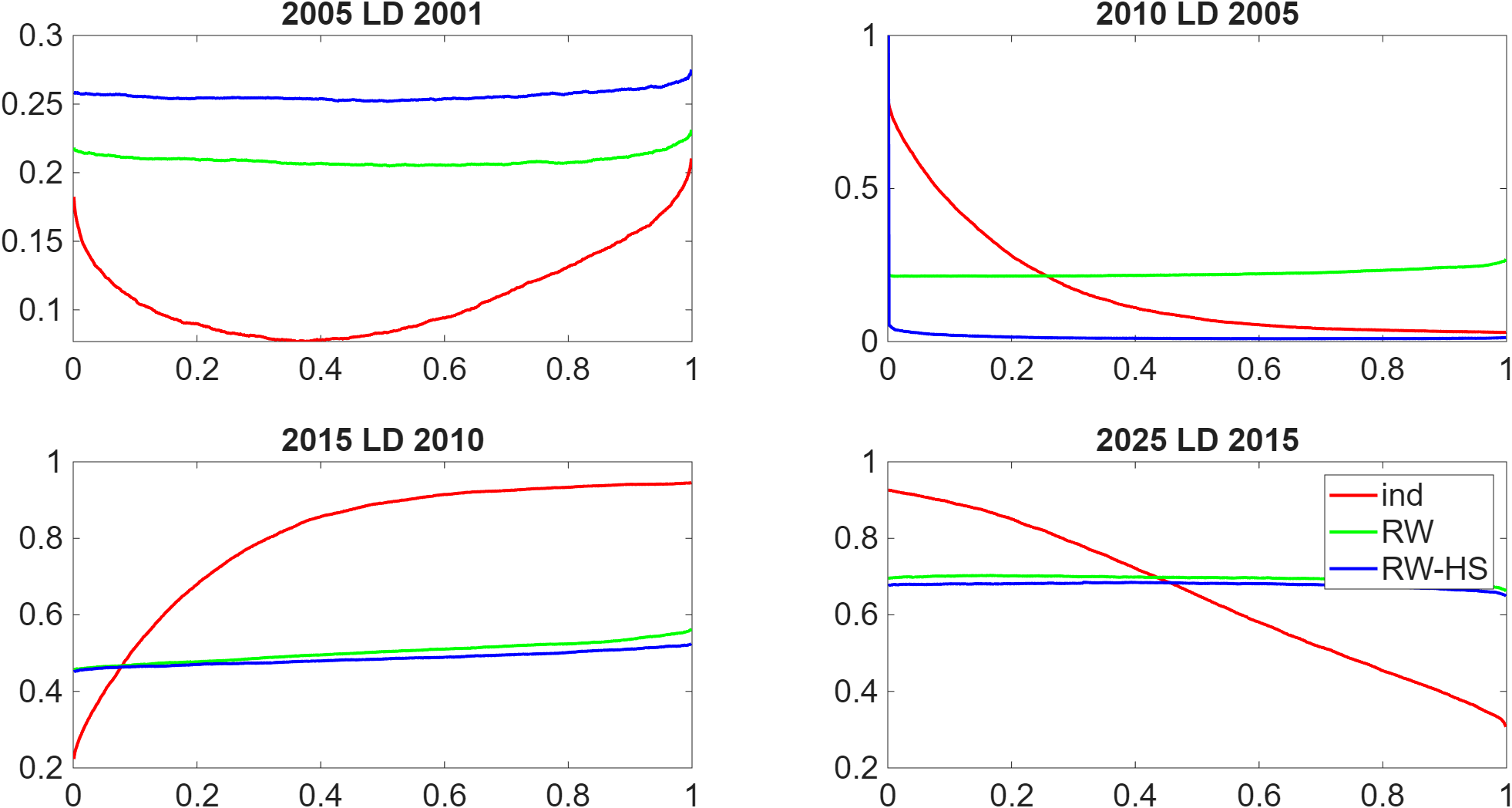}
    \caption{Estimated probability curves for Lorenz dominance obtained from the independent Dagum income model (ind), the random walk Dagum income model (RW), and the random walk Dagum income model with horseshoe priors (RW-HS) for the simulated dataset.  
}
    \label{fig:yLDx_prob_sim1}
\end{figure}

Figures~\ref{fig:yFSDx_prob_sim1}--\ref{fig:yLDx_prob_sim1} show that the probability curves differ systematically across the independent, RW, and RW-HS specifications, and these differences help explain the dominance probabilities reported in Tables~\ref{tab:tableprobsim1} and~\ref{tab:tableprobsim1_poorest}. For FSD and GLD, the independent model generally produces much more irregular and more extreme pointwise probability curves, with pronounced U-shaped or hump-shaped patterns across the support. This is especially evident for the 2010--2005 and 2025--2015 comparisons, where the independent model assigns relatively high pointwise dominance probabilities near the tails but much lower probabilities in the middle of the distribution. By contrast, the RW and RW-HS curves are markedly smoother and less erratic, reflecting the stabilising effect of temporal pooling. In several cases, such as 2010--2005, the dynamic models produce probability curves that remain close to zero over most of the support, indicating that once year-to-year noise is smoothed out there is little evidence of dominance. More generally, these figures show that the independent model is much more sensitive to local fluctuations in the estimated income distributions, whereas the RW and RW-HS models deliver more coherent probability profiles over the whole range of population shares.

The LD curves reveal a somewhat different pattern. Here, the RW and RW-HS models often lie above the independent model over large parts of the support, particularly for the 2005--2001, 2015--2010, and 2025--2015 comparisons, which is consistent with their substantially higher posterior probabilities of LD in Table~\ref{tab:tableprobsim1}. The RW and RW-HS curves are again very similar, showing that the horseshoe prior mainly refines the degree of smoothing rather than changing the substantive conclusions. At the same time, the figures also illustrate why quite high pointwise probabilities do not necessarily translate into high overall dominance probabilities: dominance must hold simultaneously for all evaluation points, so even a relatively small dip in the curve can substantially reduce the joint posterior probability. 

In summary, the simulation results show that the RW and RW-HS models substantially outperform the independent model. Both dynamic specifications track the true parameter and distributional trajectories much more closely, with smoother estimates and much narrower credible intervals, while the independent model produces noisy estimates and inflated uncertainty. These improvements also carry over to the dominance analysis, where the RW and RW-HS models yield more stable and coherent probability profiles. Overall, modelling temporal dependence yields more accurate and reliable inference, with RW-HS providing an additional gain over RW.


\section{Empirical applications\label{sec:empiricalapplications}}
Section \ref{sec:RealDataApplication} briefly describes HILDA data. Section \ref{sec:aboriginalpopulationsubgroup} discusses the empirical results for the  Aboriginal population subgroup. 

\subsection{Data\label{sec:RealDataApplication}}
We use HILDA data from 2001 to 2021 to study the income distributions over time of Aboriginal subpopulations in Section \ref{sec:aboriginalpopulationsubgroup} and residents of the Australian Capital Territory (ACT) in Section \ref{sec:ACTpopulationsubgroup} of the online supplement. 
The welfare measure considered is household disposable income converted to a per-individual basis. Net disposable household income is obtained by subtracting ``total disposable income negative per household'' from ``total disposable income positive per household''. Following \citet{SilaDugain2019}, equivalised income is constructed by dividing net disposable household income by the square root of household size, and this equivalised value is then assigned to each household member. The analysis is restricted to individuals aged 15 years. However, children aged below 15 are still included in the household size used to calculate equivalised income. Values were adjusted using the Consumer Price Index, treating 2021/2022 as the base.

\subsection{Aboriginal population subgroup\label{sec:aboriginalpopulationsubgroup}}
This section presents the empirical application of our modelling framework to Aboriginal population subgroups. Using HILDA data from 2001 to 2021, we compare a range of parametric income models to determine which specification best captures the evolution of the income distributions over time. Specifically, we consider the Dagum, Singh--Maddala, Beta 2, and GB2 distributions, described in Section \ref{sec:incomedistributions} of the online supplement,  under an independent model, a random walk model, and a random walk model with horseshoe shrinkage priors. 

Table~\ref{tab:predictivescoreabo} reports the log predictive scores
\citep{GneitingRaftery2007} obtained from 10-fold cross-validation for the
Aboriginal population subgroup, where larger values indicate better
out-of-sample predictive performance. For each year, the observations are
randomly divided into ten approximately equal-sized folds. For a given fold,
the model is fitted using the observations in the remaining nine folds from
each year, and the fitted model is then used to evaluate the predictive density
of the held-out observations. This procedure is repeated ten times so that each
fold is used once as the validation sample. The log predictive score is then
computed by averaging the log of the predictive densities of all held-out observations across all years. Several conclusions emerge from the table. First, allowing the income distribution parameters to evolve over time generally improves predictive accuracy relative to the independent model. This is evident for the Dagum, Beta 2, and GB2 specifications, for which both the RW and RW-HS models achieve higher log predictive scores than the corresponding independent model. 

Second, among the four candidate income distributions, the GB2 model performs best overall, with the RW-HS specification attaining the highest log predictive score of $-4009.22$, followed very closely by the RW version at $-4009.32$. This suggests that the random walk GB2 model provides the most flexible and accurate representation of the income distributions for this subgroup, while the addition of horseshoe shrinkage yields a further small improvement. Third, the Dagum model also performs competitively, with RW and RW-HS log predictive scores of $-4009.70$ and $-4009.80$, respectively, although both are slightly worse than the corresponding GB2 specifications. 

In contrast, the Beta 2 model performs substantially worse than the other candidates under all three model structures, with log predictive scores around $-4032$ to $-4035$, indicating poor predictive fit. Finally, the Singh--Maddala model shows almost no difference across the independent, RW, and RW-HS specifications, suggesting that introducing temporal dependence or shrinkage provides little benefit for this distribution in the present application. 
Overall, the cross-validation results support selecting the RW-HS GB2 model as the preferred specification for the Aboriginal population subgroup, and it is therefore used as the main model in the discussion that follows.


\begin{table}[H]
\caption{Log predictive scores for the Aboriginal population subgroup obtained from 10-fold cross-validation for model selection. The competing models are the independent income model (ind), the random walk (RW) income model, and the random walk income model with horseshoe priors (RW-HS), each fitted using the Dagum, Singh--Maddala, Beta 2, and generalised Beta 2 (GB2) distributions.\label{tab:predictivescoreabo}}

\centering{}%
\begin{tabular}{cccc}
\hline 
 & ind & RW & RW-HS\tabularnewline
\hline 
Dagum & -4011.62 & -4009.70 & -4009.80\tabularnewline
Beta2 & -4034.68 & -4031.82 & -4031.88\tabularnewline
Singh-Maddala & -4012.63 & -4012.59 & -4012.62\tabularnewline
GB2 & -4012.83 & -4009.32 & -4009.22\tabularnewline
\hline 
\end{tabular}
\end{table}

\begin{figure}[H]
    \centering
    \includegraphics[width=0.8\linewidth]{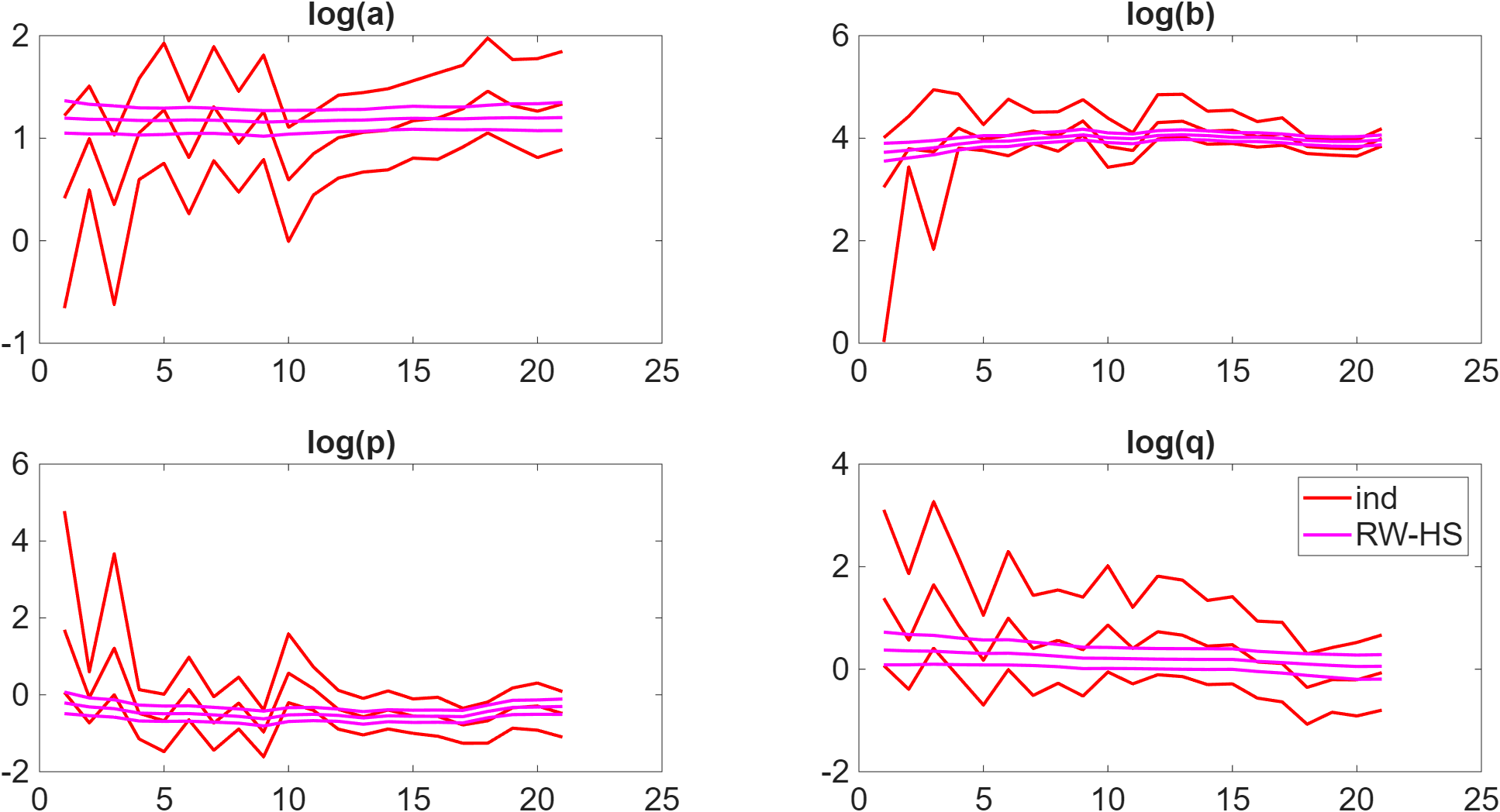}
    \caption{The posterior means (with 95\% credible intervals) of model parameters over time obtained from the independent GB2 income model (ind) and the random walk GB2 income model with horseshoe priors (RW-HS) for the Aboriginal population subgroups.  
}
    \label{fig:Figure_param_GB2_abo}
\end{figure}

\begin{figure}[H]
    \centering
    \includegraphics[width=0.8\linewidth]{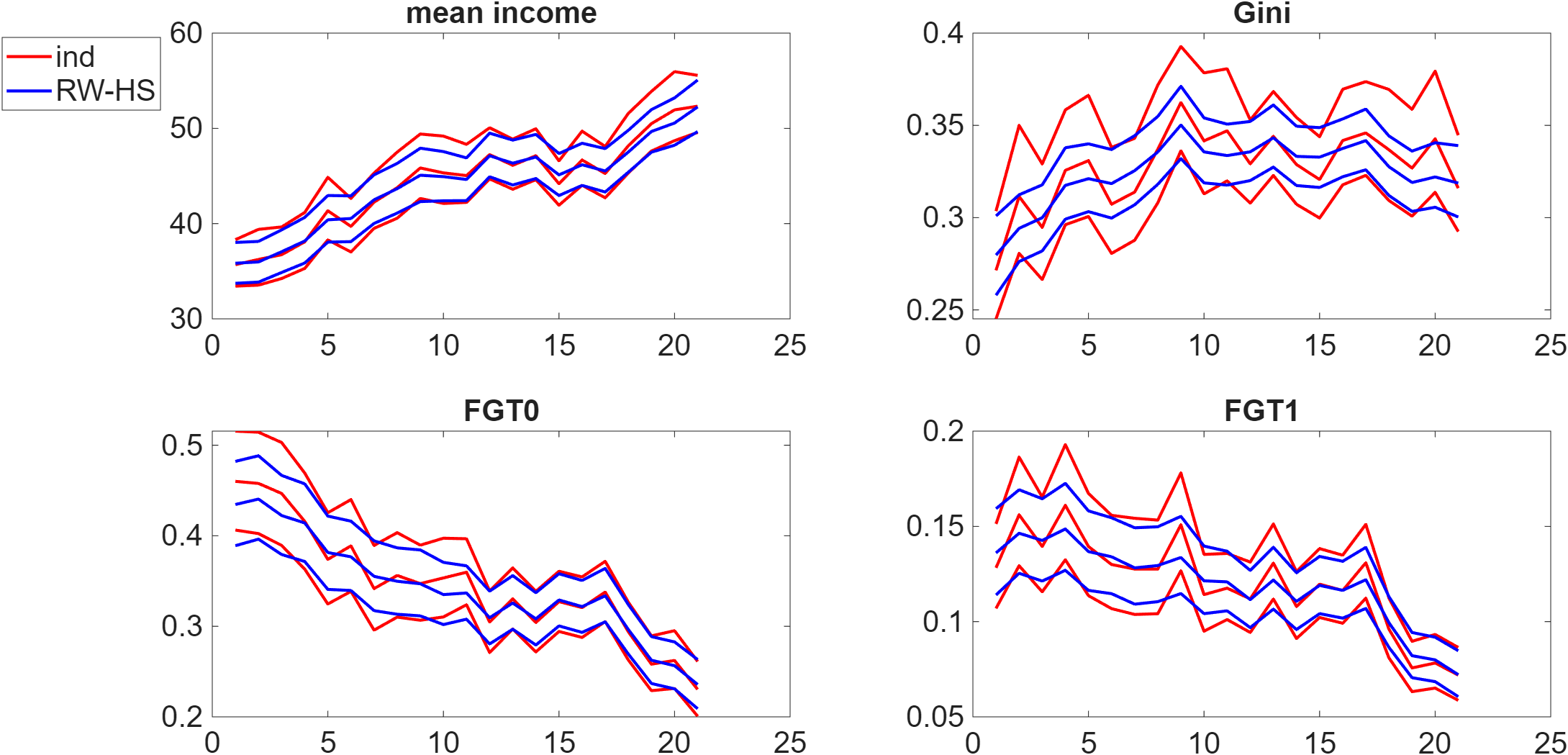}
    \caption{The posterior means (with 95\% credible intervals) of the mean income, Gini index, FGT0, and FGT1 over time obtained from the independent GB2 income model (ind) and the random walk GB2 income model with horseshoe priors (RW-HS) for the Aboriginal population subgroup. 
}
    \label{fig:Figure_GB2_mean_gini_FGT0_FGT1_abo}
\end{figure}

Figures~\ref{fig:Figure_param_GB2_abo} and~\ref{fig:Figure_GB2_mean_gini_FGT0_FGT1_abo} present the estimated GB2 parameters and the corresponding welfare measures over time for the Aboriginal population subgroup under the independent model (ind) and the random walk model with horseshoe shrinkage priors (RW-HS). The most notable difference between the two specifications is that the RW-HS model yields considerably smoother and more stable temporal trajectories, whereas the independent model exhibits substantial year-to-year fluctuations, particularly in the estimates of \(\log(a)\) and \(\log(p)\).  Despite these differences in smoothness, both models imply similar broad distributional trends. In particular, mean income increases over time, especially in the later years of the sample, while the Gini coefficient indicates a moderate rise in inequality in the earlier period, followed by relative stability thereafter. The FGT0 and FGT1 measures both display a general downward trend, pointing to declines in the incidence and intensity of poverty over the study period for the Aboriginal population subgroup. 

Figures~\ref{fig:Figure_PDF_GB2_abo}--\ref{fig:Figure_LC_GB2_abo} in Section \ref{sec:additionalfiguresaboriginalsubgroup} of the online supplement present the posterior means and 99\% credible intervals for the fitted GB2 income PDFs, CDFs, GLCs, and LCs for the Aboriginal population subgroup at time \(t=21\) under the independent model and the random walk model with horseshoe priors. The posterior mean CDFs, LCs, GLCs, and PDFs under the independent model and the random walk model with horseshoe priors are very close to one another, indicating that both models deliver essentially the same overall distributional picture. The differences in posterior means are generally small rather than substantial: the RW-HS LC lies slightly above that of the independent model over most population shares, suggesting marginally lower inequality, while the corresponding GLC is slightly lower, reflecting a somewhat smaller fitted mean income. Likewise, the posterior mean CDFs and PDFs are very similar across the support, with only minor deviations in the lower and middle parts of the income distribution. Importantly, the RW-HS model tends to produce narrower 99\% credible intervals than the independent model, indicating greater estimation precision and a more stable characterisation of uncertainty.


Tables~\ref{tab:FSDGLDLDOVERALL} and~\ref{tab:FSDGLDLDPOOREST} show that the posterior probabilities of Lorenz and stochastic dominance can differ substantially between the independent model and the random walk model with horseshoe priors, especially for intermediate year-to-year comparisons and for the poorest 10\% of the population. For FSD, both models agree that there is essentially no evidence that the 2005 income distribution dominates that of 2001, either overall or for the poorest 10\%, and both models also agree that the strongest welfare improvement occurs from 2015 to 2021, with posterior probabilities close to one overall and effectively one for the poorest 10\%. The main differences arise in the middle of the sample. From 2005 to 2010, the independent model suggests moderate evidence of FSD overall and very strong evidence for the poorest 10\%, whereas RW-HS yields much weaker support. By contrast, from 2010 to 2015, the independent model assigns essentially zero probability to FSD, while RW-HS gives noticeably larger probabilities, particularly for the poorest 10\%, indicating some improvement at the lower end of the distribution. 

The GLD results tell a similar story: there is almost no evidence of welfare improvement from 2001 to 2005, very strong evidence of improvement from 2015 to 2021, and markedly different conclusions across models for 2005--2010 and 2010--2015, with the independent model favouring improvement in the former period and RW-HS giving relatively more support in the latter. For LD, which focuses on inequality comparisons, the evidence is generally weaker at the overall population level, suggesting that changes in welfare were not driven primarily by uniform reductions in inequality across the whole distribution. However, the poorest 10\% display a more nuanced pattern: the independent model suggests strong inequality improvement from 2005 to 2010, RW-HS instead gives stronger support for inequality improvement from 2010 to 2015, and both models indicate very strong inequality improvement for the poorest 10\% from 2015 to 2021. Overall, these results suggest that the most robust welfare gain for the Aboriginal population subgroup occurs between 2015 and 2021.

The probability curves in Figures \ref{fig:yFSDx_prob_abo} to \ref{fig:yLDx_prob_abo} in Section \ref{sec:additionalfiguresaboriginalsubgroup} of the online supplement reinforce these differences between the independent model and RW-HS by showing that the two specifications can imply very different pointwise dominance behaviour, even when the resulting overall dominance probabilities are similar in broad direction. Across the FSD, GLD, and LD panels, the RW-HS curves are generally smoother and less extreme, whereas the independent model often produces sharply varying, and hump-shaped profiles. This is particularly evident for the 2015--2010 comparison, where the independent model yields highly uneven probability curves across the support, while RW-HS gives flatter and more regular profiles. For 2021 versus 2015, by contrast, both models produce probability curves that are close to one throughout most of the support for FSD and GLD, consistent with the strong dominance probabilities reported in the tables. Another important feature is that pointwise dominance probabilities can be high over large parts of the support while the joint dominance probability remains low, because dominance must hold simultaneously at all evaluation points. This is especially clear for the 2005--2001 and several LD comparisons, where the curves may be moderately large over part of the domain but still fail to imply a large overall probability of dominance. 


\begin{table}[h]
\caption{Estimated probabilities of Lorenz dominance, generalised Lorenz dominance, and first-order stochastic dominance for the Aboriginal population subgroup, based on the independent GB2 income model (ind), the random walk GB2 income model (RW), and the random walk GB2 income model with horseshoe priors (RW-HS)\label{tab:FSDGLDLDOVERALL}}

\centering{}%
\begin{tabular}{ccc}
\hline 
 & ind & RW-HS\tabularnewline
\hline 
2005 FSD 2001 & 0.0000 & 0.0057\tabularnewline
2005 GLD 2001 & 0.0000 & 0.0069\tabularnewline
2005 LD 2001 & 0.0000 & 0.0001\tabularnewline
\hline 
2010 FSD 2005 & 0.5277 & 0.4194\tabularnewline
2010 GLD 2005 & 0.8307 & 0.4378\tabularnewline
2010 LD 2005 & 0.1319 & 0.0363\tabularnewline
\hline 
2015 FSD 2010 & 0.0001 & 0.1740\tabularnewline
2015 GLD 2010 & 0.0024 & 0.3274\tabularnewline
2015 LD 2010 & 0.0026 & 0.3042\tabularnewline
\hline 
2021 FSD 2015 & 0.9614 & 0.9865\tabularnewline
2021 GLD 2015 & 0.9938 & 1.0000\tabularnewline
2021 LD 2015 & 0.0912 & 0.1364\tabularnewline
\hline 
\end{tabular}
\end{table}

\begin{table}[h]
\caption{Estimated probabilities of Lorenz dominance, generalised Lorenz dominance, and first-order stochastic dominance over the poorest 10\% of the population for the Aboriginal population subgroup, based on the independent GB2 income model (ind), the random walk GB2 income model (RW), and the random walk GB2 income model with horseshoe priors (RW-HS)\label{tab:FSDGLDLDPOOREST}}

\centering{}%
\begin{tabular}{ccc}
\hline 
 & ind & RW-HS\tabularnewline
\hline 
2005 FSD 2001 & 0.0000 & 0.0057\tabularnewline
2005 GLD 2001 & 0.0000 & 0.0069\tabularnewline
2005 LD 2001 & 0.0000 & 0.0003\tabularnewline
\hline 
2010 FSD 2005 & 0.9667 & 0.4295\tabularnewline
2010 GLD 2005 & 0.9857 & 0.4393\tabularnewline
2010 LD 2005 & 0.9510 & 0.2244\tabularnewline
\hline 
2015 FSD 2010 & 0.0033 & 0.4656\tabularnewline
2015 GLD 2010 & 0.0034 & 0.4820\tabularnewline
2015 LD 2010 & 0.0030 & 0.4881\tabularnewline
\hline 
2021 FSD 2015 & 0.9954 & 1.0000\tabularnewline
2021 GLD 2015 & 0.9938 & 1.0000\tabularnewline
2021 LD 2015 & 0.9756 & 0.9986\tabularnewline
\hline 
\end{tabular}
\end{table}

Figures \ref{fig:pred_density_Abo}--\ref{fig:pred_LC_Abo} in Section \ref{sec:additionalfiguresaboriginalsubgroup} of the online supplement present the posterior predictive distributions for the Aboriginal population subgroup in 2022 and 2025, obtained by projecting the RW-HS GB2 model beyond the observed 2001--2021 period. Figure \ref{fig:pred_density_Abo} shows that the 2025 predictive density is more dispersed than that for 2022, with a noticeably wider range of plausible incomes and a more pronounced upper tail, indicating greater uncertainty about the future shape of the income distribution. This pattern is also reflected in Figure \ref{fig:pred_CDF_Abo}, which shows the predictive CDFs for 2022 and 2025. Figures \ref{fig:pred_GLC_Abo} and \ref{fig:pred_LC_Abo} suggest that the predicted generalised Lorenz and Lorenz curves for 2002 closely resemble those for 2025. Importantly, however, the prediction intervals are clearly wider in 2025 than in 2022 across all four curves, showing that forecast uncertainty accumulates substantially as the prediction horizon moves further away from the sample used for estimation.


Figures \ref{fig:pred_means_Abo}--\ref{fig:pred_FGT1_Abo} in Section \ref{sec:additionalfiguresaboriginalsubgroup} of the online supplement summarise the predictive distributions of key welfare measures for the Aboriginal population subgroup over the out-of-sample period 2022--2025 under the RW-HS GB2 model fitted to the 2001--2021 data. Figure \ref{fig:pred_means_Abo} shows that the predicted mean income remains centred in a broadly similar range across the forecast horizon, although the predictive density becomes progressively flatter and more dispersed from 2022 to 2025, indicating that uncertainty about future mean income increases substantially for the later years. A similar pattern is evident in Figure \ref{fig:pred_GINI_Abo}, where the predictive densities for the Gini coefficient remain concentrated around comparable values, suggesting no dramatic change in relative inequality, but the later-year densities are clearly wider and exhibit more tail mass, so any apparent movement should be interpreted cautiously. The poverty measures in Figures \ref{fig:pred_FGT0_Abo} and \ref{fig:pred_FGT1_Abo} suggest a modest tendency towards lower poverty by 2025, as the predictive mass shifts slightly toward smaller values for both the headcount ratio and the poverty gap. However, these improvements are accompanied by much greater dispersion and longer right tails in the later years, especially for 2024 and 2025, reflecting the build-up of forecast uncertainty as the prediction horizon extends further beyond the observed sample. Overall, the model points to broadly stable or slightly improving welfare outcomes for the Aboriginal subgroup, but the substantially wider predictive densities in the later years make clear that these longer-horizon forecasts are much less precise than those for 2022.

\section{Conclusions\label{sec:conclusions}}

This paper develops a flexible Bayesian time-varying parametric model for income distributions in which the parameters of income distributions are allowed to evolve over time, rather than being estimated separately and independently for each year. By embedding the income distribution parameters in a latent random-walk state process, the proposed framework borrows strength across adjacent years and thereby produces more stable and coherent inference for inequality, poverty, and welfare comparisons. This is especially important for population subgroups with relatively small numbers of observations, where independent year-by-year estimation can generate noisy parameter paths, wider credible intervals, and spurious year-to-year variation in derived welfare measures. In contrast, the proposed model preserves the interpretability and tractability of parametric income modelling while substantially improving estimation precision and allowing coherent prediction of future income distributions and related welfare, inequality, and poverty summaries. The simulation results reinforce these advantages, showing that the dynamic specification tracks the underlying time-varying parameters more closely and delivers tighter uncertainty quantification than the corresponding independent income models, particularly when the true distribution evolves smoothly over time.
A further contribution of the paper is the use of horseshoe shrinkage priors within the dynamic income modelling framework. The horseshoe prior provides smoother temporal trajectories for the income distribution parameters and for the associated inequality and poverty measures. 

In the empirical application, this benefit is clearly visible for both the Aboriginal population subgroup and the Australian Capital Territory (ACT) subgroup, where the random walk GB2 model with horseshoe priors is selected by 10-fold cross-validation as the preferred specification. For the Aboriginal subgroup, the results indicate rising mean income over time, a moderate increase in inequality in the earlier years followed by relative stability, and an overall decline in the incidence and intensity of poverty. The dominance analysis further shows that the strongest and most robust welfare improvement occurs between 2015 and 2021, particularly for the poorest 10\% of the population. For the ACT subgroup, the preferred model similarly yields smoother parameter estimates and more stable welfare summaries, with mean income generally increasing over time, inequality rising in the earlier part of the sample and then stabilising or declining slightly, and poverty measures exhibiting an overall downward trend. The dominance results suggest that the clearest welfare gains for the ACT subgroup also occur in the later part of the sample, especially at the lower end of the income distribution. Overall, the findings demonstrate that the proposed time-varying income model with horseshoe shrinkage priors provides a practical and useful alternative to independent income models for analysing changes in income distributions over time. For longer time series, this framework could be extended further by using dynamic shrinkage priors, which can capture both smooth parameter evolution and occasional abrupt changes or jumps in the income distribution parameters \citep{kowal2019dynamic,knaus2023dynamic}.

\renewcommand{\thealgorithm}{S\arabic{algorithm}}
\renewcommand{\theequation}{S\arabic{equation}}
\renewcommand{\thesection}{S\arabic{section}}
\renewcommand{\thepage}{S\arabic{page}}
\renewcommand{\thetable}{S\arabic{table}}
\renewcommand{\thefigure}{S\arabic{figure}}
\setcounter{page}{1}
\setcounter{section}{0}
\setcounter{equation}{0}
\setcounter{algorithm}{0}
\setcounter{table}{0}
\setcounter{figure}{0}

\pagebreak

\section*{Online Supplement for ``Flexible Bayesian Models for Time-Varying Income Distributions''}	

We use the following notation in the online supplement. Equation (1), Algorithm~1,
Section~1, etc, refer to the main paper, while Equation (S1),
Algorithm~S1, Section~S1, etc, refer to the supplement. All the acronyms used without definition in the supplement, are defined in the main paper.

Section~\ref{sec:incomedistributions} describes several flexible parametric models for income data. 
Section~\ref{sec:additionaldetailrandomwalkincome} discusses additional details for the random walk income models.
Section~\ref{sec:additionaldetailshorseshoe} describes additional details for the random walk income models with horseshoe priors.
Section~\ref{sec:additionaldetailLorenzstochastic} describes additional details for computing posterior probabilities of Lorenz and stochastic dominance.
Section~\ref{sec:additionalfiguresimulationstudy} provides additional results for the simulation study discussed in Section \ref{sec:simulationstudy}.
Section~\ref{sec:additionalfiguresaboriginalsubgroup} provides additional results for the Aboriginal population subgroup. Section~\ref{sec:ACTpopulationsubgroup} discusses empirical results for the ACT population subgroup.

\section{Parametric income distributions\label{sec:incomedistributions}}

In this section, we consider several flexible parametric models for positive income data, namely the GB2 family and three important special cases: the Singh--Maddala (Burr XII), Dagum (Burr III), and Beta 2 distributions. For each model, we collect the distribution parameters into the {parameter vector} $\boldsymbol{\phi}$ and write the marginal income density generically as $p_Y(y\mid \boldsymbol{\phi})$ for $y>0$. In this section, we suppress the subscript $t$ for ease of notation.

Since the Singh--Maddala and Dagum distributions have been shown to fit income data well \citep{Dagum1977,SinghMaddala1976}, they are natural choices for income densities. Both belong to the Burr family of distributions (Burr, 1942) and can be viewed as special cases of the Generalised Beta distribution of the second kind (GB2) introduced by \citet{McDonald1984}. The Beta 2 distribution (also known as the Beta-prime distribution) is another closely related member of this family and is widely used for modelling income distributions.

Let $Y$ denote income on $(0,\infty)$ with density $p_Y(y\mid\boldsymbol{\phi})$ and cumulative distribution
function $F_Y(y\mid\boldsymbol{\phi})$. For any real $k$ such that $\mu^{(k)}=\mathbb{E}(Y^k)$ exists,
define the {$k$th moment distribution function} by
\begin{equation}
F_Y^{(k)}(y|\boldsymbol{\phi})
=
\frac{1}{\mu^{(k)}}\int_{0}^{y} t^{k} p_Y(t|\boldsymbol{\phi})\,dt,
\qquad y>0.
\label{eq:mdf_def}
\end{equation}
In particular, $F_Y^{(0)}(y|\boldsymbol{\phi})=F_Y(y;\boldsymbol{\phi})$ is the cdf and
$F_Y^{(1)}(y;\boldsymbol{\phi})$ is the {first moment distribution function}.

Let $\mathrm{B}(p,q)=\int_{0}^{1} t^{p-1}(1-t)^{q-1}\,dt$ be the beta function and define the
{regularised incomplete beta function} (i.e.\ the Beta$(p,q)$ cdf) by
\begin{equation}
\mathrm{B}_{x}(p,q)
=
\frac{1}{\mathrm{B}(p,q)}\int_{0}^{x} t^{p-1}(1-t)^{q-1}\,dt,
\qquad 0\le x \le 1.
\label{eq:reg_inc_beta}
\end{equation}

\noindent For a given population share $u\in[0,1]$, the Lorenz curve can be written as
\begin{equation}
L(u;\boldsymbol{\phi})
=
F_Y^{(1)}\!\left(F_Y^{-1}(u;\boldsymbol{\phi});\boldsymbol{\phi}\right),
\qquad 0\le u\le 1,
\label{eq:lorenz_quantile_theta}
\end{equation}
where $F_Y^{-1}(u;\boldsymbol{\phi})$ denotes the quantile function. Intuitively,
$L(u;\boldsymbol{\phi})$ is the share of total income received by the poorest $100u\%$
of the population. The generalised Lorenz curve is obtained by multiplying Lorenz curves by the mean $\mu$.

\subsubsection*{GB2 distribution}

For the GB2 distribution, the {parameter vector} is
\[
\boldsymbol{\phi}=(a,b,p,q)^\top.
\]
Its density is
\begin{equation}
p_Y(y\mid \boldsymbol{\phi})
=
\frac{a\,y^{ap-1}}
{b^{ap}B(p,q)\,[1+(y/b)^a]^{p+q}},
\qquad y>0,
\end{equation}
where $b>0$ is a scale parameter and $a>0$, $p>0$, and $q>0$ are shape parameters.

The cumulative distribution function can be written in terms of the regularised incomplete beta function as
\begin{equation}
F_Y(y\mid \boldsymbol{\phi})
=
\frac{1}{B(p,q)}
\int_{0}^{w(y)} t^{p-1}(1-t)^{q-1}\,dt
=
B_{\,w(y)}(p,q),
\qquad y>0,
\end{equation}
where
\begin{equation}
w(y)=\frac{(y/b)^a}{1+(y/b)^a},
\qquad 0<w(y)<1.
\label{eq:gb2_wy}
\end{equation}

\noindent The mean exists when $q>1/a$, and is given by
\begin{equation}
\mu
=
b\,\frac{B\!\left(p+\frac{1}{a},\,q-\frac{1}{a}\right)}{B(p,q)}
=
b\,\frac{\Gamma\!\left(p+\frac{1}{a}\right)\Gamma\!\left(q-\frac{1}{a}\right)}
{\Gamma(p)\Gamma(q)}.
\label{eq:gb2_mean}
\end{equation}
A nonzero mode exists when $ap>1$, in which case
\begin{equation}
m
=
b\left(\frac{ap-1}{aq+1}\right)^{1/a}.
\label{eq:gb2_mode}
\end{equation}

\noindent More generally, for $k<aq$, the $k$th moment is
\begin{equation}
\mu^{(k)}
=
E(Y^k)
=
b^k\,
\frac{B\!\left(p+\frac{k}{a},\,q-\frac{k}{a}\right)}{B(p,q)}.
\label{eq:gb2_moment}
\end{equation}
Hence the $k$th moment distribution functions are
\begin{equation}
F_Y^{(k)}(y\mid \boldsymbol{\phi})
=
\mathrm{B}_{w(y)}\!\left(p+\frac{k}{a},\,q-\frac{k}{a}\right),
\qquad
F_Y^{(1)}(y\mid \boldsymbol{\phi})
=
\mathrm{B}_{w(y)}\!\left(p+\frac{1}{a},\,q-\frac{1}{a}\right),
\label{eq:gb2_mdf}
\end{equation}
where $\mathrm{B}_{x}(\alpha,\beta)$ denotes the regularised incomplete beta function.

The quantile function may be written as
\begin{equation}
F_Y^{-1}(u\mid \boldsymbol{\phi})
=
b\left(
\frac{\mathrm{B}^{-1}_{u}(p,q)}
{1-\mathrm{B}^{-1}_{u}(p,q)}
\right)^{1/a},
\qquad 0<u<1,
\label{eq:gb2_quantile}
\end{equation}
where $\mathrm{B}^{-1}_{u}(p,q)$ denotes the inverse of the regularised incomplete beta function. Thus, the GB2 distribution provides a very flexible four-parameter family that nests several important income distributions, including the Singh--Maddala, Dagum, and Beta 2 models, as special cases.

\subsubsection*{Singh--Maddala distribution (Burr XII)}
For the Singh--Maddala distribution, the {parameter vector} is $\boldsymbol{\phi}=(a,b,q)^\top$. Its density can be written as
\begin{equation}
p_Y(y\mid \boldsymbol{\phi})=\frac{a q\, y^{a-1}}{b^{a}[1+(y/b)^a]^{1+q}}, \qquad y>0,
\end{equation}
with cumulative distribution function
\begin{equation}
F_Y(y\mid \boldsymbol{\phi})=1-\left[1+(y/b)^a\right]^{-q}, \qquad y>0,
\end{equation}
where $a$, $b$, and $q$ are strictly positive. The mean and mode are
\begin{equation}
\mu=\frac{b\,\Gamma(1+1/a)\Gamma(q-1/a)}{\Gamma(q)}, \qquad 
m=b\left(\frac{a-1}{aq+1}\right)^{1/a},
\end{equation}
with $a>1$ required for the mode to exist (and $q>1/a$ for the mean to exist). 

\noindent For $k<aq$, the moment distribution functions are
\begin{equation}
F_Y^{(k)}(y\mid\boldsymbol{\phi})
=
\mathrm{B}_{w(y)}\!\left(1+\frac{k}{a},\,q-\frac{k}{a}\right),
\qquad
F_Y^{(1)}(y\mid\boldsymbol{\phi})
=
\mathrm{B}_{w(y)}\!\left(1+\frac{1}{a},\,q-\frac{1}{a}\right).
\label{eq:sm_mdf}
\end{equation}

\noindent A convenient quantile function is
\begin{equation}
F_Y^{-1}(u\mid\boldsymbol{\phi})
=
b\left[(1-u)^{-1/q}-1\right]^{1/a},
\qquad 0<u<1.
\label{eq:sm_quantile}
\end{equation}

\subsubsection*{Dagum distribution (Burr III)}
For the Dagum distribution, the {parameter vector} is $\boldsymbol{\phi}=(a,b,p)^\top$. Its density is
\begin{equation}
p_Y(y\mid \boldsymbol{\phi})=\frac{a p\, y^{ap-1}}{b^{ap}[1+(y/b)^a]^{p+1}}, \qquad y>0,
\end{equation}
and its cumulative distribution function is
\begin{equation}
F_Y(y\mid \boldsymbol{\phi})=\left[1+(y/b)^{-a}\right]^{-p}, \qquad y>0,
\end{equation}
where $a$, $b$, and $p$ are strictly positive. The mean and mode can be written as
\begin{equation}
\mu=b\,\frac{\Gamma(p+1/a)\Gamma(1-1/a)}{\Gamma(p)}, \qquad 
m=b\left(\frac{ap-1}{a+1}\right)^{1/a},
\end{equation}
with $ap>1$ required for the mode to exist (and $a>1$ for the mean to exist). 
For $k<a$, the moment distribution functions are
\begin{equation}
F_Y^{(k)}(y\mid\boldsymbol{\phi})
=
\mathrm{B}_{w(y)}\!\left(p+\frac{k}{a},\,1-\frac{k}{a}\right),
\qquad
F_Y^{(1)}(y\mid\boldsymbol{\phi})
=
\mathrm{B}_{w(y)}\!\left(p+\frac{1}{a},\,1-\frac{1}{a}\right).
\label{eq:dagum_mdf}
\end{equation}

The quantile function is
\begin{equation}
F_Y^{-1}(u\mid\boldsymbol{\phi})
=
b\left(u^{-1/p}-1\right)^{-1/a},
\qquad 0<u<1.
\label{eq:dagum_quantile}
\end{equation}

\subsubsection*{Beta 2 distribution}
For the beta 2 distribution, the {parameter vector} is
$\boldsymbol{\phi}=(b,p,q)^\top$. Its density is
\begin{equation}
p_Y(y\mid \boldsymbol{\phi})
=
\frac{y^{p-1}}{b^{p}B(p,q)\,[1+y/b]^{p+q}},
\qquad y>0,
\end{equation}
where $b>0$, $p>0$, and $q>0$. A nonzero mode requires $p>1$, in which case
\begin{equation}
{m}
=
\frac{b(p-1)}{q+1}.
\end{equation}
The mean exists when $q>1$, in which case
\begin{equation}
\mu=\frac{b p}{q-1}.
\end{equation}
The cumulative distribution function can be expressed using the regularised incomplete beta function:
\begin{equation}
F_Y(y\mid \boldsymbol{\phi})=\frac{1}{B(p,q)}\int_{0}^{y/(b+y)} t^{p-1}(1-t)^{q-1}\,dt
= B_{\,y/(b+y)}(p,q),
\end{equation}
where $B_x(p,q)$ denotes the regularised incomplete beta function, which is readily computed in standard statistical software. For $k<q$, the corresponding moment distribution functions are
\begin{equation}
F_Y^{(k)}(y\mid\boldsymbol{\phi})
=
\mathrm{B}_{u(y)}(p+k,\,q-k),
\qquad
u(y)=\frac{y}{b+y},
\label{eq:beta2_mdf}
\end{equation}
provided that $q>k$. In particular, for $k=1$,
\begin{equation}
F_Y^{(1)}(y\mid\boldsymbol{\phi})
=
\mathrm{B}_{u(y)}(p+1,\,q-1).
\end{equation}

\section{Additional details for the random walk income model\label{sec:additionaldetailrandomwalkincome}}
This section provides additional details for the random walk income model. 
Let
\[
\Delta \theta_{k,t}=\theta_{k,t}-\theta_{k,t-1},
\qquad t=2,\ldots,T,
\]
and define
\[
S_k=\sum_{t=2}^T (\Delta \theta_{k,t})^2,
\qquad k=1,\ldots,d.
\]
Under the half-Cauchy prior on each $\sigma_k$, the conditional density of $\sigma_k^2$ is proportional to an inverse-gamma kernel multiplied by the half-Cauchy correction term. We generate a proposal
\[
\sigma_k^{2\star}\sim \mathrm{IG}(\alpha_k,\beta_k),
\]
where
\[
\alpha_k=\frac{T-2}{2},
\qquad
\beta_k=\frac{S_k}{2},
\]
and accept with probability
\[
\alpha_{\sigma_k^2}
=
\min\left\{
1,\,
\frac{1+\sigma_k^2}{1+\sigma_k^{2\star}}
\right\}.
\]
Equivalently, this step targets
\[
p(\sigma_k^2\mid \boldsymbol\theta_{1:T})
\propto
\mathrm{IG}(\sigma_k^2\mid \alpha_k,\beta_k)\,
\frac{1}{1+\sigma_k^2}.
\]

\section{Additional details for the random walk income model with horseshoe shrinkage priors\label{sec:additionaldetailshorseshoe}}
This section provides additional details for the random walk income model with horseshoe priors.

Assume that $\phi_{1,1},\ldots,\phi_{d,1}>0$ and let
\[
\boldsymbol{\theta}_1
=
\bigl(\log(\phi_{1,1}),\ldots,\log(\phi_{d,1})\bigr)^\top,
\]
with
\[
\phi_{1,1},\ldots,\phi_{d,1}
\stackrel{\mathrm{ind}}{\sim}
\mathrm{Half\text{-}Cauchy}(0,1).
\]
Under this log-transformation, the induced prior density of the initial state
$\boldsymbol{\theta}_1$ is given by
\begin{equation}
p(\boldsymbol{\theta}_1)
=
\prod_{k=1}^d
\frac{2\exp(\theta_{k,1})}{\pi\bigl(1+\exp(2\theta_{k,1})\bigr)},
\label{eq:initialstate}
\end{equation}
where $\phi_{k,1}=\exp(\theta_{k,1})$, $k=1,\ldots,d$.

For posterior computation, define the increments
\[
\Delta \theta_{k,t} = \theta_{k,t}-\theta_{k,t-1},
\qquad t=2,\ldots,T,\quad k=1,\ldots,d,
\]
and let
\[
S_k = \sum_{t=2}^T (\Delta \theta_{k,t})^2,
\qquad k=1,\ldots,d.
\]
Let $n=T-1$ denote the number of state increments. Under the inverse-gamma representation proposed by \citet{makalic2015simple}, the full conditional distributions of the shrinkage parameters are
\[
\lambda_k^2 \mid \cdot
\sim
\mathrm{IG}\!\left(\frac{n+1}{2},\,\frac{1}{\nu_k}+\frac{S_k}{2\tau^2}\right),
\qquad k=1,\ldots,d,
\]
\[
\nu_k \mid \cdot
\sim
\mathrm{IG}\!\left(1,\,1+\frac{1}{\lambda_k^2}\right),
\qquad k=1,\ldots,d,
\]
\[
\tau^2 \mid \cdot
\sim
\mathrm{IG}\!\left(\frac{nd+1}{2},\,\frac{1}{\xi}+\sum_{k=1}^d \frac{S_k}{2\lambda_k^2}\right),
\]
and
\[
\xi \mid \cdot
\sim
\mathrm{IG}\!\left(1,\,1+\frac{1}{\tau^2}\right).
\]
These closed-form updates can be combined with Metropolis-within-Gibbs updates for the latent states $\boldsymbol\theta_{1:T}$. In practice, the horseshoe structure yields smoother temporal trajectories than the independent model, while remaining flexible enough to capture genuine changes in the income distribution over time.

The state updates are similar to the baseline random walk model, except that the covariance matrix $\mathbf Q$ is replaced by
\[
\mathbf Q_{\mathrm{HS}}
=
\tau^2\mathbf\Lambda,
\qquad
\mathbf\Lambda=\mathrm{diag}(\lambda_1^2,\ldots,\lambda_d^2).
\]
Define
\[
\log \pi_t^{\mathrm{RW\mbox{-}HS}}(\boldsymbol\theta_t)=
\begin{cases}
\ell_1(\boldsymbol\theta_1)
+\log p(\boldsymbol\theta_1)
+\log \mathcal N(\boldsymbol\theta_2\mid \boldsymbol\theta_1,\mathbf Q_{\mathrm{HS}}),
& t=1,
\\[0.6em]
\ell_t(\boldsymbol\theta_t)
+\log \mathcal N(\boldsymbol\theta_t\mid \boldsymbol\theta_{t-1},\mathbf Q_{\mathrm{HS}})
+\log \mathcal N(\boldsymbol\theta_{t+1}\mid \boldsymbol\theta_t,\mathbf Q_{\mathrm{HS}}),
& 1<t<T,
\\[0.6em]
\ell_T(\boldsymbol\theta_T)
+\log \mathcal N(\boldsymbol\theta_T\mid \boldsymbol\theta_{T-1},\mathbf Q_{\mathrm{HS}}),
& t=T.
\end{cases}
\]
The acceptance probability is then
\[
\alpha_t
=
\min\Bigl\{1,
\exp\bigl(
\log \pi_t^{\mathrm{RW\mbox{-}HS}}(\boldsymbol\theta_t^\star)
-
\log \pi_t^{\mathrm{RW\mbox{-}HS}}(\boldsymbol\theta_t)
\bigr)
\Bigr\}.
\]

These updates show the key computational advantage of the horseshoe representation: conditional on the latent states, all shrinkage parameters can be sampled by direct Gibbs steps, so the only Metropolis-Hastings moves are those for the latent states themselves.

\begin{algorithm}[H]
\caption{Metropolis-within-Gibbs sampler for the random walk income model with horseshoe shrinkage priors}
\label{alg:mwg_rw_hs_income_model}
\begin{algorithmic}[1]
\Require Data $\{\mathbf y_t\}_{t=1}^T$, initial values $(\boldsymbol\theta_{1:T}^{(0)},\tau^{2(0)},\boldsymbol\lambda^{2(0)},\xi^{(0)},\boldsymbol\nu^{(0)})$
\For{$m=1,\ldots,M$}
    \State \textbf{(A) Update latent states $\boldsymbol\theta_{1:T}$}
    \State Set $\mathbf\Lambda^{(m-1)}=\mathrm{diag}(\lambda_1^{2(m-1)},\ldots,\lambda_d^{2(m-1)})$
    \State Set $\mathbf Q_{\mathrm{HS}}^{(m-1)}=\tau^{2(m-1)}\mathbf\Lambda^{(m-1)}$
    \For{$t=1,\ldots,T$}
        \State Propose $\boldsymbol\theta_t^\star \sim \mathcal N\!\left(\boldsymbol\theta_t^{(m-1)},\,\kappa_t\boldsymbol\Sigma_t\right)$
        \State Compute $\log \pi_t^{\mathrm{RW\mbox{-}HS}}(\boldsymbol\theta_t^\star)$ and $\log \pi_t^{\mathrm{RW\mbox{-}HS}}(\boldsymbol\theta_t^{(m-1)})$
        \State Set
        \[
        \alpha_t=
        \min\left\{
        1,\exp\!\left(
        \log \pi_t^{\mathrm{RW\mbox{-}HS}}(\boldsymbol\theta_t^\star)
        -
        \log \pi_t^{\mathrm{RW\mbox{-}HS}}(\boldsymbol\theta_t^{(m-1)})
        \right)
        \right\}
        \]
        \State With probability $\alpha_t$, set $\boldsymbol\theta_t^{(m)}=\boldsymbol\theta_t^\star$; otherwise set $\boldsymbol\theta_t^{(m)}=\boldsymbol\theta_t^{(m-1)}$
    \EndFor

    \State \textbf{(B) Compute state increment sums of squares}
    \For{$k=1,\ldots,d$}
        \State Compute $S_k=\sum_{t=2}^T (\theta_{k,t}^{(m)}-\theta_{k,t-1}^{(m)})^2$
    \EndFor

    \State \textbf{(C) Update local shrinkage parameters}
    \For{$k=1,\ldots,d$}
        \State Draw
        \[
        \lambda_k^{2(m)}
        \sim
        \mathrm{IG}\!\left(
        \frac{T}{2},
        \frac{1}{\nu_k^{(m-1)}}+\frac{S_k}{2\tau^{2(m-1)}}
        \right)
        \]
        \State Draw
        \[
        \nu_k^{(m)}
        \sim
        \mathrm{IG}\!\left(
        1,\,
        1+\frac{1}{\lambda_k^{2(m)}}
        \right)
        \]
    \EndFor

    \State \textbf{(D) Update global shrinkage parameters}
    \State Draw
    \[
    \tau^{2(m)}
    \sim
    \mathrm{IG}\!\left(
    \frac{(T-1)d+1}{2},
    \frac{1}{\xi^{(m-1)}}+\sum_{k=1}^d \frac{S_k}{2\lambda_k^{2(m)}}
    \right)
    \]
    \State Draw
    \[
    \xi^{(m)}
    \sim
    \mathrm{IG}\!\left(
    1,\,
    1+\frac{1}{\tau^{2(m)}}
    \right)
    \]

    \State Store $\bigl(\boldsymbol\theta_{1:T}^{(m)},\tau^{2(m)},\boldsymbol\lambda^{2(m)},\xi^{(m)},\boldsymbol\nu^{(m)}\bigr)$
\EndFor
\end{algorithmic}
\end{algorithm}

\section{Additional details for Lorenz and stochastic dominance \label{sec:additionaldetailLorenzstochastic}}

This section provides additional details to compute the posterior probabilities of Lorenz and stochastic dominance proposed by \citet{lander2020bayesian}.

Having obtained $M$ MCMC draws from the posterior distributions of the two income distributions at time $t$,
say
\[
\boldsymbol{\phi}_{A,t}^{(1)},\ldots,\boldsymbol{\phi}_{A,t}^{(M)}
\qquad\text{and}\qquad
\boldsymbol{\phi}_{B,t}^{(1)},\ldots,\boldsymbol{\phi}_{B,t}^{(M)},
\]
we can compute, for each draw and each $u$, the corresponding Lorenz curves
$L_{A,t}(u;\boldsymbol{\phi}_{A,t}^{(m)})$ and $L_{B,t}(u;\boldsymbol{\phi}_{B,t}^{(m)})$, generalised Lorenz
curves
\[
\mathrm{GL}_{A,t}(u;\boldsymbol{\phi}_{A,t}^{(m)})
=
\mu_{A,t}\,L_{A,t}(u;\boldsymbol{\phi}_{A,t}^{(m)}),
\qquad
\mathrm{GL}_{B,t}(u;\boldsymbol{\phi}_{B,t}^{(m)})
=
\mu_{B,t}\,L_{B,t}(u;\boldsymbol{\phi}_{B,t}^{(m)}),
\]
and quantile functions
\[
F_{Y_{A,t}}^{-1}(u\mid\boldsymbol{\phi}_{A,t}^{(m)})
\qquad\text{and}\qquad
F_{Y_{B,t}}^{-1}(u\mid\boldsymbol{\phi}_{B,t}^{(m)}).
\]

Let $C_t(u;\boldsymbol{\phi}_t)$ denote a generic curve at time $t$, representing any one of the following:
the Lorenz curve $L_t(u;\boldsymbol{\phi}_t)$, the generalised Lorenz curve
$\mathrm{GL}_t(u;\boldsymbol{\phi}_t)$, or the quantile function
$F_t^{-1}(u\mid\boldsymbol{\phi}_t)$. Write $C_{A,t}(u;\boldsymbol{\phi}_{A,t})$ and
$C_{B,t}(u;\boldsymbol{\phi}_{B,t})$ for the pair of curves to be compared. For each
$u$ and each draw $m=1,\ldots,M$, let
\[
C_{A,t}^{(m)}(u)=C_t(u;\boldsymbol{\phi}_{A,t}^{(m)}),
\qquad
C_{B,t}^{(m)}(u)=C_t(u;\boldsymbol{\phi}_{B,t}^{(m)}).
\]
We evaluate these curves on the dense grid
\[
u\in\{0.001,0.002,\ldots,0.999\},
\qquad
u_i=0.001\,i,\quad i=1,\ldots,999.
\]

Following \citet{gunawan2021posterior}, we estimate the posterior probability that distribution $A$
dominates distribution $B$ as the proportion of MCMC draws for which
$C_{A,t}^{(m)}(u)\ge C_{B,t}^{(m)}(u)$ holds for {all} grid points $u$. Let
$\mathbb{I}[\cdot]$ denote the indicator function. Then
\[
P(A\ \text{dominates}\ B)
=
\frac{1}{M}\sum_{m=1}^M \prod_{i=1}^{999}
\mathbb{I}\!\left[C_{A,t}^{(m)}(u_i)\ge C_{B,t}^{(m)}(u_i)\right],
\]
\[
P(B\ \text{dominates}\ A)
=
\frac{1}{M}\sum_{m=1}^M \prod_{i=1}^{999}
\mathbb{I}\!\left[C_{B,t}^{(m)}(u_i)\ge C_{A,t}^{(m)}(u_i)\right],
\]
and
\[
P(\text{neither distribution dominates})
=
1-P(A\ \text{dominates}\ B)-P(B\ \text{dominates}\ A).
\]



A by-product of this procedure is the {probability curve}
\[
P_{AB}(u)
=
\frac{1}{M}\sum_{m=1}^M
\mathbb{I}\!\left[C_{A,t}^{(m)}(u)\ge C_{B,t}^{(m)}(u)\right],
\]
plotted against $u$. These curves give the posterior probability of pointwise dominance at
each population share $u$. Over any range of $u$, the probability of dominance on that
range can be no larger than the minimum value of $P_{AB}(u)$ within the range. This makes
$P_{AB}(u)$ a useful device for identifying which parts of the distribution, such as the tails
or the middle, drive the overall dominance probability. In particular, if dominance is largely
determined by tail behaviour, one can assess sensitivity by omitting extreme values of $u$.
Likewise, if interest focuses on a particular segment of the population, such as the poor,
one can examine how the dominance probability changes when attention is restricted to the
corresponding range of $u$.

\section{Additional figures for the simulation study}\label{sec:additionalfiguresimulationstudy}
This section provides additional figures for the simulation study discussed in Section \ref{sec:simulationstudy}. 

\begin{figure}[H]
    \centering
    \includegraphics[width=0.8\linewidth]{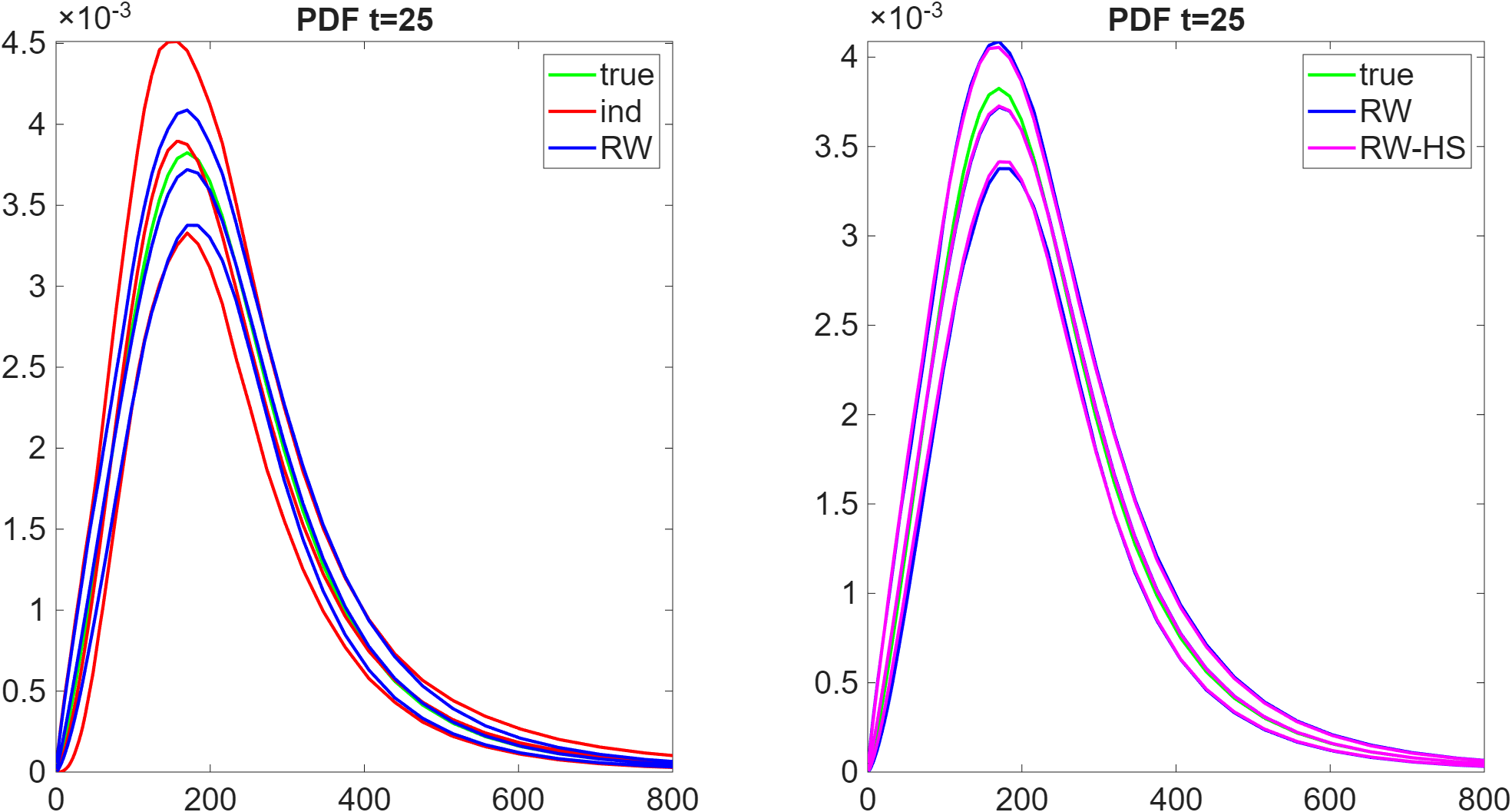}
    \caption{The posterior means (with 99\% credible intervals) of the income PDF at $t=25$ obtained from the independent Dagum income model (ind), the random walk Dagum income model (RW), and the random walk Dagum income model with horseshoe priors (RW-HS) for the simulated dataset. The true PDF is also plotted. 
}
    \label{fig:Figure_PDF_sim1}
\end{figure}

\begin{figure}[H]
    \centering
    \includegraphics[width=0.8\linewidth]{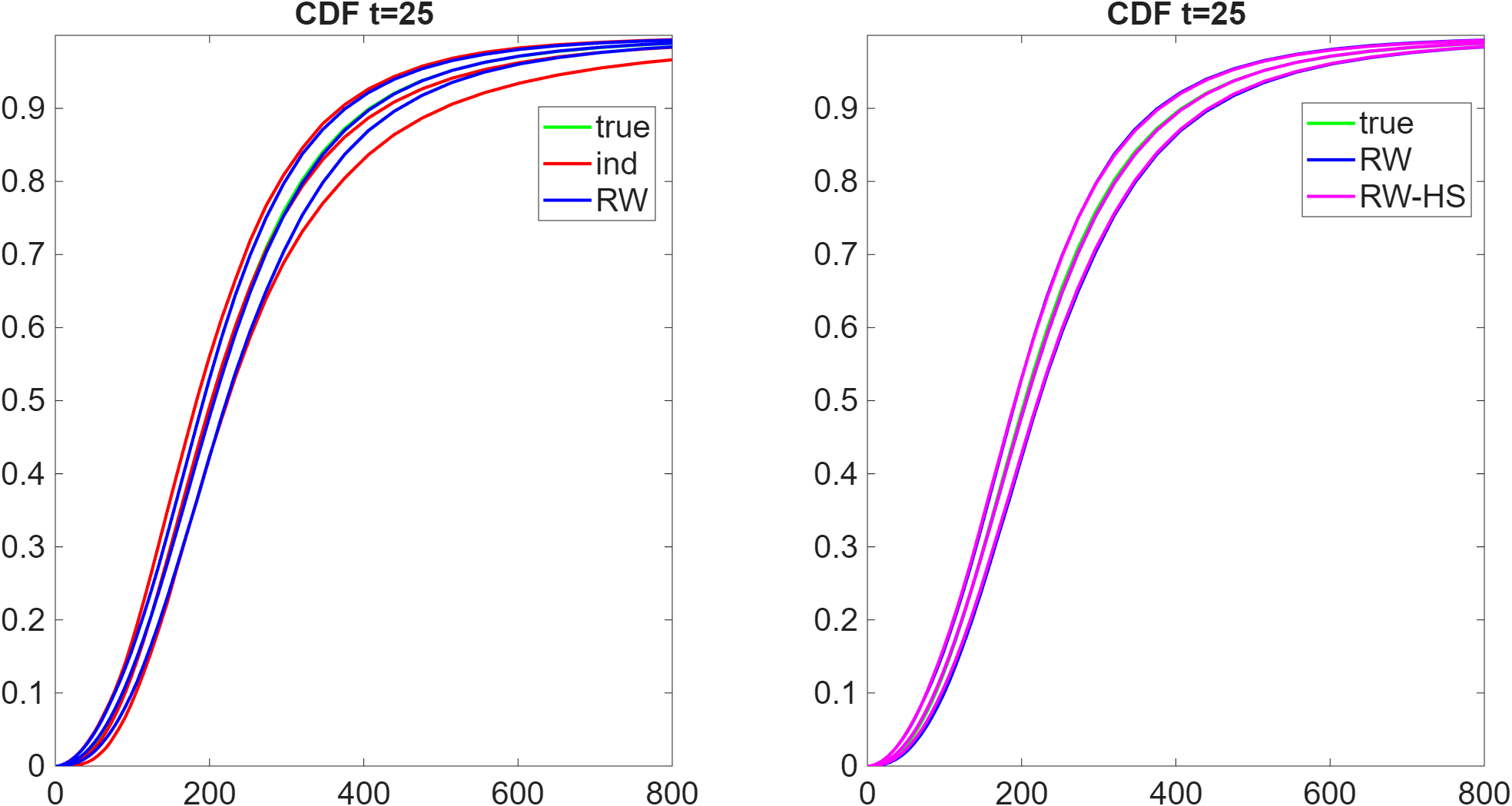}
    \caption{The posterior means (with 99\% credible intervals) of the income CDF at $t=25$ obtained from the independent Dagum income model (ind), the random walk Dagum income model (RW), and the random walk Dagum income model with horseshoe priors (RW-HS) for the simulated dataset. The true CDF is also plotted. 
}
    \label{fig:Figure_CDF_sim1}
\end{figure}

The PDF and CDF plots in Figures \ref{fig:Figure_PDF_sim1} and \ref{fig:Figure_CDF_sim1} (with 99\% credible intervals) show that all models capture the shape of the true density and CDF. The independent model produces wide bands and large variations in peak height and location, indicating that year-by-year estimation with $n=250$ can translate sampling noise into spurious changes in the density and CDF. The RW and RW-HS models yield narrower credible bands and posterior means that are closer to the true density curve. The RW-HS model provides slightly tighter credible intervals than the RW model.

\begin{figure}[H]
    \centering
    \includegraphics[width=0.8\linewidth]{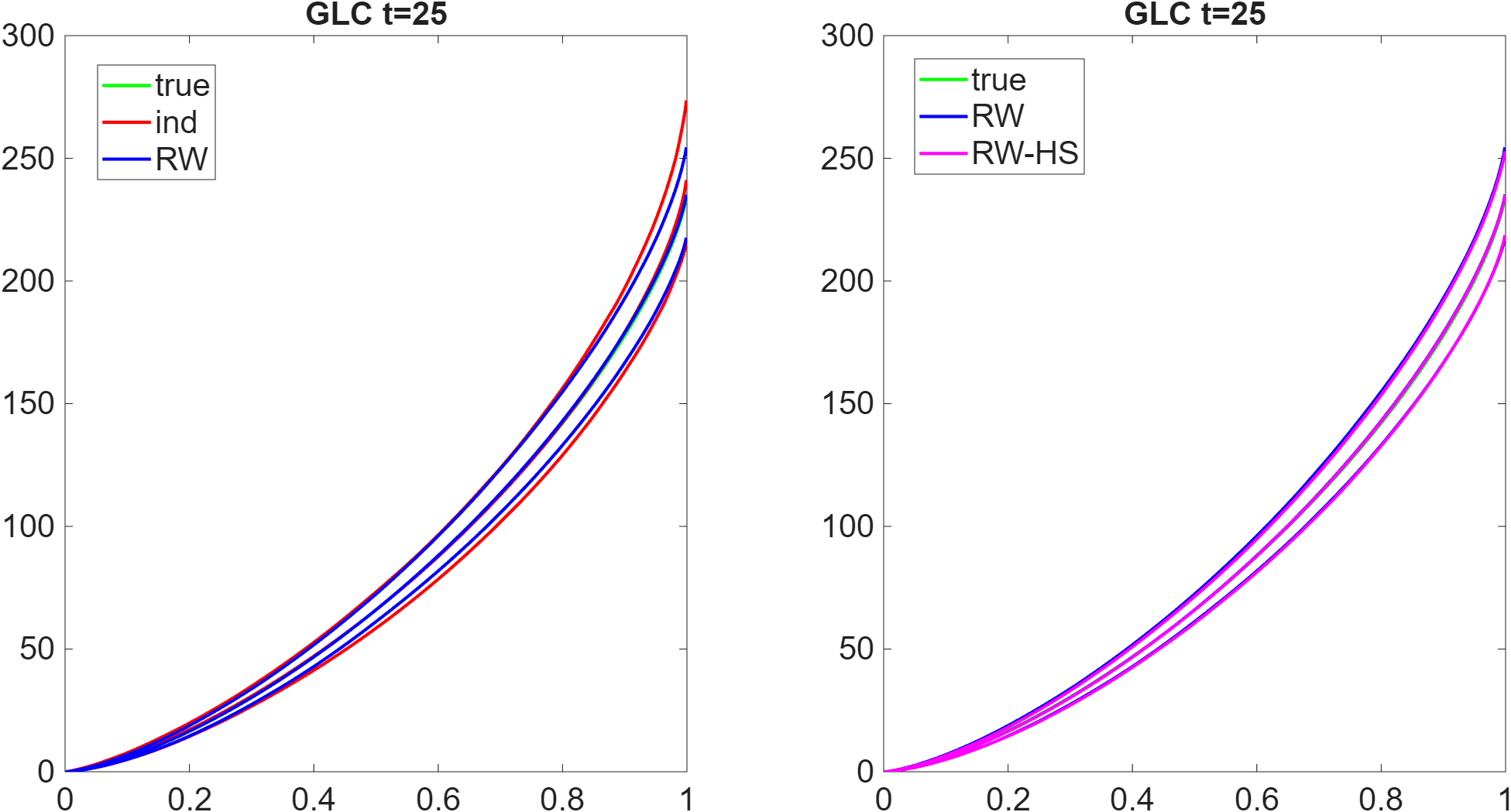}
    \caption{The posterior means (with 99\% credible intervals) of the Generalised Lorenz curve at $t=25$ obtained from the independent Dagum income model (ind), the random walk Dagum income model (RW), and the random walk Dagum income model with horseshoe priors (RW-HS) for the simulated dataset. The true Generalised Lorenz curve is also plotted. 
}
    \label{fig:Figure_GLC_sim1_t25_099}
\end{figure}
\begin{figure}[H]
    \centering
    \includegraphics[width=0.8\linewidth]{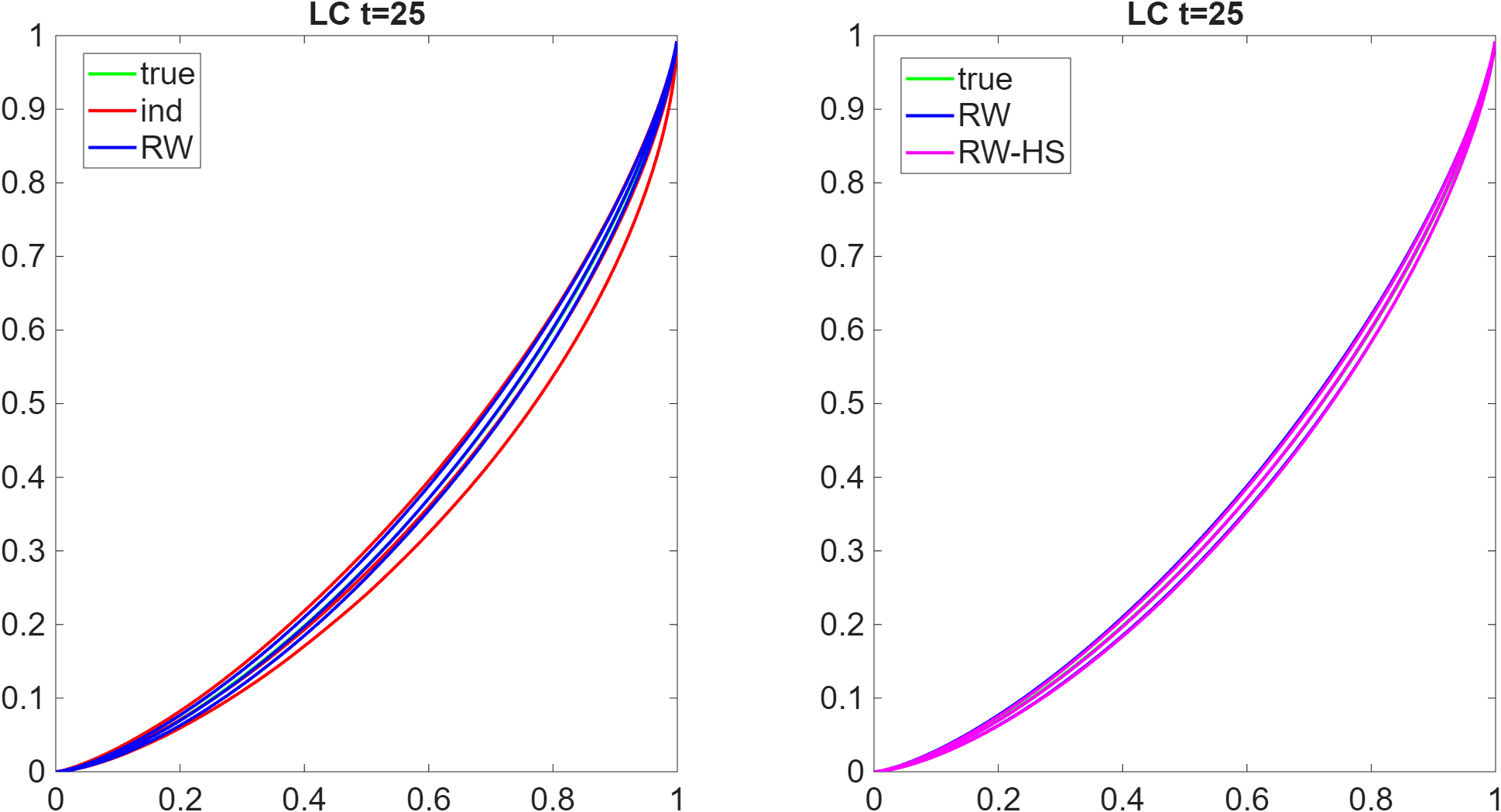}
    \caption{The posterior means (with 99\% credible intervals) of the Lorenz curve at $t=25$ obtained from the independent Dagum income model (ind), the random walk Dagum income model (RW), and the random walk Dagum income model with horseshoe priors (RW-HS) for the simulated dataset. The true Lorenz curve is also plotted.  
}
    \label{fig:Figure_LC_sim1_t25_099}
\end{figure}

Figures~\ref{fig:Figure_GLC_sim1_t25_099} and~\ref{fig:Figure_LC_sim1_t25_099} show the GLCs and LCs for time period $t=25$, obtained under the independent, RW, and RW-HS models. These figures highlight the efficiency gains from temporal pooling when the data-generating process evolves smoothly. For the GLCs, uncertainty is smallest at low population shares and increases toward the upper tail; accordingly, the independent model exhibits a substantially wider 99\% credible band near the top of the distribution. The RW and RW-HS models deliver markedly tighter bands over the entire curve, especially for the upper deciles, implying more precise inference. The same conclusion holds for the LCs at $t=25$: the independent fit implies much greater uncertainty about income shares, particularly beyond the median and into the top quantiles, whereas the RW and RW-HS models produce narrower credible regions and smoother implied inequality profiles. Overall, these comparisons confirm that the RW and RW-HS specifications not only stabilise point estimates but also substantially improve uncertainty quantification.


\section{Additional figures for Aboriginal population subgroups\label{sec:additionalfiguresaboriginalsubgroup}}

This section provides additional figures for the Aboriginal population subgroup discussed in Section \ref{sec:aboriginalpopulationsubgroup}.

\begin{figure}[H]
    \centering
    \includegraphics[width=0.8\linewidth]{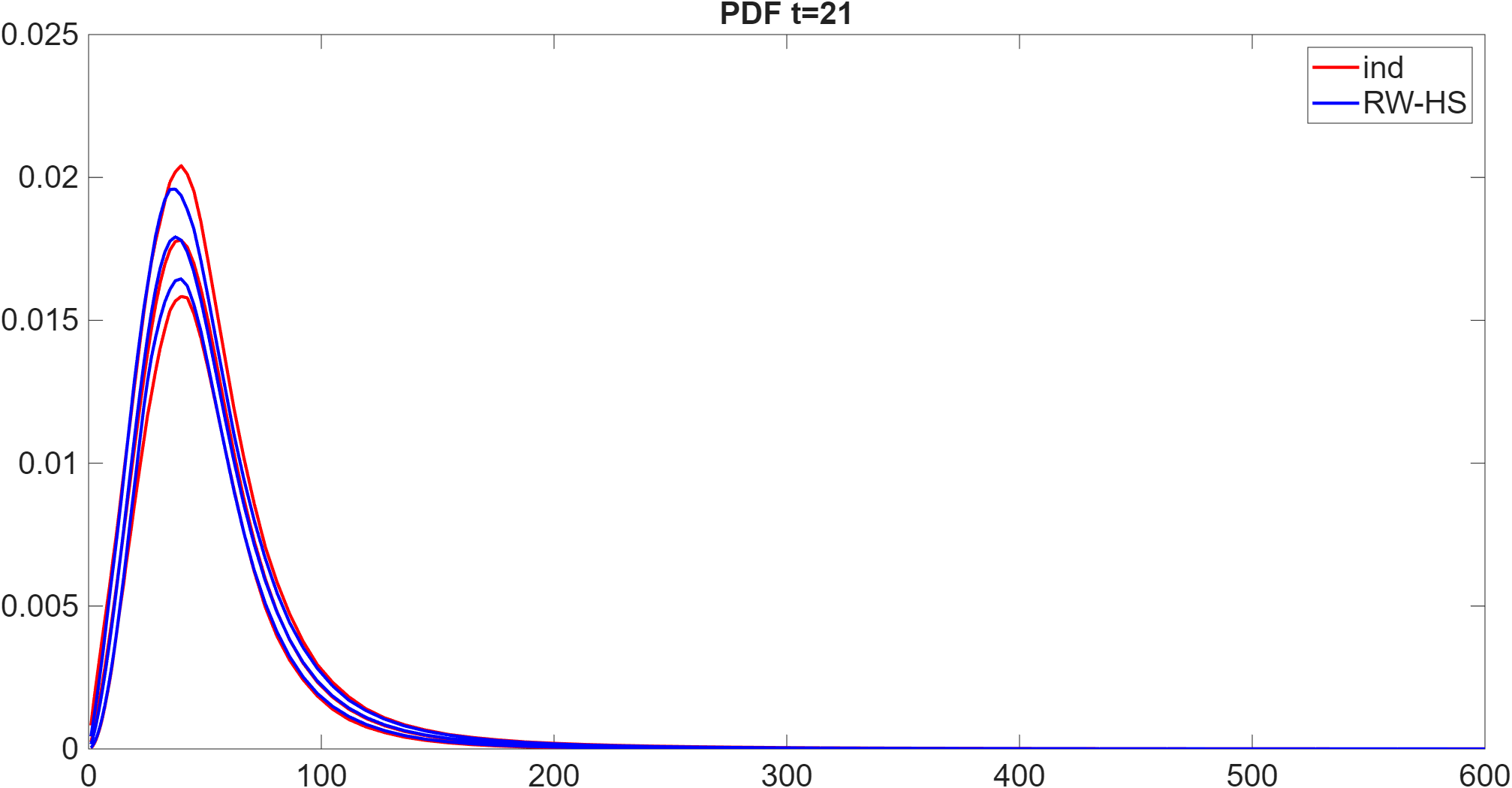}
    \caption{The posterior means (with 99\% credible intervals) of the income density obtained from the independent GB2 income model (ind) and the random walk GB2 income model with horseshoe priors (RW-HS) for the Aboriginal population subgroup for the year 2021. 
}
    \label{fig:Figure_PDF_GB2_abo}
\end{figure}

\begin{figure}[H]
    \centering
    \includegraphics[width=0.8\linewidth]{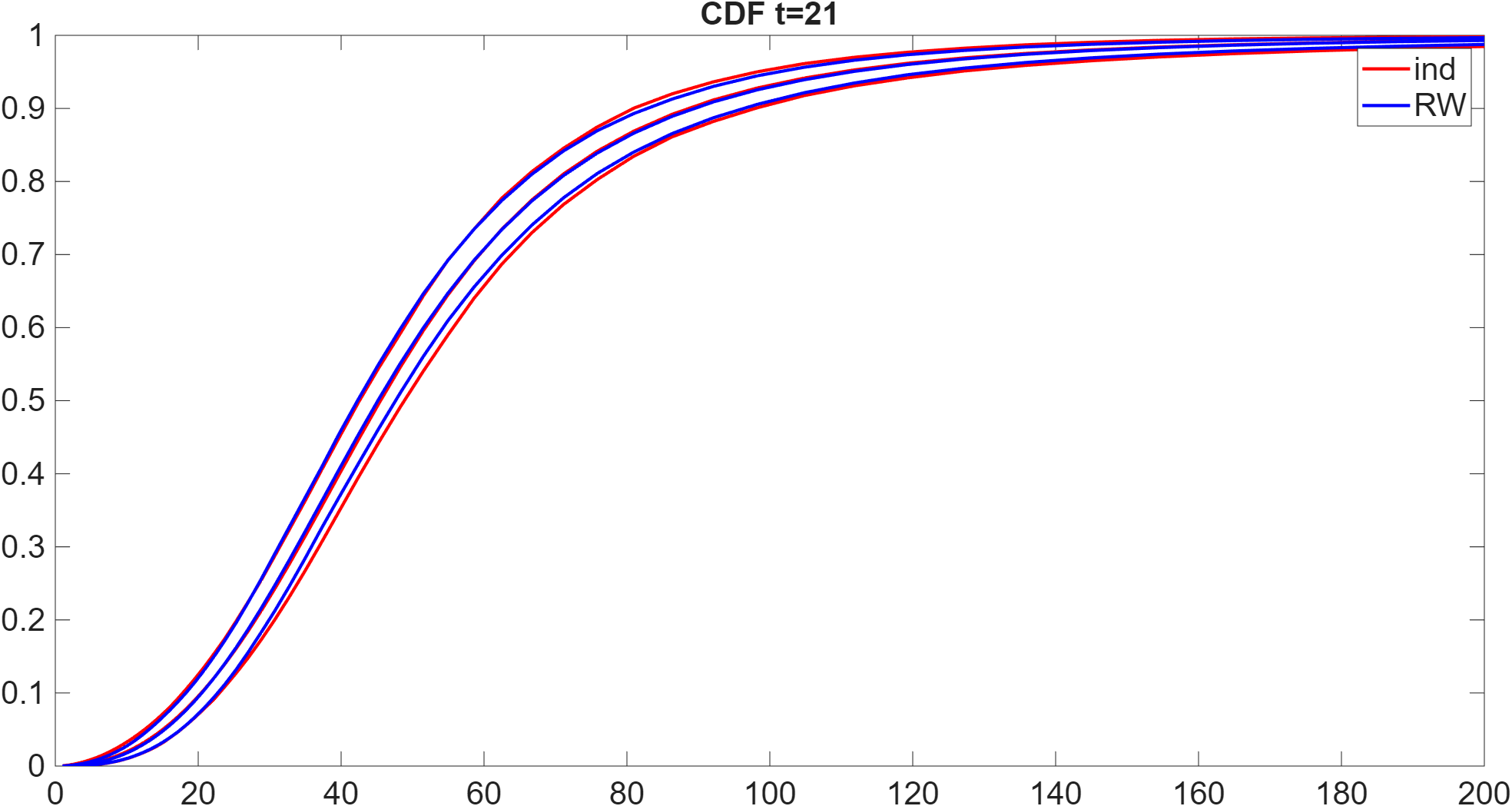}
    \caption{The posterior means (with 99\% credible intervals) of the income CDF obtained from the independent GB2 income model (ind) and the random walk GB2 income model with horseshoe priors (RW-HS) for the Aboriginal population subgroup for the year 2021. 
}
    \label{fig:Figure_CDF_GB2_abo}
\end{figure}

\begin{figure}[H]
    \centering
    \includegraphics[width=0.8\linewidth]{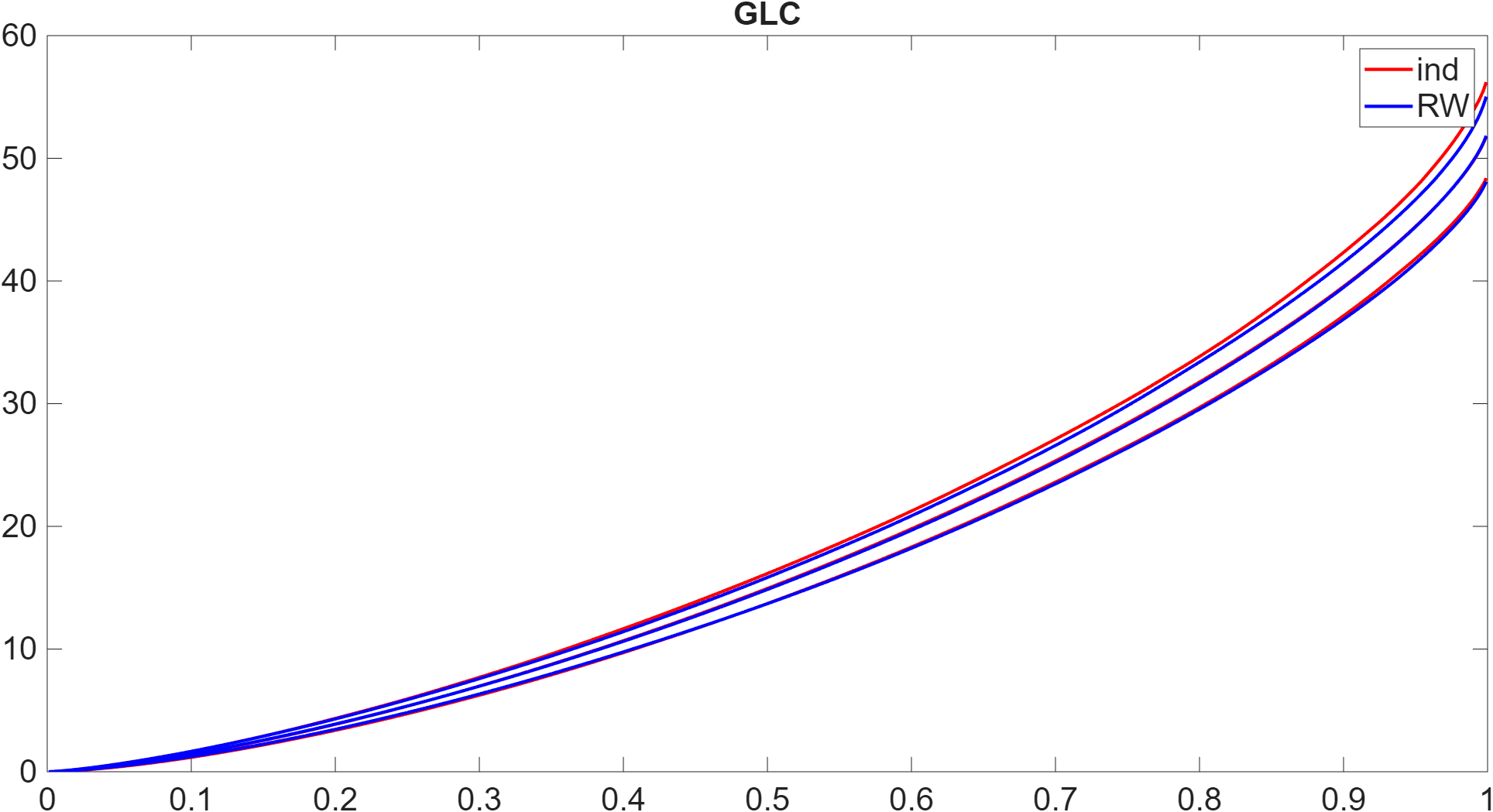}
    \caption{The posterior means (with 99\% credible intervals) of the Generalised Lorenz curve obtained from the independent GB2 income model (ind) and the random walk GB2 income model with horseshoe priors (RW-HS) for the Aboriginal population subgroup for the year 2021.  
}
    \label{fig:Figure_GLC_GB2_abo}
\end{figure}

\begin{figure}[H]
    \centering
    \includegraphics[width=0.8\linewidth]{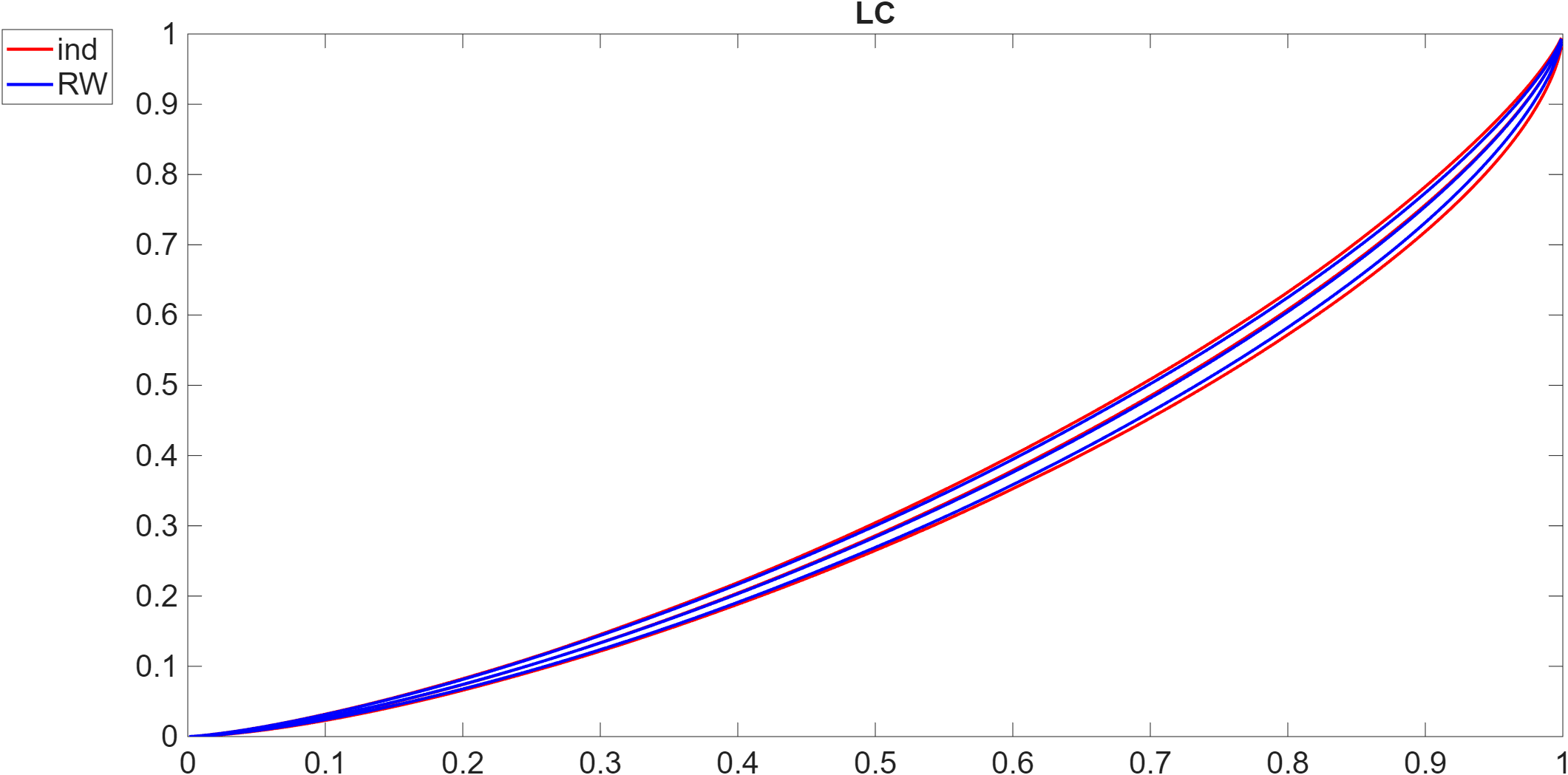}
    \caption{The posterior means (with 99\% credible intervals) of the Lorenz curve obtained from the independent GB2 income model (ind) and the random walk GB2 income model with horseshoe priors (RW-HS) for the Aboriginal population subgroup for the year 2021.  
}
    \label{fig:Figure_LC_GB2_abo}
\end{figure}

Figures~\ref{fig:Figure_PDF_GB2_abo}--\ref{fig:Figure_LC_GB2_abo} present the posterior means and 99\% credible intervals for the fitted GB2 income PDFs, CDFs, GLCs, and LCs for the Aboriginal population subgroup at time \(t=21\) under the independent model and the random walk model with horseshoe priors. The posterior mean CDFs, LCs, GLCs, and PDFs under the independent model and the random walk model with horseshoe priors are very close to one another, indicating that both models deliver essentially the same overall distributional picture. The differences in posterior means are generally small rather than substantial: the RW-HS LC lies slightly above that of the independent model over most population shares, suggesting marginally lower inequality, while the corresponding GLC is slightly lower, reflecting a somewhat smaller fitted mean income. Likewise, the posterior mean CDFs and PDFs are very similar across the support, with only minor deviations in the lower and middle parts of the income distribution. Importantly, the RW-HS model tends to produce narrower 99\% credible intervals than the independent model, indicating greater estimation precision and a more stable characterisation of uncertainty. Although the 99\% credible intervals from the two models overlap substantially in all panels, the tighter bands under RW-HS show the benefit of borrowing strength across adjacent years through temporal smoothing and horseshoe shrinkage.


\begin{figure}[H]
    \centering
    \includegraphics[width=0.8\linewidth]{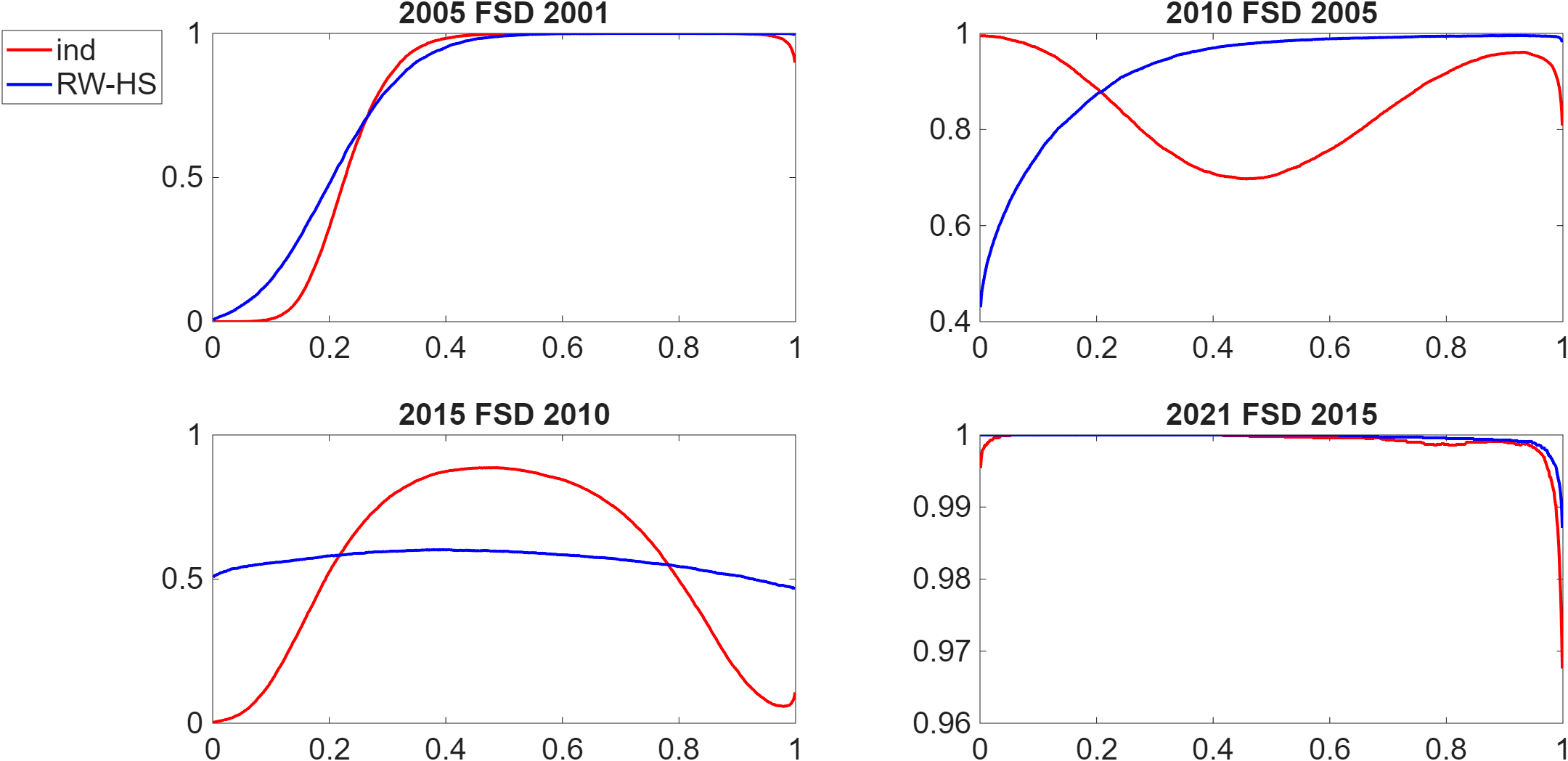}
    \caption{Estimated probability curves for first order stochastic dominance obtained from the independent GB2 income model (ind) and the random walk GB2 income model with horseshoe priors (RW-HS) for the Aboriginal population subgroup.  
}
    \label{fig:yFSDx_prob_abo}
\end{figure}

\begin{figure}[H]
    \centering
    \includegraphics[width=0.8\linewidth]{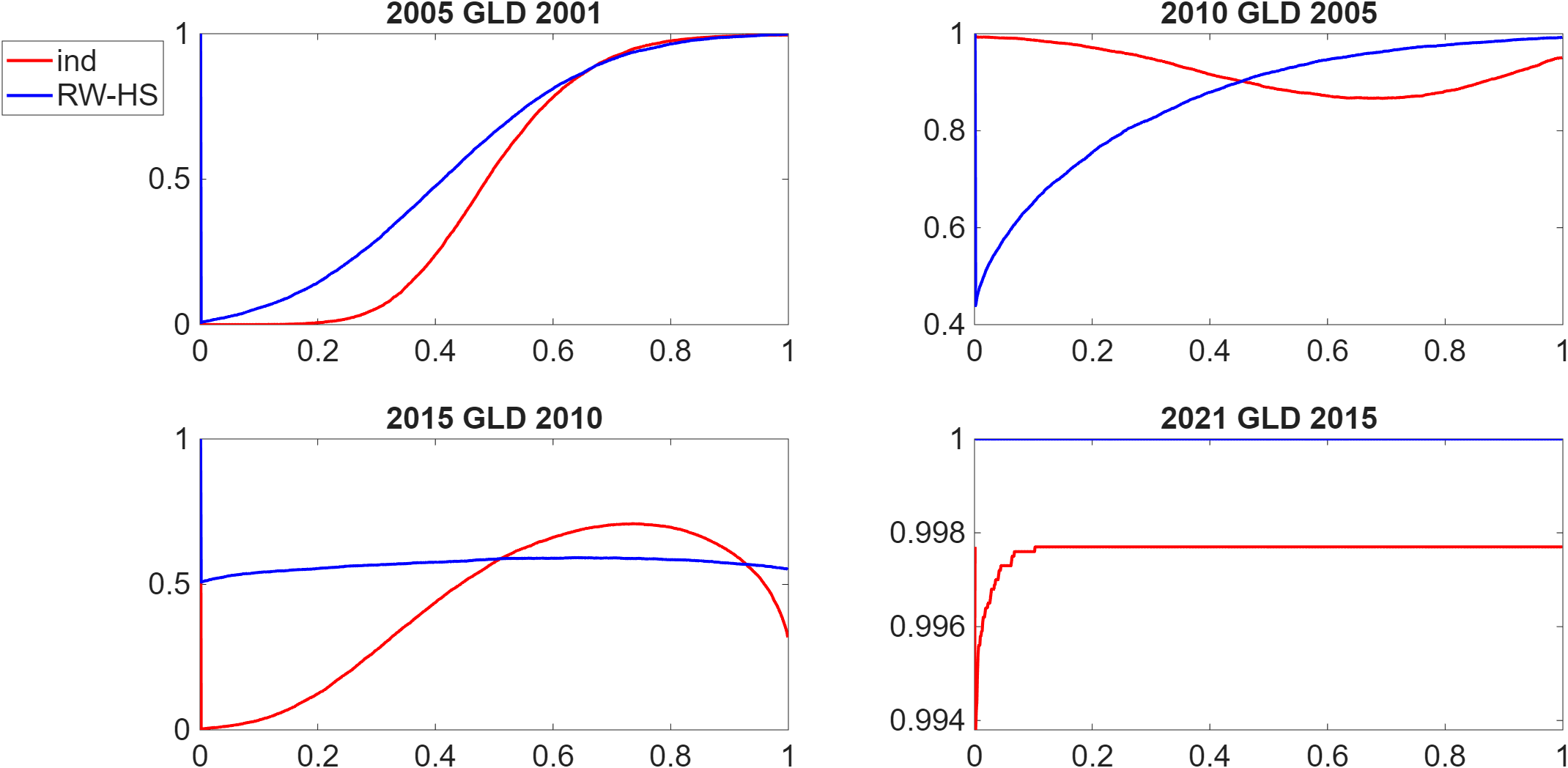}
    \caption{Estimated probability curves for generalised Lorenz dominance obtained from the independent GB2 income model (ind) and the random walk GB2 income model with horseshoe priors (RW-HS) for the Aboriginal population subgroup.  
}
    \label{fig:yGLDx_prob_abo}
\end{figure}

\begin{figure}[H]
    \centering
    \includegraphics[width=0.8\linewidth]{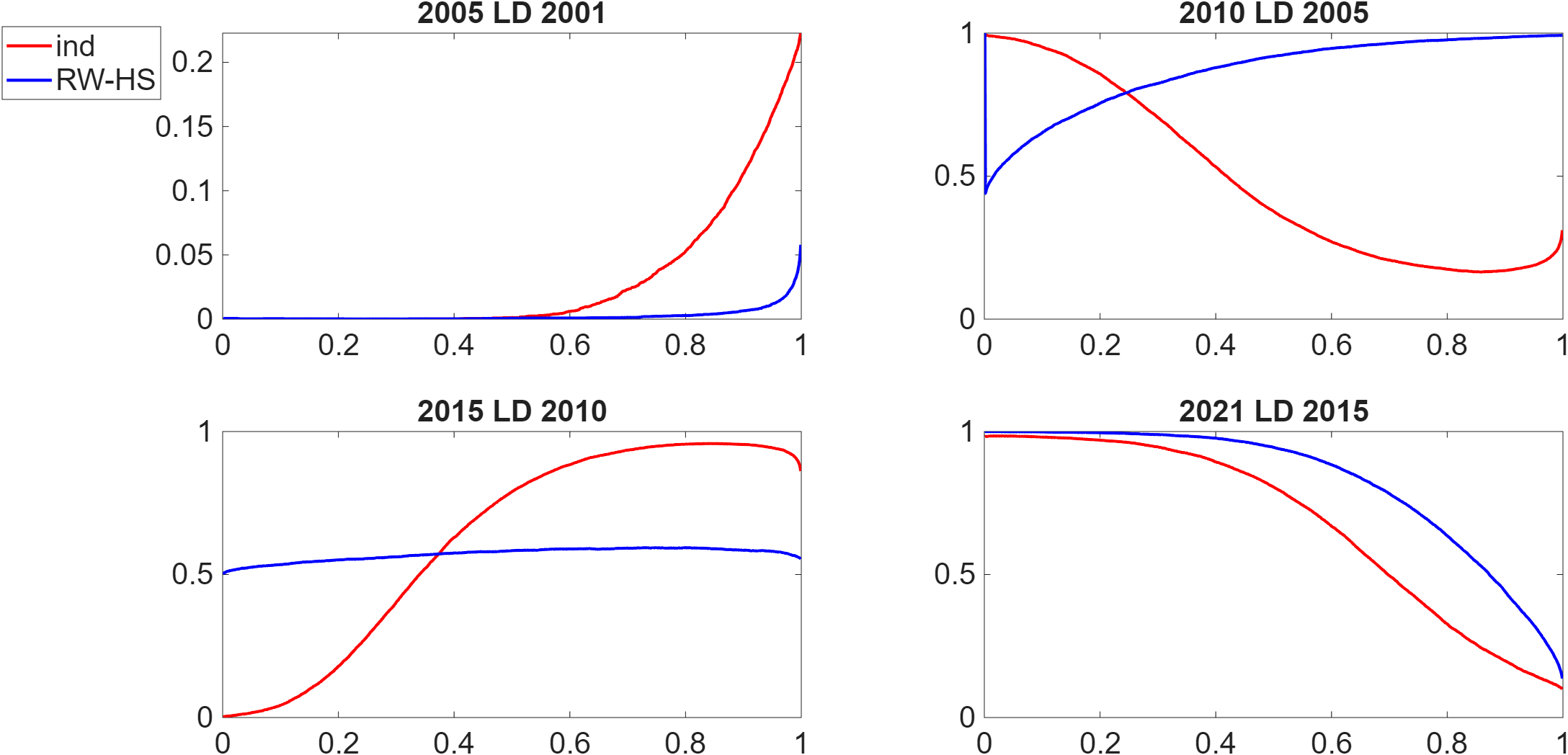}
    \caption{Estimated probability curves for Lorenz dominance obtained from the independent GB2 income model (ind) and the random walk GB2 income model with horseshoe priors (RW-HS) for the Aboriginal population subgroup.  
}
    \label{fig:yLDx_prob_abo}
\end{figure}

The probability curves in Figures \ref{fig:yFSDx_prob_abo} to \ref{fig:yLDx_prob_abo} reinforce these differences between the independent model and RW-HS by showing that the two specifications can imply very different pointwise dominance behaviour, even when the resulting overall dominance probabilities are similar in broad direction. Across the FSD, GLD, and LD panels, the RW-HS curves are generally smoother and less extreme, whereas the independent model often produces sharply varying, and hump-shaped profiles. This is particularly evident for the 2015--2010 comparison, where the independent model yields highly uneven probability curves across the support, while RW-HS gives flatter and more regular profiles. For 2021 versus 2015, by contrast, both models produce probability curves that are close to one throughout most of the support for FSD and GLD, consistent with the strong dominance probabilities reported in the tables. Another important feature is that pointwise dominance probabilities can be high over large parts of the support while the joint dominance probability remains low, because dominance must hold simultaneously at all evaluation points. This is especially clear for the 2005--2001 and several LD comparisons, where the curves may be moderately large over part of the domain but still fail to imply a large overall probability of dominance.


\begin{figure}[H]
    \centering
    \includegraphics[width=0.8\linewidth]{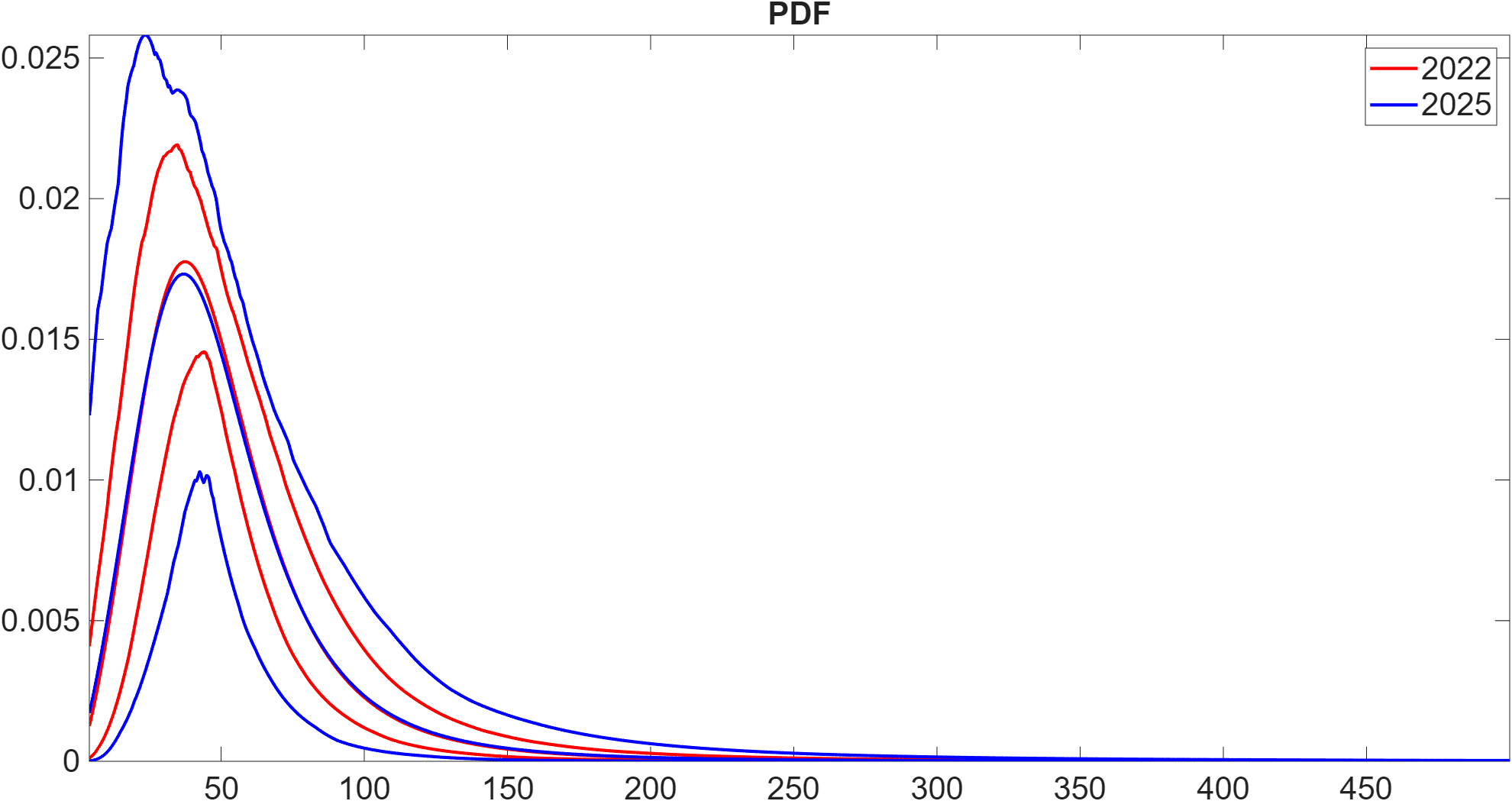}
    \caption{Posterior predictive densities (with 95\% prediction intervals) for the years 2022 and 2025 obtained from the random walk GB2 income model with horseshoe priors (RW-HS) for the Aboriginal population subgroup.  
}
    \label{fig:pred_density_Abo}
\end{figure}

\begin{figure}[H]
    \centering
    \includegraphics[width=0.8\linewidth]{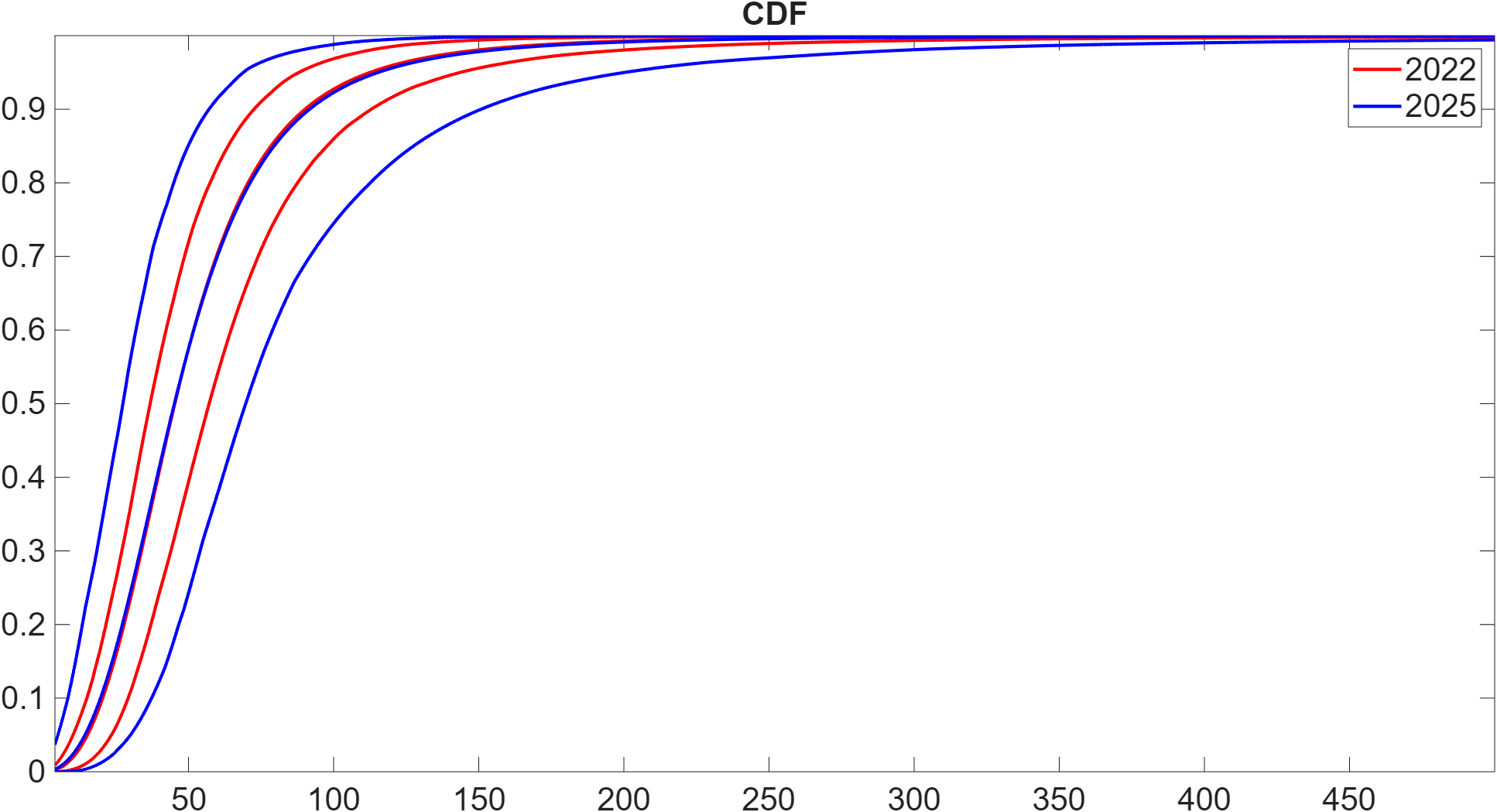}
    \caption{Posterior predictive CDFs (with 95\% prediction intervals) for the years 2022 and 2025 obtained from the random walk GB2 income model with horseshoe priors (RW-HS) for the Aboriginal population subgroup.  
}
    \label{fig:pred_CDF_Abo}
\end{figure}

\begin{figure}[H]
    \centering
    \includegraphics[width=0.8\linewidth]{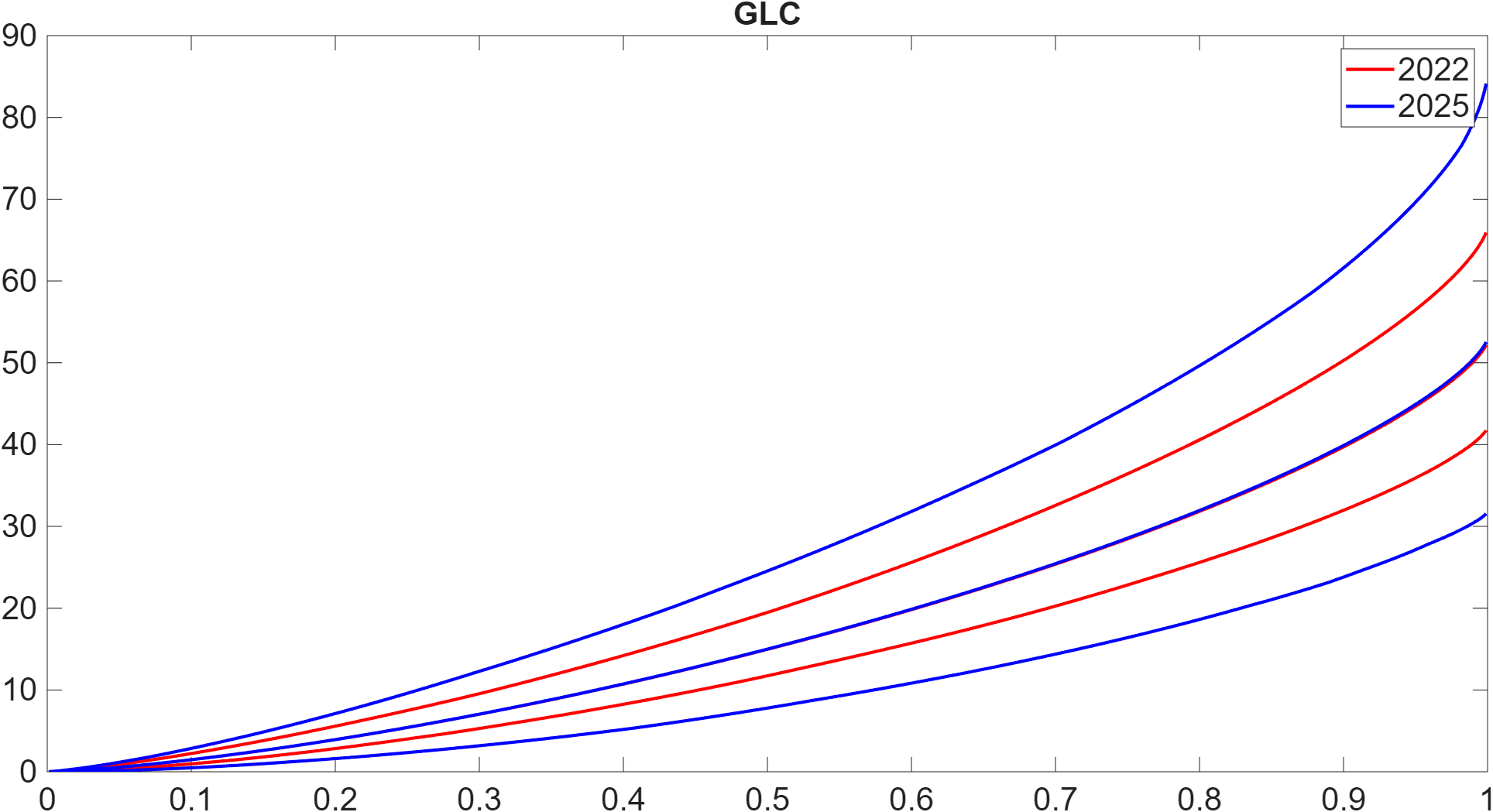}
    \caption{Posterior predictive generalised Lorenz curves (GLCs) (with 95\% prediction intervals) for the years 2022 and 2025 obtained from the random walk GB2 income model with horseshoe priors (RW-HS) for the Aboriginal population subgroup.  
}
    \label{fig:pred_GLC_Abo}
\end{figure}

\begin{figure}[H]
    \centering
    \includegraphics[width=0.8\linewidth]{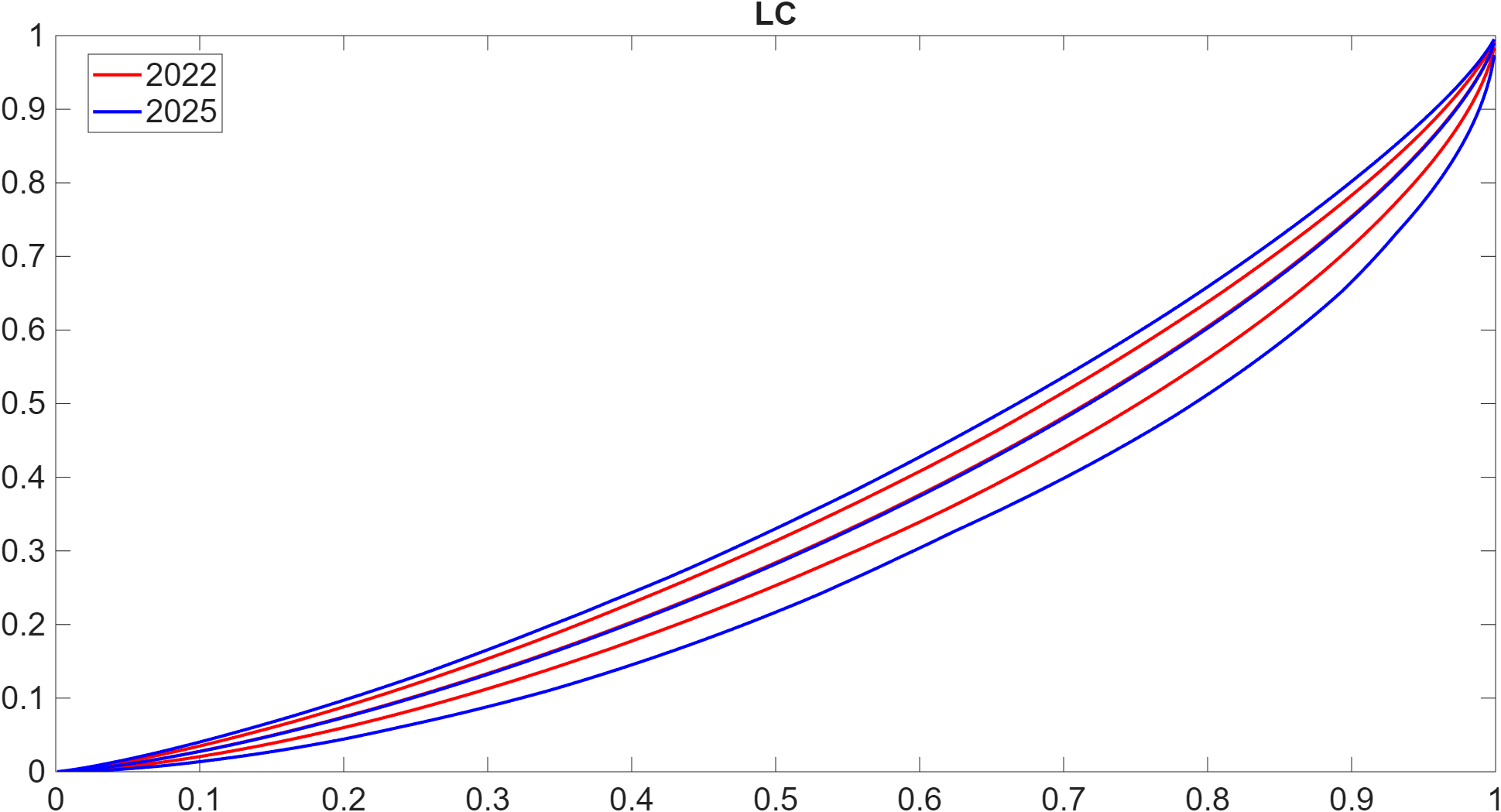}
    \caption{Posterior predictive Lorenz curves (GLCs) (with 95\% prediction intervals) for the years 2022 and 2025 obtained from the random walk GB2 income model with horseshoe priors (RW-HS) for the Aboriginal population subgroup.  
}
    \label{fig:pred_LC_Abo}
\end{figure}

Figures \ref{fig:pred_density_Abo}--\ref{fig:pred_LC_Abo} in Section \ref{sec:additionalfiguresaboriginalsubgroup} of the online supplement present the posterior predictive distributions for the Aboriginal population subgroup in 2022 and 2025, obtained by projecting the RW-HS GB2 model beyond the observed 2001--2021 period. Figure \ref{fig:pred_density_Abo} shows that the 2025 predictive density is more dispersed than that for 2022, with a noticeably wider range of plausible incomes and a more pronounced upper tail, indicating greater uncertainty about the future shape of the income distribution. This pattern is also reflected in Figure \ref{fig:pred_CDF_Abo}, which shows the predictive CDFs for 2022 and 2025. Figures \ref{fig:pred_GLC_Abo} and \ref{fig:pred_LC_Abo} suggest that the predicted generalised Lorenz and Lorenz curves for 2002 closely resemble those for 2025. Importantly, however, the prediction intervals are clearly wider in 2025 than in 2022 across all four curves, showing that forecast uncertainty accumulates substantially as the prediction horizon moves further away from the sample used for estimation.


\begin{figure}[H]
    \centering
    \includegraphics[width=0.8\linewidth]{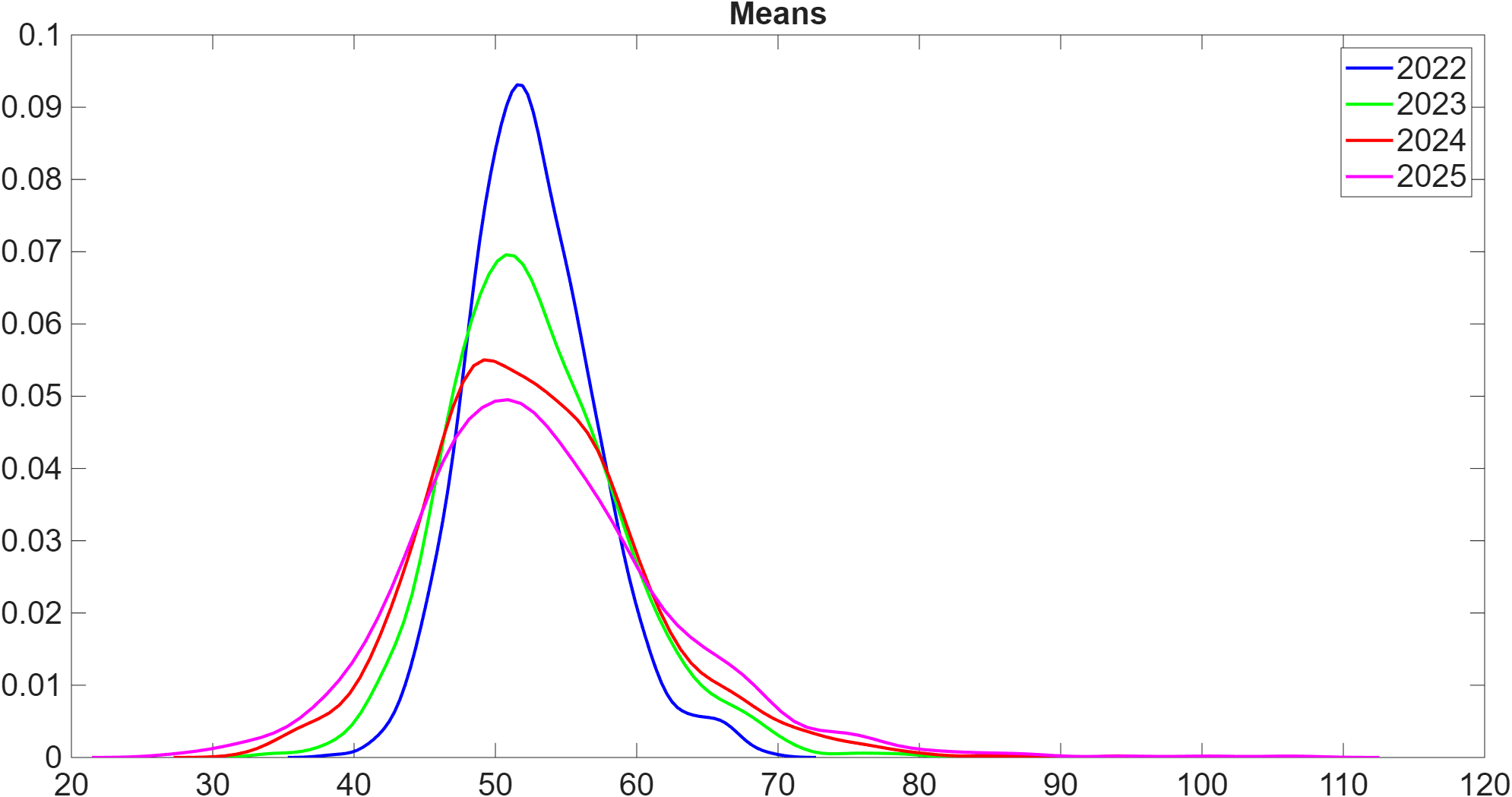}
    \caption{Kernel density estimates of the predicted mean incomes for the Aboriginal population subgroup from 2022 to 2025, obtained using the random walk GB2 income model with horseshoe priors (RW-HS).  
}
    \label{fig:pred_means_Abo}
\end{figure}

\begin{figure}[H]
    \centering
    \includegraphics[width=0.8\linewidth]{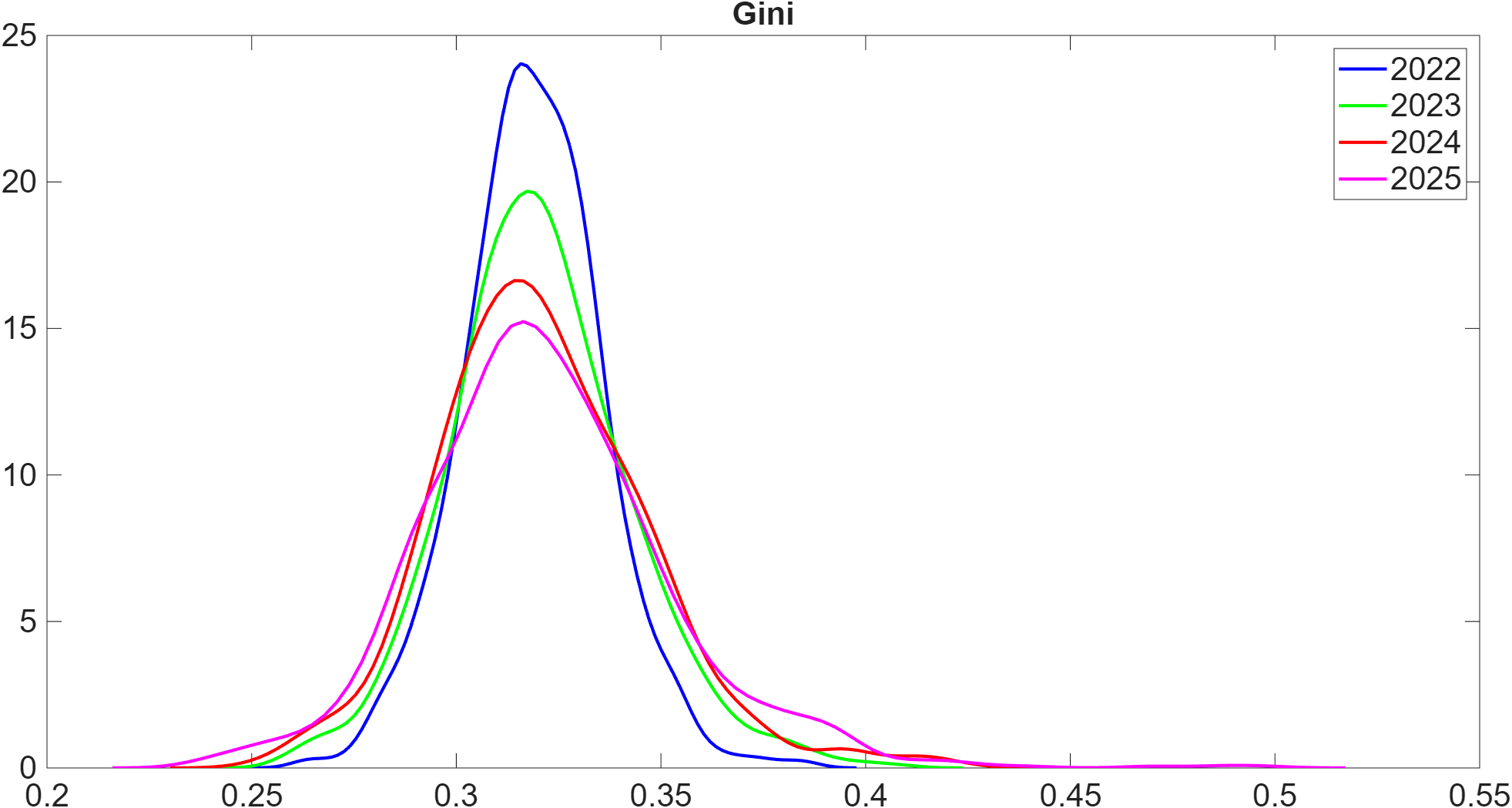}
    \caption{Kernel density estimates of the predicted Gini coefficients for the Aboriginal population subgroup from 2022 to 2025, obtained using the random walk GB2 income model with horseshoe priors (RW-HS).  
}
    \label{fig:pred_GINI_Abo}
\end{figure}

\begin{figure}[H]
    \centering
    \includegraphics[width=0.8\linewidth]{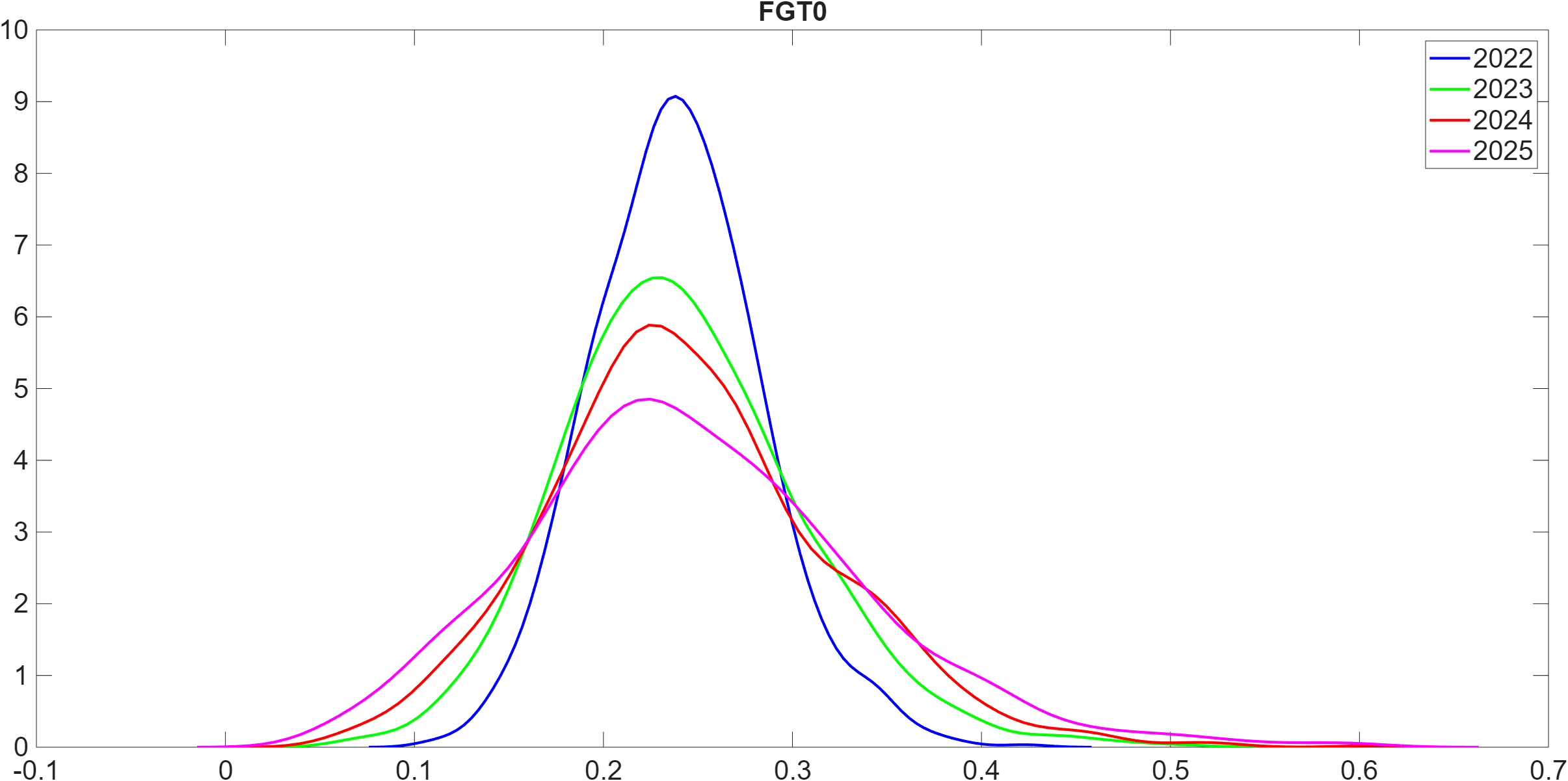}
    \caption{Kernel density estimates of the predicted FGT0 indices for the Aboriginal population subgroup from 2022 to 2025, obtained using the random walk GB2 income model with horseshoe priors (RW-HS).  
}
    \label{fig:pred_FGT0_Abo}
\end{figure}

\begin{figure}[H]
    \centering
    \includegraphics[width=0.8\linewidth]{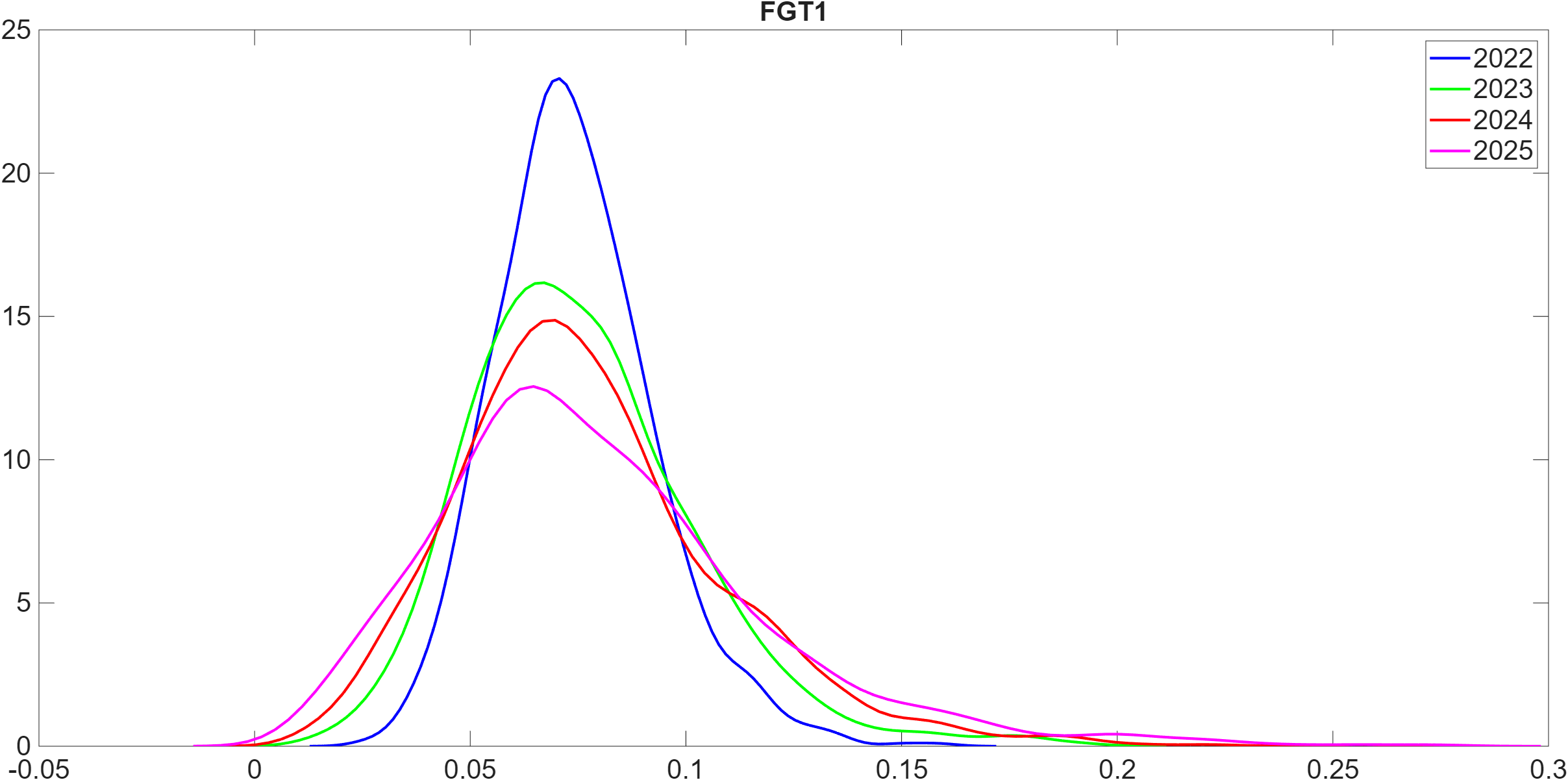}
    \caption{Kernel density estimates of the predicted FGT1 indices for the Aboriginal population subgroup from 2022 to 2025, obtained using the random walk GB2 income model with horseshoe priors (RW-HS).  
}
    \label{fig:pred_FGT1_Abo}
\end{figure}

Figures \ref{fig:pred_means_Abo}--\ref{fig:pred_FGT1_Abo} summarise the predictive distributions of key welfare measures for the Aboriginal population subgroup over the out-of-sample period 2022--2025 under the RW-HS GB2 model fitted to the 2001--2021 data. Figure \ref{fig:pred_means_Abo} shows that the predicted mean income remains centred in a broadly similar range across the forecast horizon, although the predictive density becomes progressively flatter and more dispersed from 2022 to 2025, indicating that uncertainty about future mean income increases substantially for the later years. A similar pattern is evident in Figure \ref{fig:pred_GINI_Abo}, where the predictive densities for the Gini coefficient remain concentrated around comparable values, suggesting no dramatic change in relative inequality, but the later-year densities are clearly wider and exhibit more tail mass, so any apparent movement should be interpreted cautiously. The poverty measures in Figures \ref{fig:pred_FGT0_Abo} and \ref{fig:pred_FGT1_Abo} suggest a modest tendency towards lower poverty by 2025, as the predictive mass shifts slightly toward smaller values for both the headcount ratio and the poverty gap. However, these improvements are accompanied by much greater dispersion and longer right tails in the later years, especially for 2024 and 2025, reflecting the build-up of forecast uncertainty as the prediction horizon extends further beyond the observed sample. Overall, the model points to broadly stable or slightly improving welfare outcomes for the Aboriginal subgroup, but the substantially wider predictive densities in the later years make clear that these longer-horizon forecasts are much less precise than those for 2022.

\section{Australian Capital Territory population subgroup\label{sec:ACTpopulationsubgroup}}

This section presents the empirical application of the proposed modelling framework to the residents of the Australian Capital Territory (ACT) population subgroup. Using HILDA data from 2001 to 2021, we compare a range of parametric income models to determine which specification best captures the evolution of the income distributions over time. 
Specifically, we consider the Dagum, Singh--Maddala, Beta 2, and GB2 distributions, described in Section \ref{sec:incomedistributions},  under an independent model, a random walk model, and a random walk model with horseshoe shrinkage priors. 


Table~\ref{tab:logscores_ACT} reports the log predictive scores obtained from 10-fold cross-validation for the competing income models fitted to the ACT population subgroup. Since larger log predictive scores, indicate better out-of-sample predictive performance, the table provides a principled basis for selecting the most suitable specification. Overall, the results show that allowing model parameters to evolve over time improves predictive accuracy relative to fitting each year independently, and that the additional shrinkage introduced by the horseshoe prior further enhances performance. Among all competing models, the GB2 distribution combined with the random walk and horseshoe shrinkage structure attains the highest log predictive score. This indicates that the random walk GB2 model with horseshoe priors offers the best balance of flexibility and predictive ability for modelling the income distribution of the ACT population subgroup, and it is therefore adopted in the subsequent empirical analysis.

\begin{table}[H]
\caption{Log predictive scores for the ACT population subgroup obtained from 10-fold cross-validation for model selection. The competing models are the independent income model, the random walk income model, and the random walk income model with horseshoe priors, each fitted using the Dagum, Singh--Maddala, Beta 2, and generalised Beta 2 distributions.\label{tab:logscores_ACT}}

\centering{}%
\begin{tabular}{cccc}
\hline 
 & ind & RW & RW-HS\tabularnewline
\hline 
Dagum & -2374.27 & -2372.13 & -2372.12\tabularnewline
Beta2 & -2392.85 & -2391.22 & -2391.19\tabularnewline
Singh-Maddala & -2377.02 & -2375.47 & -2375.43\tabularnewline
GB2 & -2373.26 & -2370.44 & -2370.41\tabularnewline
\hline 
\end{tabular}
\end{table}

\begin{figure}[H]
    \centering
    \includegraphics[width=0.8\linewidth]{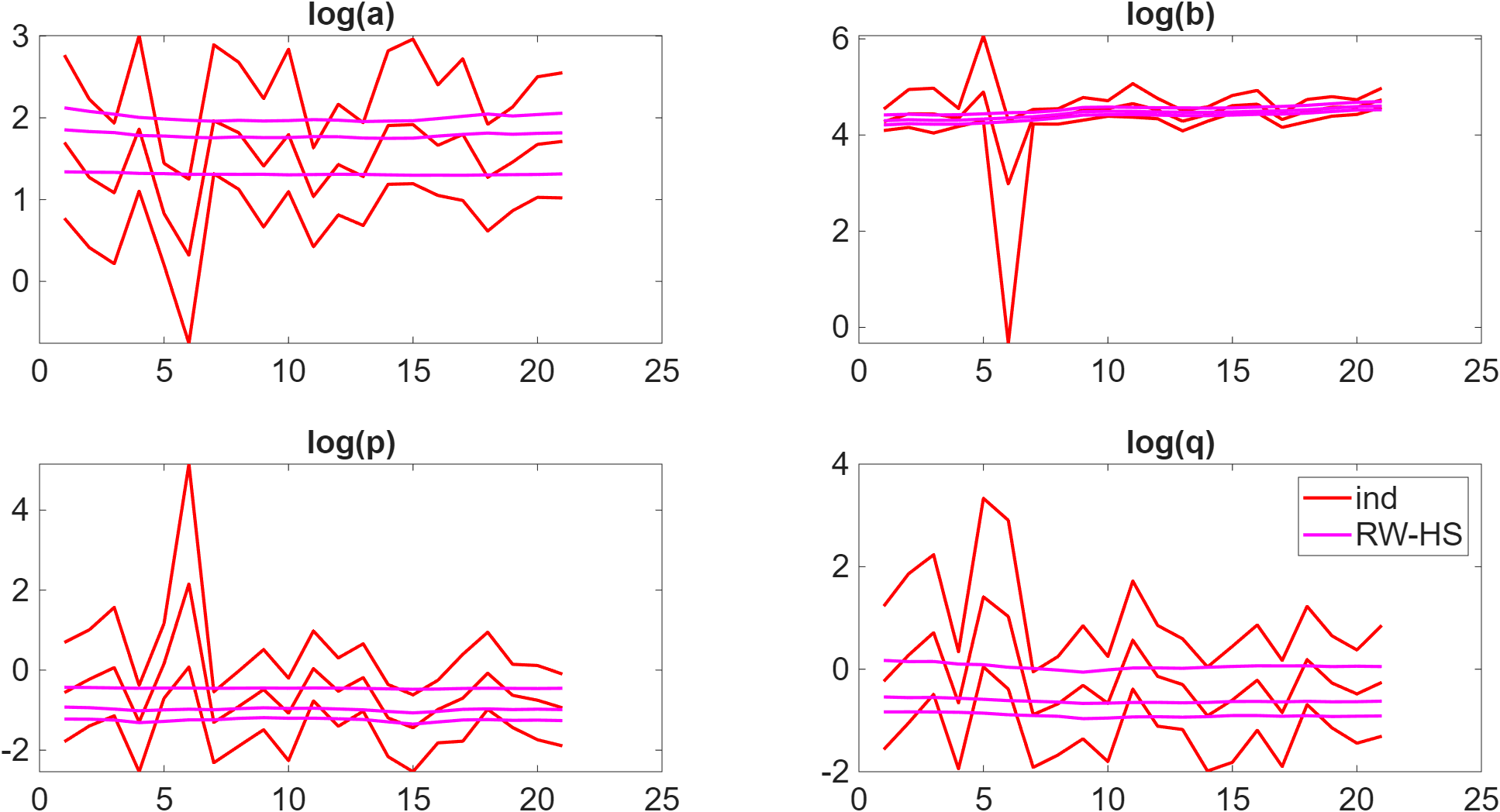}
    \caption{The posterior means (with 95\% credible intervals) of model parameters over time obtained from the independent GB2 income model (ind) and the random walk GB2 income model with horseshoe priors (RW-HS) for the ACT population subgroup.  
}
    \label{fig:Figure_param_GB2_ACT}
\end{figure}

\begin{figure}[H]
    \centering
    \includegraphics[width=0.8\linewidth]{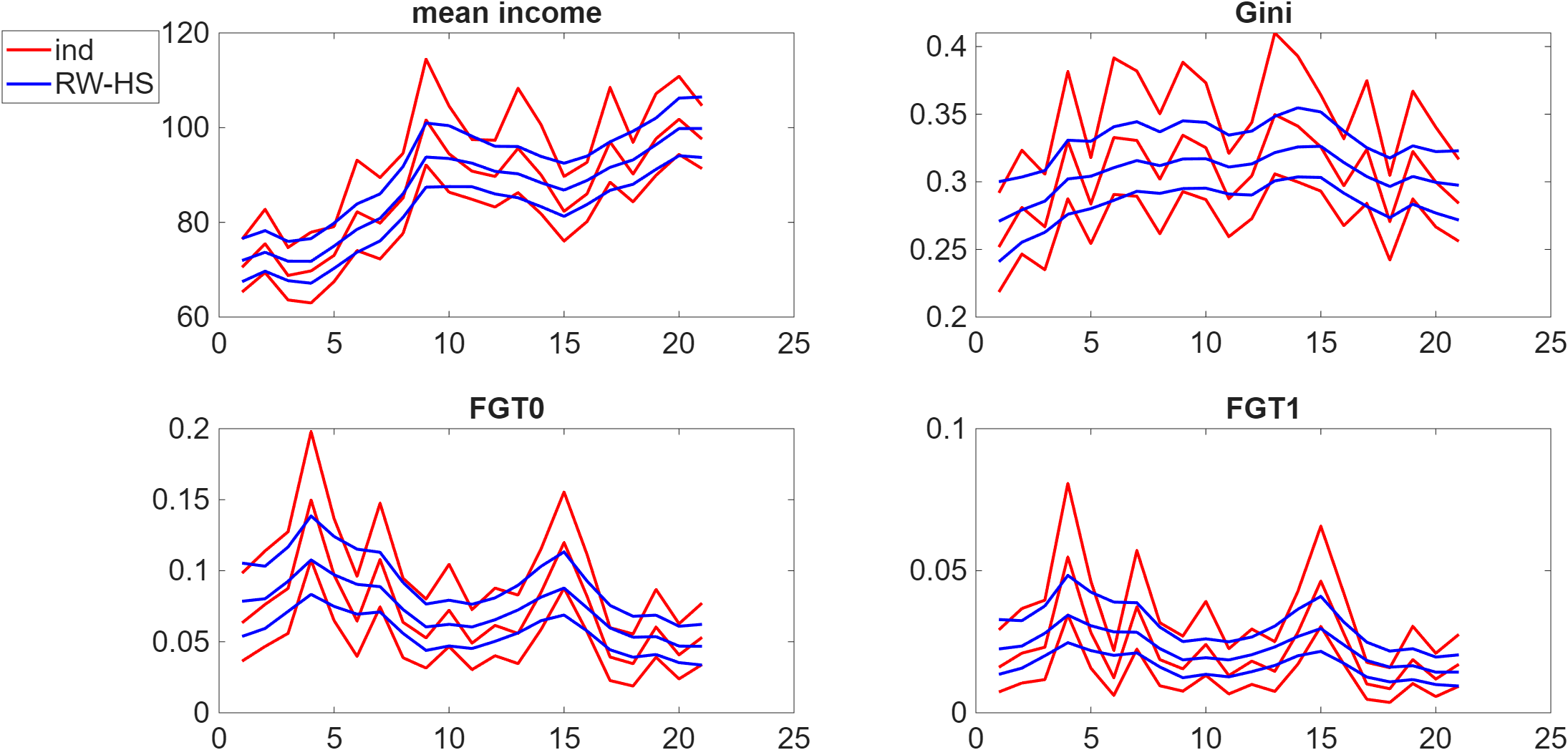}
    \caption{The posterior means (with 95\% credible intervals) of the mean income, Gini index, FGT0 and FGT1 indices over time obtained from the independent GB2 income model (ind) and the random walk GB2 income model with horseshoe priors (RW-HS) for the ACT population subgroup. 
}
    \label{fig:Figure_GB2_mean_gini_FGT0_FGT1_ACT}
\end{figure}

Figures~\ref{fig:Figure_param_GB2_ACT} and~\ref{fig:Figure_GB2_mean_gini_FGT0_FGT1_ACT} compare the posterior summaries obtained from the independent model and the random walk model with horseshoe priors (RW-HS) for the ACT population subgroup. For the GB2 parameters, the main contrast is the much greater temporal stability delivered by the RW-HS specification. The posterior means under the independent model display substantial year-to-year fluctuations, particularly for \(\log(a)\) and \(\log(p)\), with several sharp spikes and dips that suggest considerable sensitivity to annual sampling variation. By contrast, the RW-HS model produces much smoother trajectories for all three parameters, with \(\log(a)\) evolving gradually over time, \(\log(b)\) remaining relatively stable with only mild long-run movement, and \(\log(p)\) and \(\log(q)\) showing a notably more regular pattern than under the independent specification. The credible bands under RW-HS are also more stable over time, reflecting the gain from borrowing strength across adjacent years. 


The welfare, inequality, and poverty summaries implied by the fitted models show several clear empirical patterns. Mean income exhibits a broadly increasing trend over the sample period, although the independent model suggests more pronounced short-run fluctuations, whereas the RW-HS model smooths these into a steadier upward trajectory. The Gini coefficient rises in the earlier part of the sample, indicating an increase in inequality, and then remains broadly stable or declines slightly in the later years, with the RW-HS estimates again providing a less erratic profile. In contrast, the poverty measures FGT0 and FGT1 are much lower in magnitude and display an overall downward tendency over time, despite some temporary increases in the middle of the sample. This suggests that both the incidence of poverty, as measured by FGT0, and the intensity of poverty, as measured by FGT1, generally decline over time for the ACT population subgroup. Taken together, the figures suggest rising average incomes, moderately persistent inequality, and a gradual reduction in poverty, with the RW-HS model providing smoother and more reliable temporal summaries than the independent model.

The estimated posterior probabilities of dominance for the ACT population subgroup reveal several important differences between the independent model and the random walk model with horseshoe priors (RW-HS), with the largest discrepancies arising for the 2005--2010 and 2015--2021 comparisons. For FSD, both models provide little evidence that the 2005 income distribution dominates that of 2001, either overall or over the poorest 10\% of the population, and both also indicate essentially no evidence that 2015 dominates 2010. By contrast, the independent model yields only weak evidence of welfare improvement, with probabilities around 0.25, whereas RW-HS gives much stronger support, with posterior probabilities around 0.86 both overall and for the poorest 10\% for 2010 dominates 2005. For 2021 relative to 2015, overall FSD probabilities are only moderate, around 0.46 for the independent model and 0.51 for the RW-HS model, but for the poorest 10\% they are essentially one under both models, indicating a very strong improvement at the lower end of the income distribution. 

The GLD results tell a broadly similar story for welfare comparisons that combine changes in both average income and inequality. There is little evidence of improvement from 2001 to 2005 and almost none from 2010 to 2015, whereas the evidence for improvement from 2005 to 2010 is again much stronger under RW-HS than under the independent model. For 2021 relative to 2015, GLD provides strong evidence of welfare improvement, particularly under RW-HS, which assigns posterior probabilities close to one both overall and for the poorest 10\%. 

The LD, which isolates inequality comparisons, is generally weaker in the earlier periods: there is little evidence of inequality improvement from 2001 to 2005 or from 2010 to 2015, and only limited support for improvement from 2005 to 2010, although RW-HS gives somewhat more support than the independent model, especially for the poorest 10\%. The clearest result emerges for 2021 relative to 2015, where both models indicate substantial inequality improvement, with especially strong evidence for the poorest 10\%. 

The probability curves further illustrate why the independent model and RW-HS can produce different dominance probabilities. Across the FSD, GLD, and LD panels, the RW-HS curves are generally smoother and more regular, while the independent model often produces more uneven and locally volatile profiles. This difference is most striking for the 2010--2005 comparison. For both FSD and GLD, the RW-HS probability curves lie close to one over almost the entire support, whereas the corresponding curves under the independent model are substantially lower, which explains the much larger posterior dominance probabilities under RW-HS. The same pattern appears for LD, where the RW-HS curve is very high across the support, while the independent-model curve remains low, leading to sharply different conclusions about inequality improvement. For the 2015--2010 comparison, both models generate very low GLD and FSD probability curves, consistent with the near-zero posterior probabilities reported in the tables, while the LD curves remain well below one, indicating no compelling evidence of inequality reduction. For the 2021--2015 comparison, the FSD curves from both models are close to one over most of the domain, although they fall near the upper tail, which helps explain why overall FSD probabilities are moderate rather than overwhelming, even though the poorest 10\% show near-certain improvement. In contrast, the GLD and LD curves for 2021--2015 remain high throughout, especially under RW-HS, consistent with strong evidence of welfare and inequality improvement in that period. 


\begin{table}[H]
\caption{Estimated probabilities of Lorenz dominance, generalised Lorenz dominance, and first-order stochastic dominance for the ACT population subgroup, based on the independent GB2 income model (ind),  and the random walk GB2 income model with horseshoe priors (RW-HS)}

\centering{}%
\begin{tabular}{ccc}
\hline 
 & ind & RW-HS\tabularnewline
\hline 
2005 FSD 2001 & 0.0166 & 0.0467\tabularnewline
2005 GLD 2001 & 0.0530 & 0.0518\tabularnewline
2005 LD 2001 & 0.0143 & 0.0079\tabularnewline
\hline 
2010 FSD 2005 & 0.2528 & 0.8616\tabularnewline
2010 GLD 2005 & 0.2503 & 0.8720\tabularnewline
2010 LD 2005 & 0.0008 & 0.0881\tabularnewline
\hline 
2015 FSD 2010 & 0.0002 & 0.0042\tabularnewline
2015 GLD 2010 & 0.0009 & 0.0056\tabularnewline
2015 LD 2010 & 0.0429 & 0.0464\tabularnewline
\hline 
2021 FSD 2015 & 0.4635 & 0.5119\tabularnewline
2021 GLD 2015 & 0.8480 & 0.9954\tabularnewline
2021 LD 2015 & 0.6986 & 0.8365\tabularnewline
\hline 
\end{tabular}
\end{table}

\begin{table}[H]
\caption{Estimated probabilities of Lorenz dominance, generalised Lorenz dominance, and first-order stochastic dominance over the poorest 10\% of the population for the ACT population subgroup, based on the independent GB2 income model (ind), and the random walk GB2 income model with horseshoe priors (RW-HS)}

\centering{}%
\begin{tabular}{ccc}
\hline 
 & ind & RW-HS\tabularnewline
\hline 
2005 FSD 2001 & 0.0610 & 0.0520\tabularnewline
2005 GLD 2001 & 0.0716 & 0.0557\tabularnewline
2005 LD 2001 & 0.0418 & 0.0206\tabularnewline
\hline 
2010 FSD 2005 & 0.2536 & 0.8628\tabularnewline
2010 GLD 2005 & 0.2503 & 0.8720\tabularnewline
2010 LD 2005 & 0.0918 & 0.4140\tabularnewline
\hline 
2015 FSD 2010 & 0.0156 & 0.0165\tabularnewline
2015 GLD 2010 & 0.0143 & 0.0202\tabularnewline
2015 LD 2010 & 0.0572 & 0.0557\tabularnewline
\hline 
2021 FSD 2015 & 0.9942 & 0.9974\tabularnewline
2021 GLD 2015 & 0.8489 & 0.9958\tabularnewline
2021 LD 2015 & 0.9832 & 0.9777\tabularnewline
\hline 
\end{tabular}
\end{table}

\begin{figure}[H]
    \centering
    \includegraphics[width=0.8\linewidth]{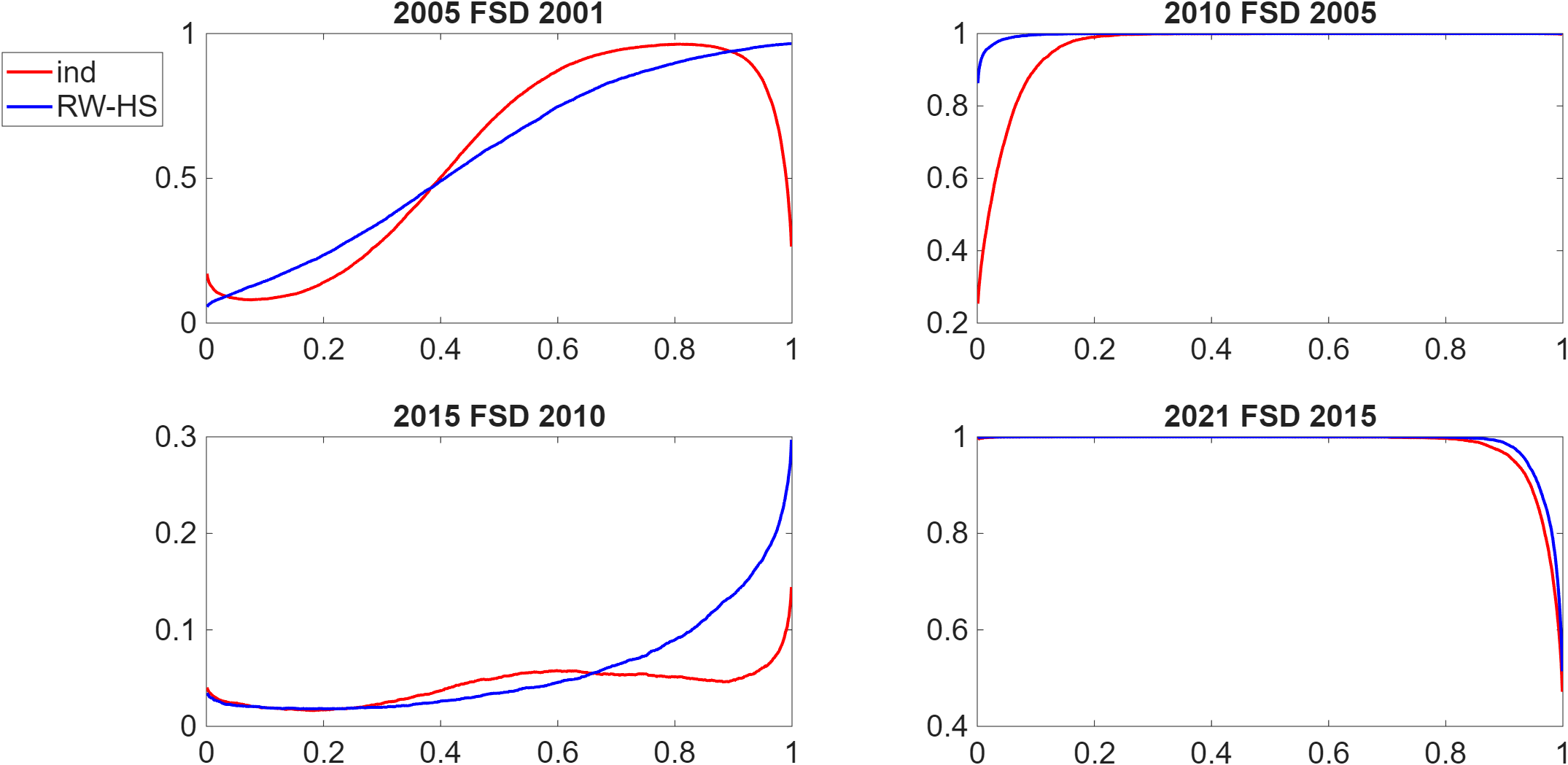}
    \caption{Estimated probability curves for first order stochastic dominance obtained from the independent GB2 income model (ind) and the random walk GB2 income model with horseshoe priors (RW-HS) for the ACT population subgroup.  
}
    \label{fig:yFSDx_prob_ACT}
\end{figure}

\begin{figure}[H]
    \centering
    \includegraphics[width=0.8\linewidth]{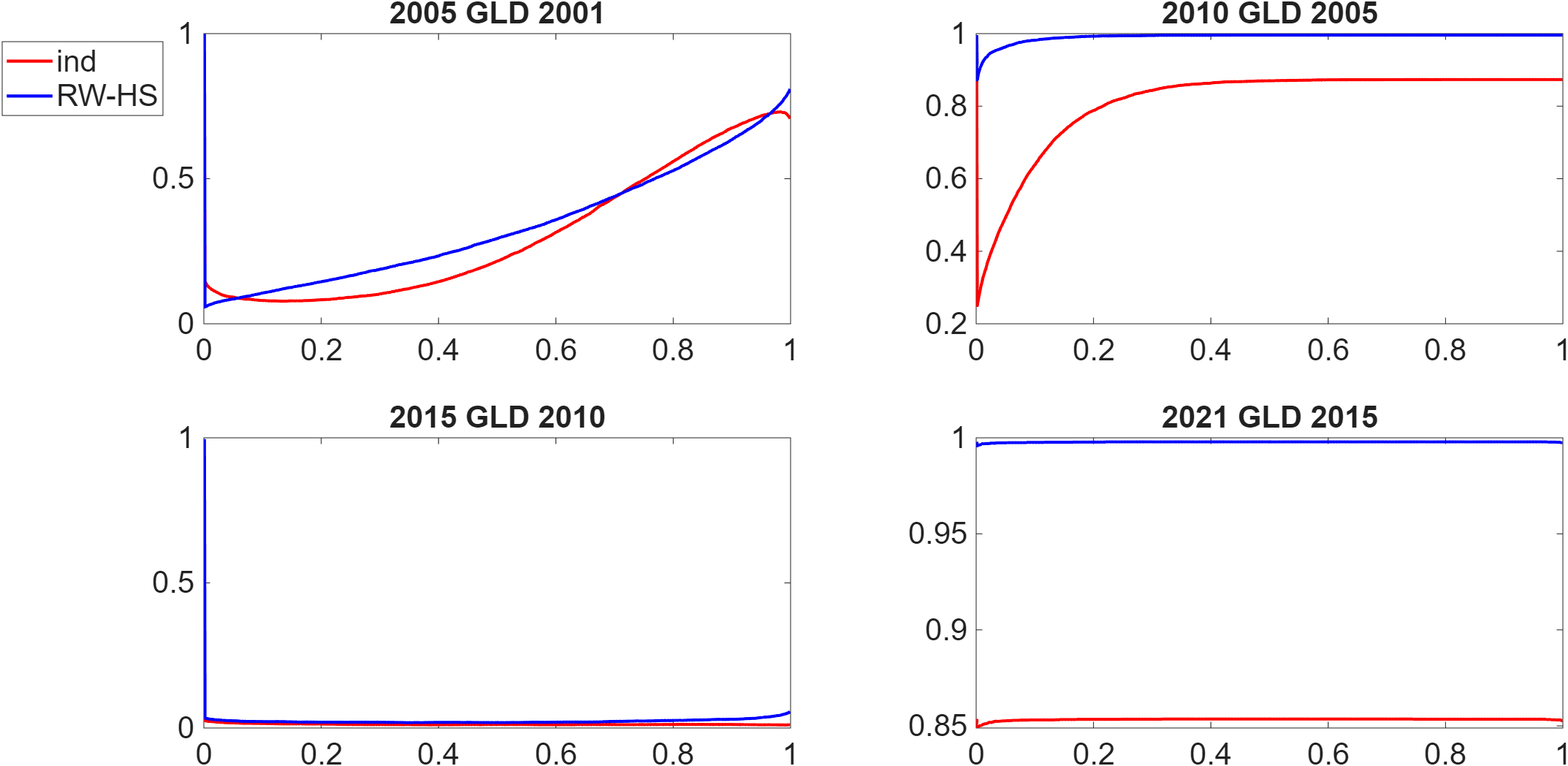}
    \caption{Estimated probability curves for generalised Lorenz dominance obtained from the independent GB2 income model (ind) and the random walk GB2 income model with horseshoe priors (RW-HS) for the ACT population subgroup.  
}
    \label{fig:yGLDx_prob_ACT}
\end{figure}

\begin{figure}[H]
    \centering
    \includegraphics[width=0.8\linewidth]{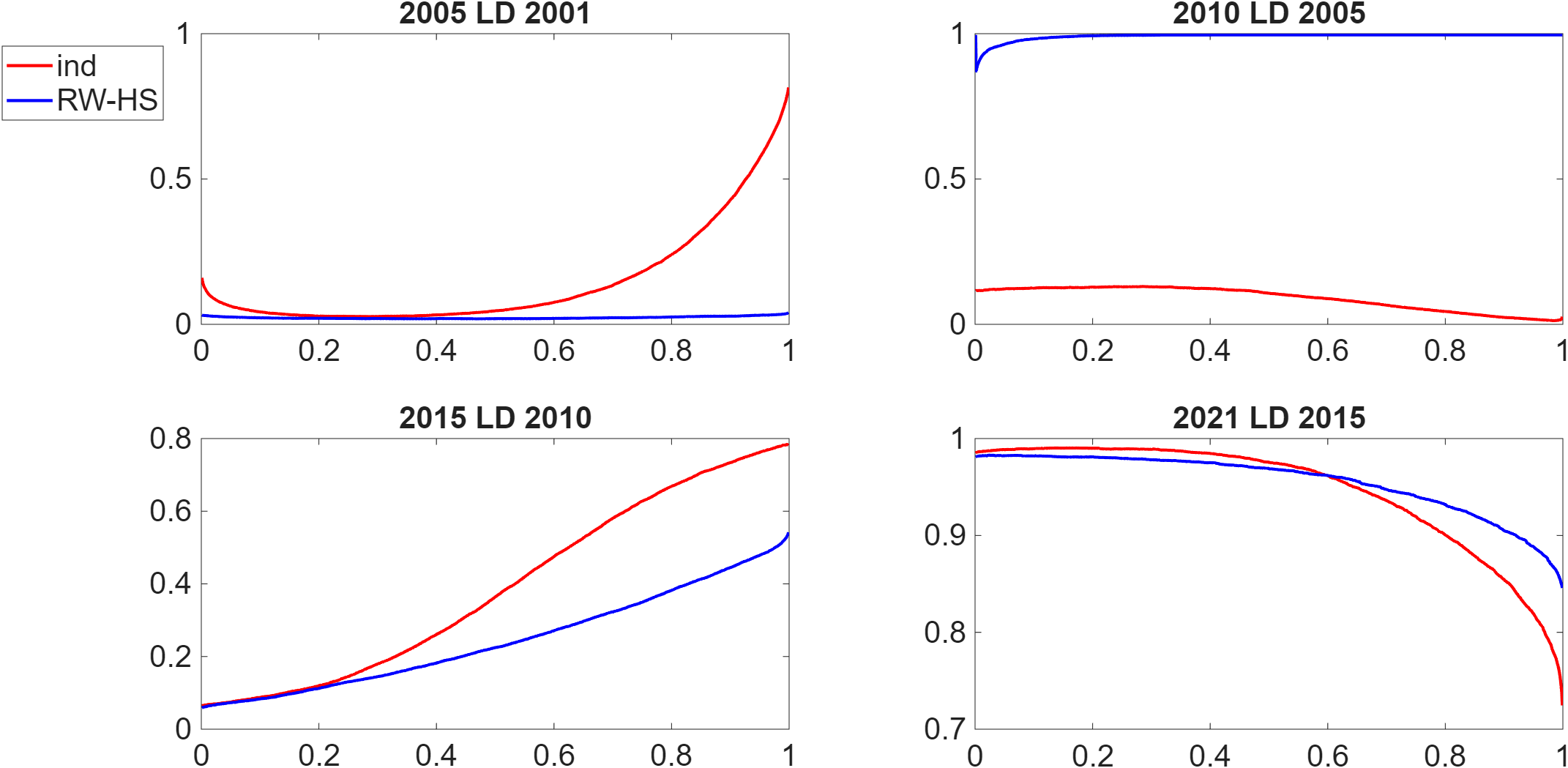}
    \caption{Estimated probability curves for Lorenz dominance obtained from the independent GB2 income model (ind) and the random walk GB2 income model with horseshoe priors (RW-HS) for the ACT population subgroup.  
}
    \label{fig:yLDx_prob_ACT}
\end{figure}



\begin{figure}[H]
    \centering
    \includegraphics[width=0.8\linewidth]{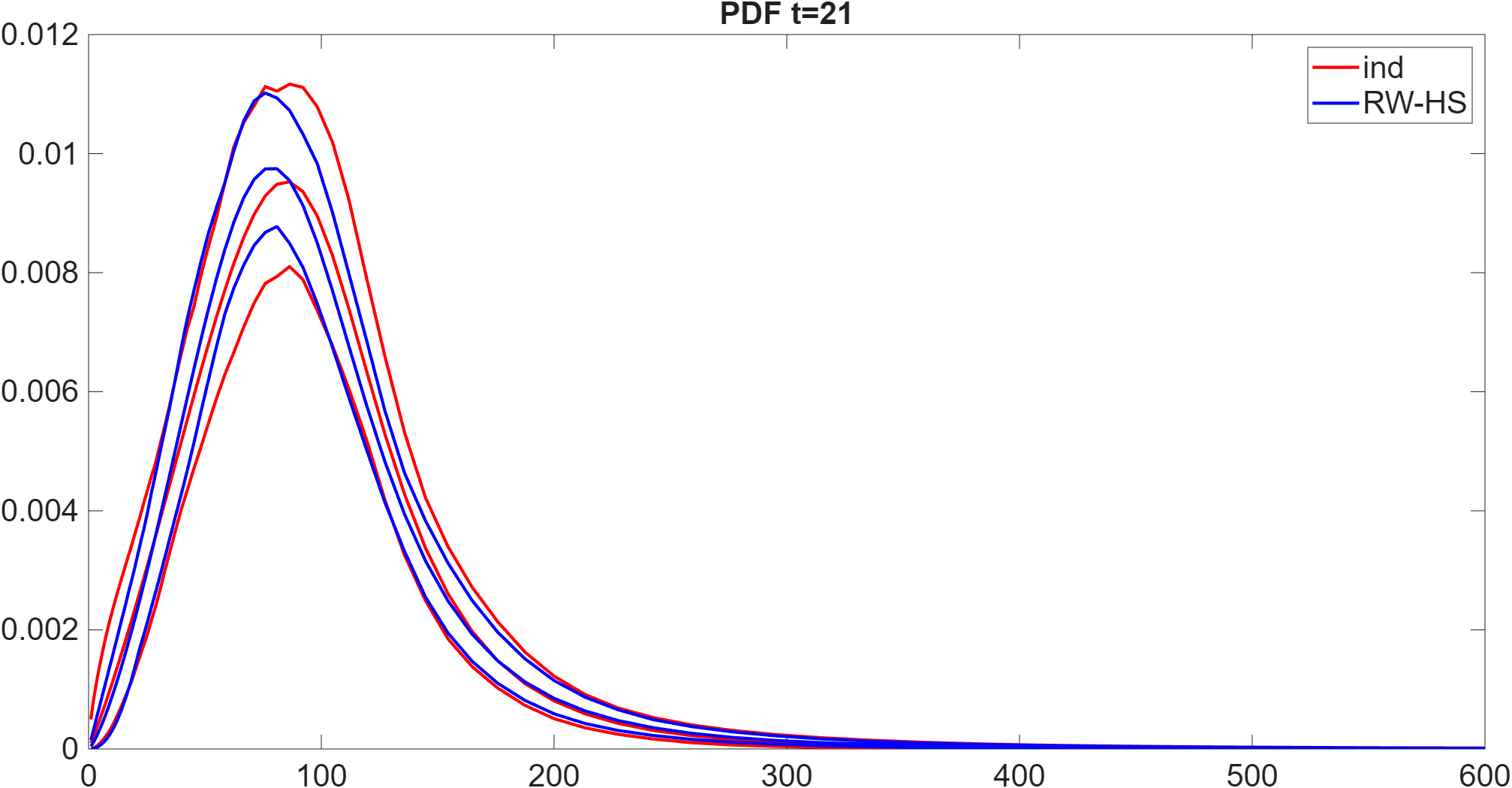}
    \caption{The posterior means (with 99\% credible intervals) of the income density  from the independent GB2 income model (ind) and the random walk GB2 income model with horseshoe priors (RW-HS) for the ACT population subgroup for the year 2021. 
}
    \label{fig:Figure_PDF_GB2_ACT}
\end{figure}

\begin{figure}[H]
    \centering
    \includegraphics[width=0.8\linewidth]{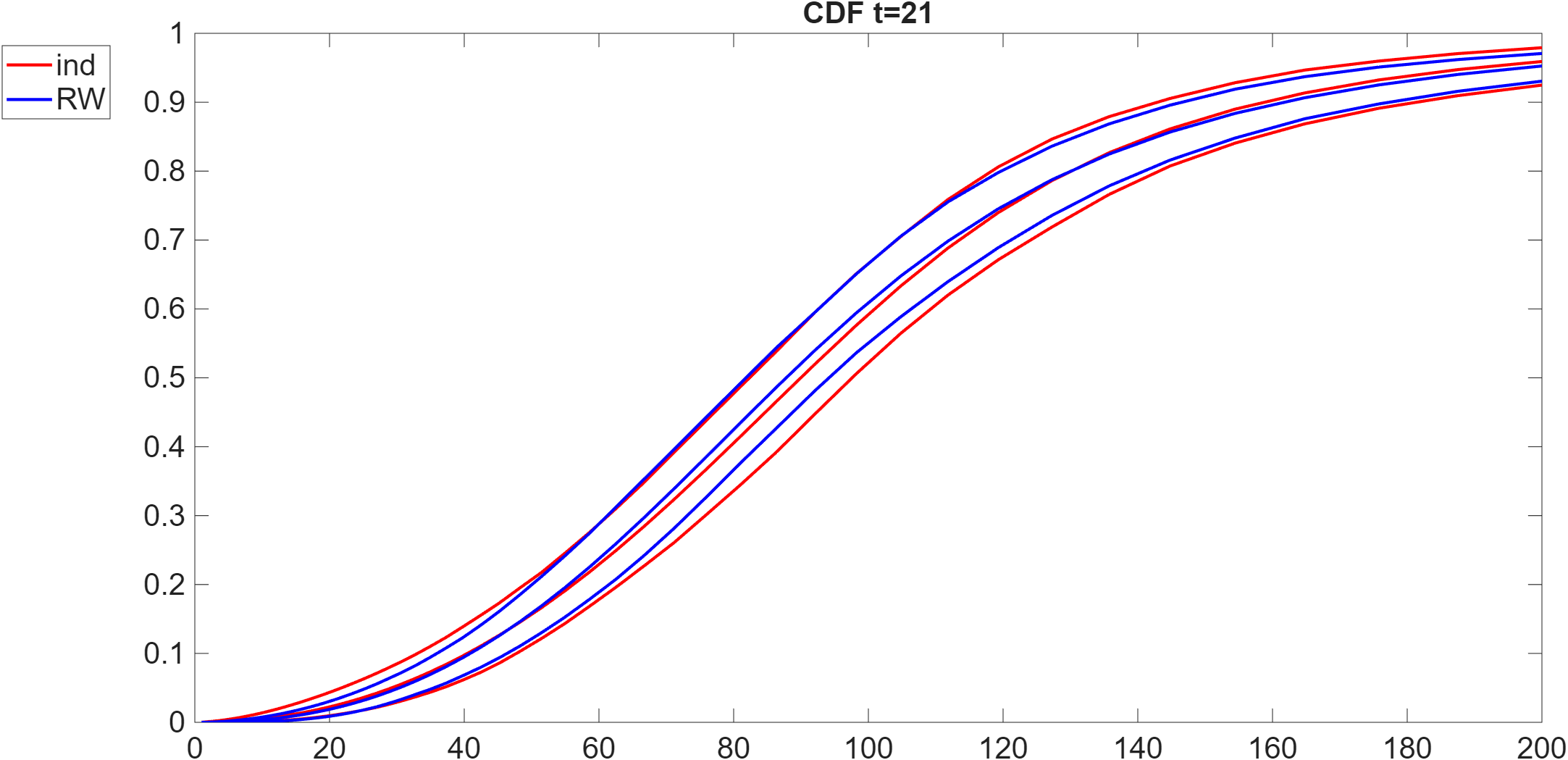}
    \caption{The posterior means (with 99\% credible intervals) of the income CDF obtained from the independent GB2 income model (ind) and the random walk GB2 income model with horseshoe priors (RW-HS) for the ACT population subgroup for the year 2021. 
}
    \label{fig:Figure_CDF_GB2_ACT}
\end{figure}

\begin{figure}[H]
    \centering
    \includegraphics[width=0.8\linewidth]{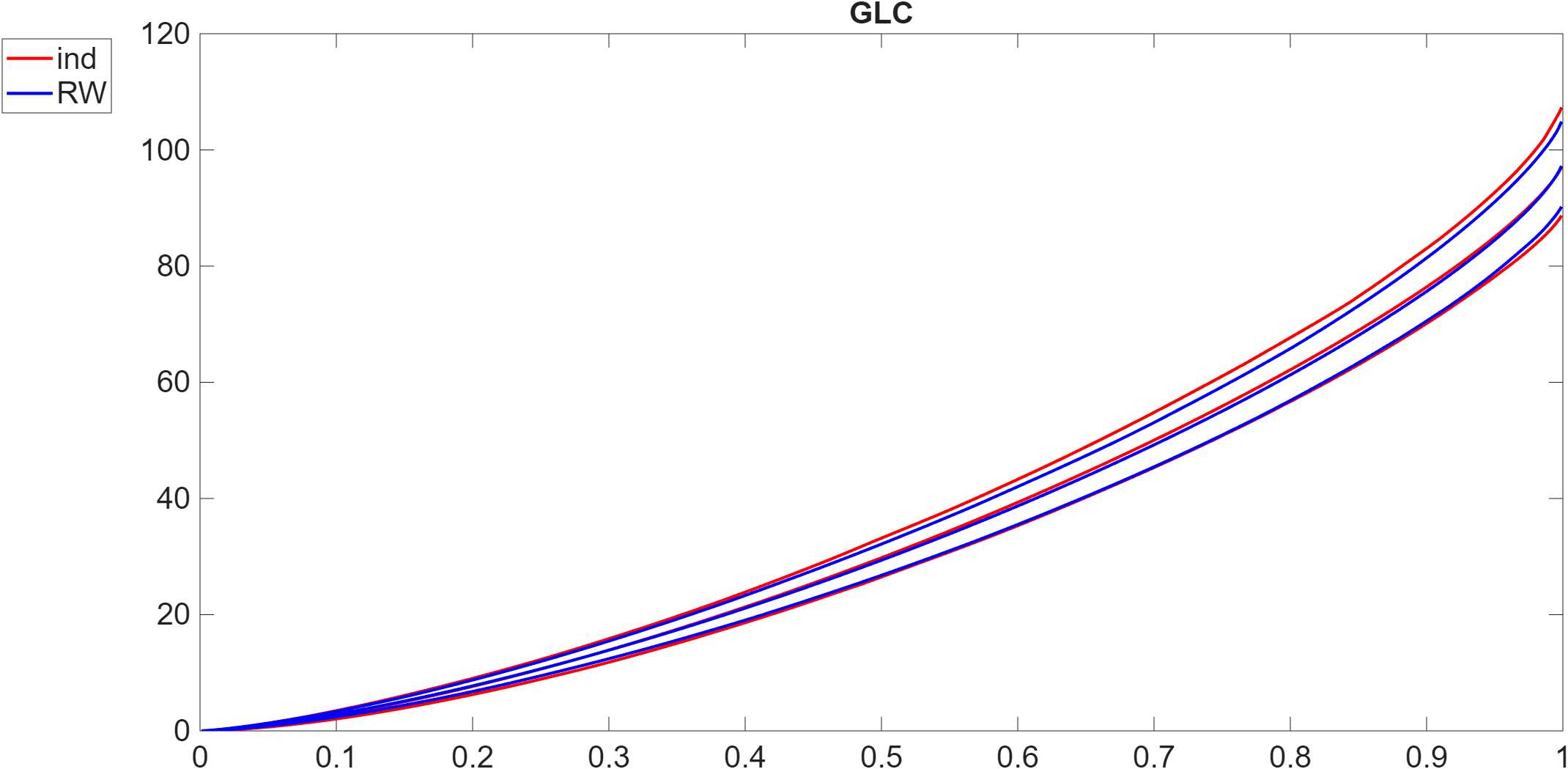}
    \caption{The posterior means (with 99\% credible intervals) of the Generalised Lorenz curve obtained from the independent GB2 income model (ind) and the random walk GB2 income model with horseshoe priors (RW-HS) for the ACT population subgroup for the year 2021.  
}
    \label{fig:Figure_GLC_GB2_ACT}
\end{figure}

\begin{figure}[H]
    \centering
    \includegraphics[width=0.8\linewidth]{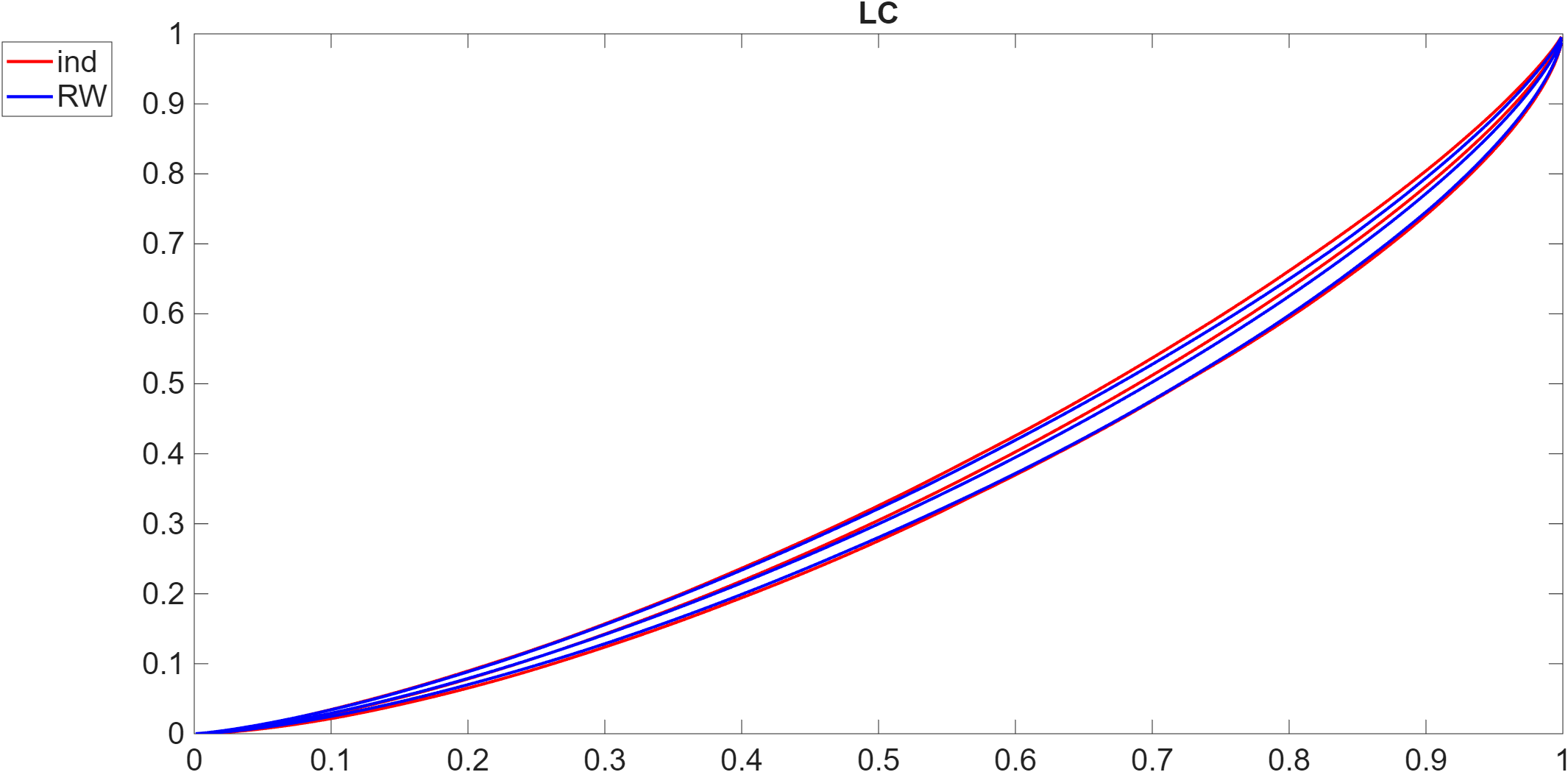}
    \caption{The posterior means (with 99\% credible intervals) of the Lorenz curve obtained from the independent GB2 income model (ind) and the random walk GB2 income model with horseshoe priors (RW-HS) for the ACT population subgroup for the year 2021.  
}
    \label{fig:Figure_LC_GB2_ACT}
\end{figure}

Figures~\ref{fig:Figure_PDF_GB2_ACT}--\ref{fig:Figure_LC_GB2_ACT} present the posterior means and 99\% credible intervals for the fitted GB2 income PDFs, CDFs, GLCs, and LCs for the ACT population subgroup at time \(t=21\) under the independent model and the random walk model with horseshoe priors. The posterior mean CDFs, LCs, GLCs, and PDFs under the independent model and the random walk model with horseshoe priors are close to one another, indicating that both models deliver essentially the same overall distributional picture. The differences in posterior means are generally small rather than substantial: the RW-HS GLC and LC lie slightly below that of the independent model over most population shares. Likewise, the posterior mean CDFs and PDFs are very similar across the support, with only minor deviations in the lower and middle parts of the income distribution. Importantly, the RW-HS model tends to produce narrower 99\% credible intervals than the independent model, indicating greater estimation precision and a more stable characterisation of uncertainty. 


\begin{figure}[H]
    \centering
    \includegraphics[width=0.8\linewidth]{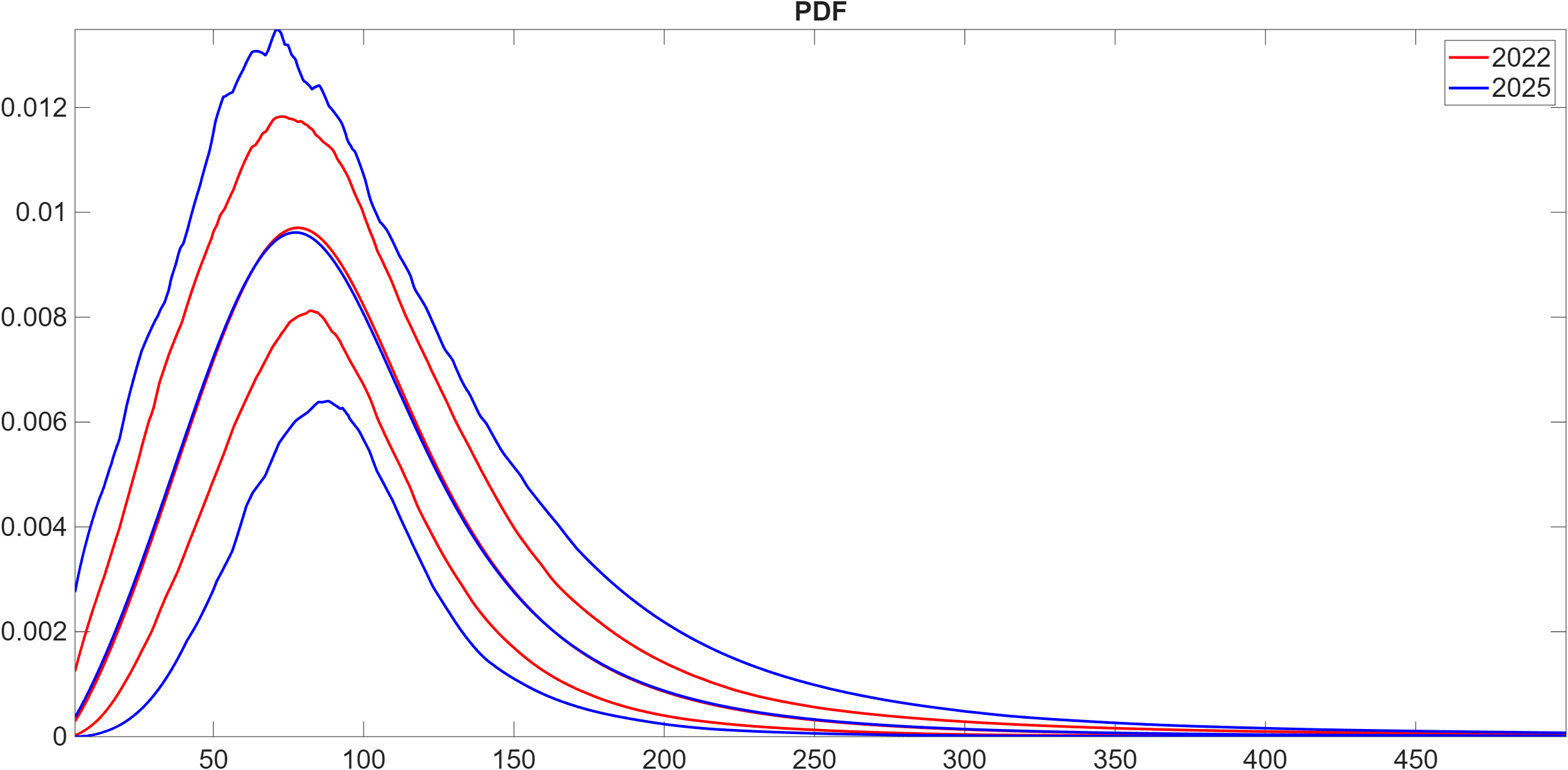}
    \caption{Posterior predictive densities (with 95\% prediction intervals) for the years 2022 and 2025 obtained from the random walk GB2 income model with horseshoe priors (RW-HS) for the ACT population subgroup.  
}
    \label{fig:pred_density_ACT}
\end{figure}

\begin{figure}[H]
    \centering
    \includegraphics[width=0.8\linewidth]{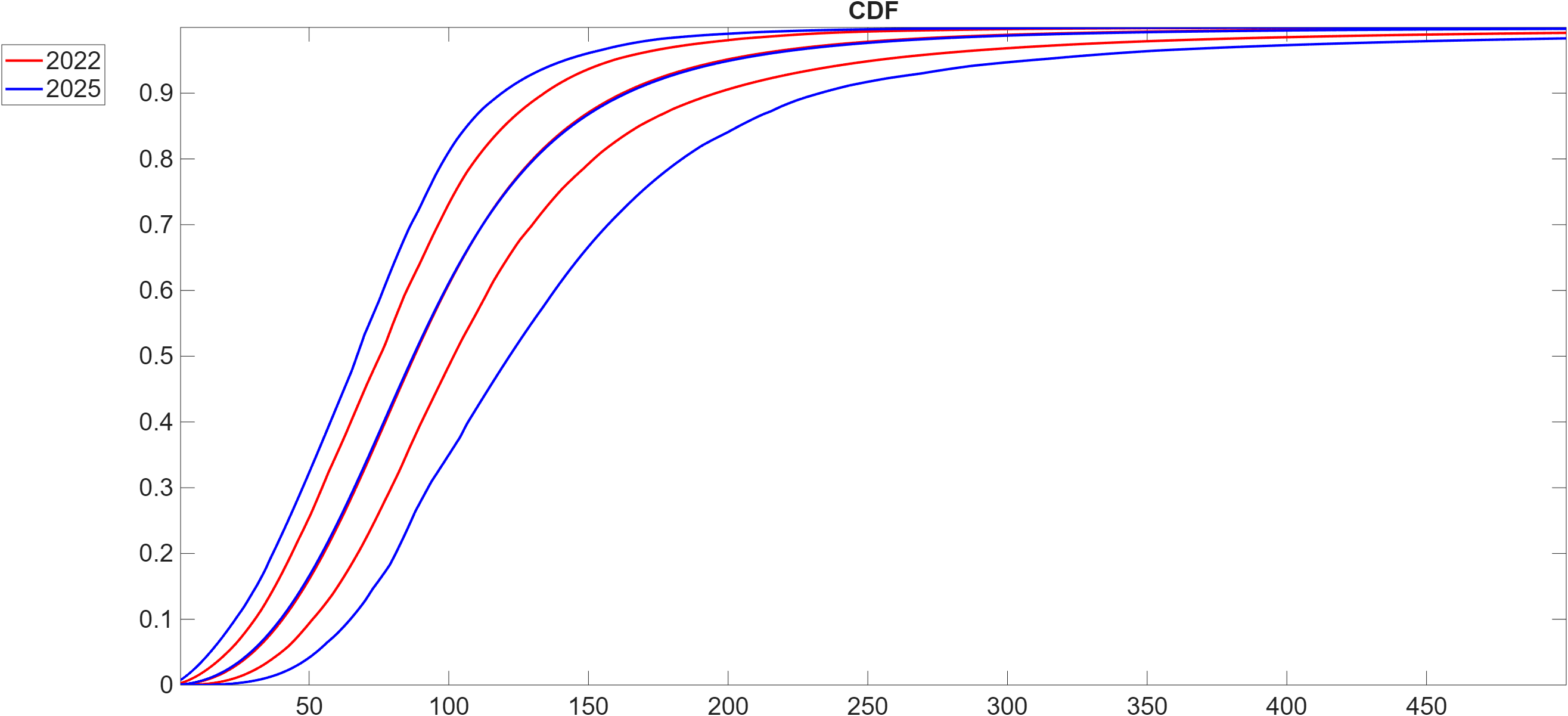}
    \caption{Posterior predictive CDFs (with 95\% prediction intervals) for the years 2022 and 2025 obtained from the random walk GB2 income model with horseshoe priors (RW-HS) for the ACT population subgroup.  
}
    \label{fig:pred_CDF_ACT}
\end{figure}

\begin{figure}[H]
    \centering
    \includegraphics[width=0.8\linewidth]{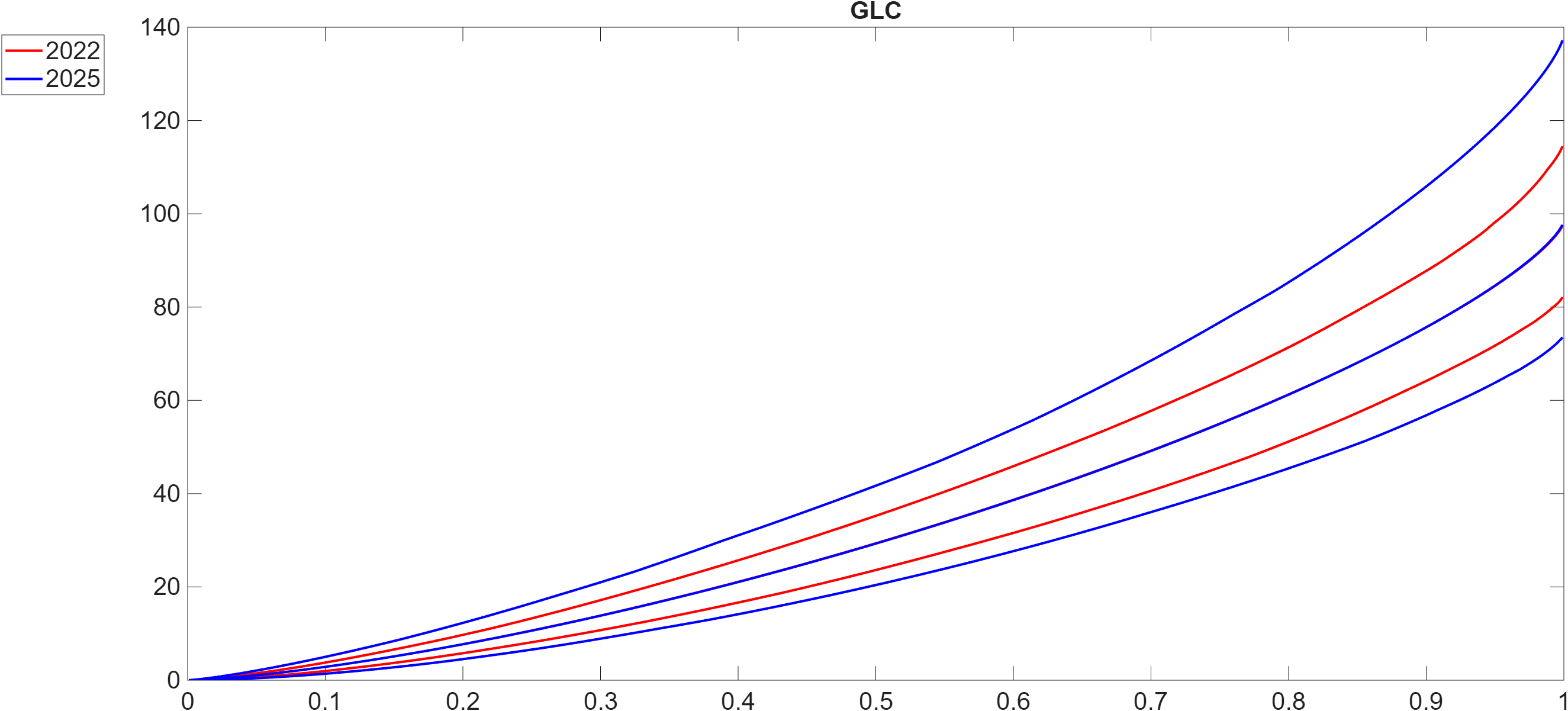}
    \caption{Posterior predictive generalised Lorenz curves (GLCs) (with 95\% prediction intervals) for the years 2022 and 2025 obtained from the random walk GB2 income model with horseshoe priors (RW-HS) for the ACT population subgroup.  
}
    \label{fig:pred_GLC_ACT}
\end{figure}

\begin{figure}[H]
    \centering
    \includegraphics[width=0.8\linewidth]{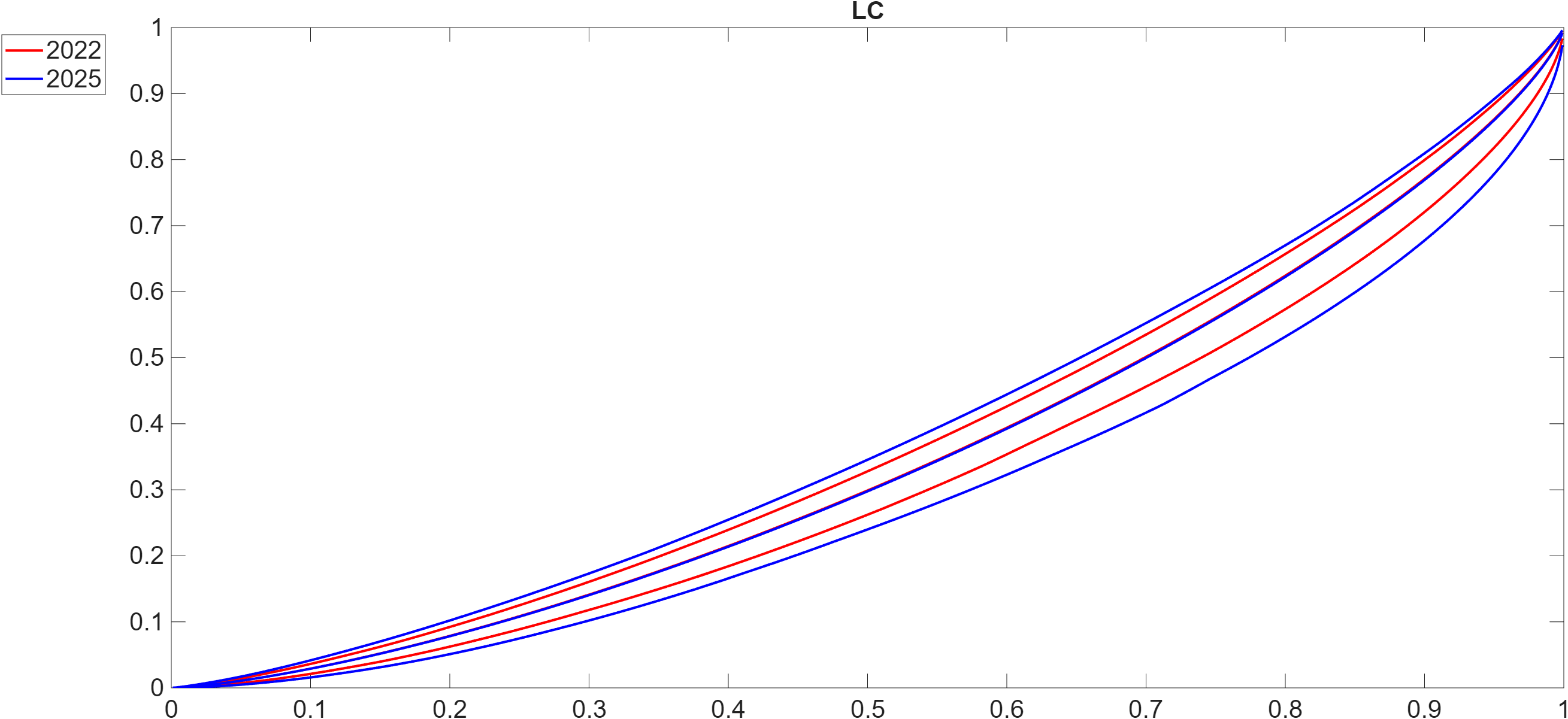}
    \caption{Posterior predictive Lorenz curves (LCs) (with 95\% prediction intervals) for the years 2022 and 2025 obtained from the random walk GB2 income model with horseshoe priors (RW-HS) for the ACT population subgroup.  
}
    \label{fig:pred_LC_ACT}
\end{figure}

Figures \ref{fig:pred_density_ACT} to \ref{fig:pred_LC_ACT} present the posterior predictive income densities, CDFs, GLCs, and LCs, respectively, for the ACT subgroup in 2022 and 2025, obtained by fitting the RW-HS GB2 model to the observed data from 2001 to 2021 and then projecting beyond the sample period. These figures show that the 95\% prediction intervals for 2025 are uniformly wider than those for 2022, particularly in the predictive density and CDF plots, reflecting the natural build-up of forecast uncertainty as the prediction horizon moves further away from the observed 2001--2021 period. 


\begin{figure}[H]
    \centering
    \includegraphics[width=0.8\linewidth]{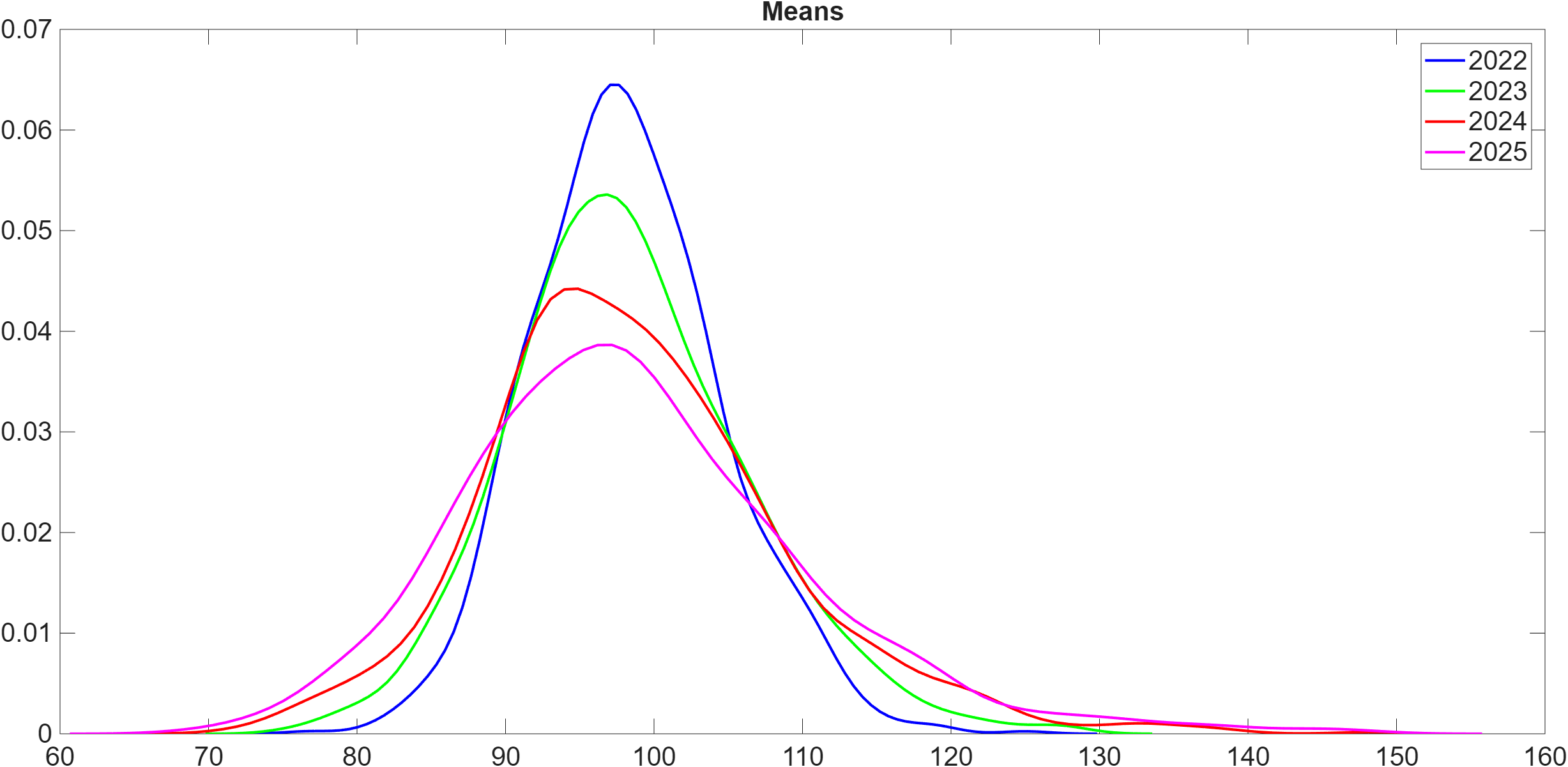}
    \caption{Kernel density estimates of the predicted mean incomes for the ACT population subgroup from 2022 to 2025, obtained using the random walk GB2 income model with horseshoe priors (RW-HS).  
}
    \label{fig:pred_means_ACT}
\end{figure}

\begin{figure}[H]
    \centering
    \includegraphics[width=0.8\linewidth]{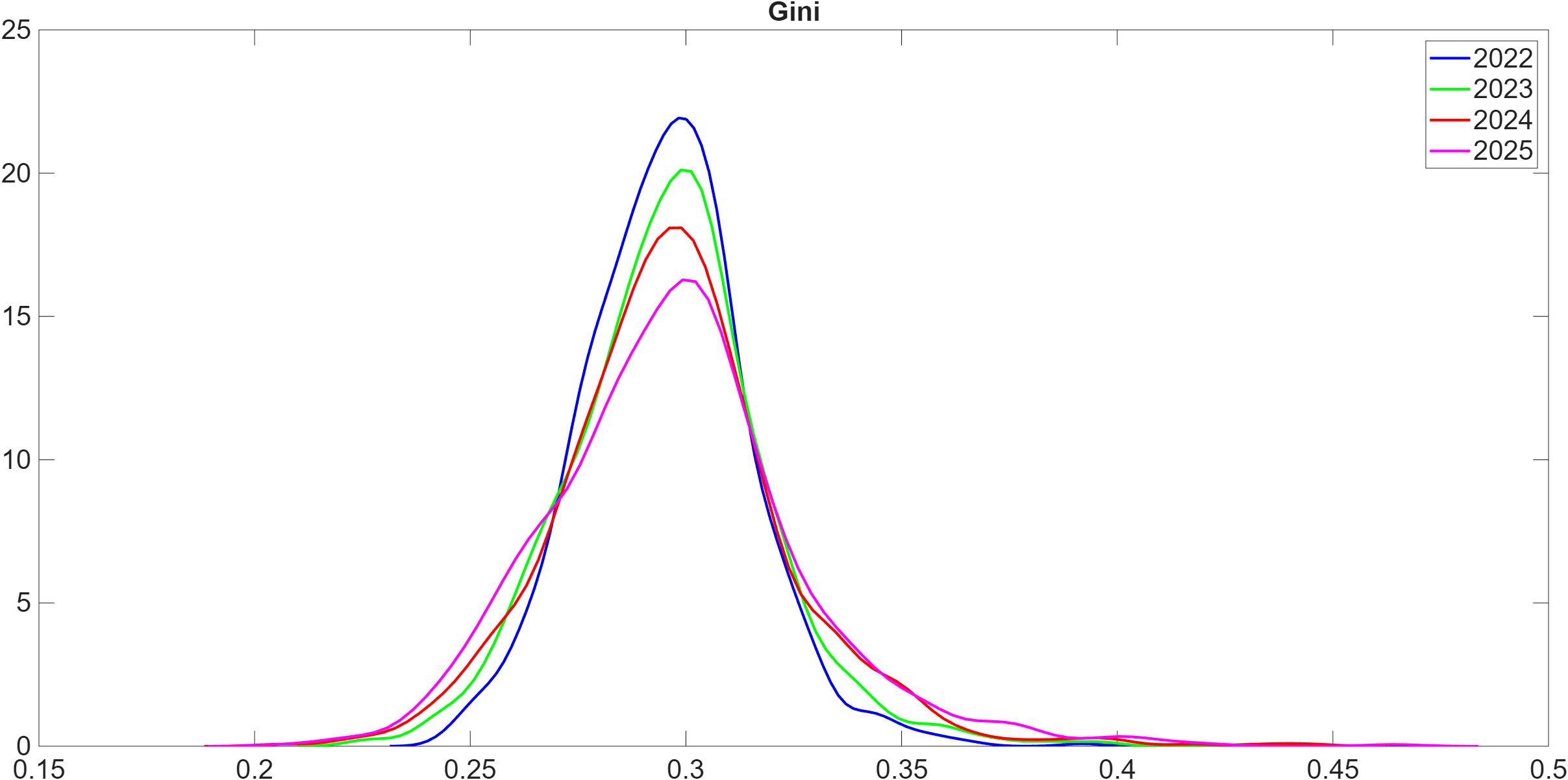}
    \caption{Kernel density estimates of the predicted Gini coefficients for the ACT population subgroup from 2022 to 2025, obtained using the random walk GB2 income model with horseshoe priors (RW-HS).  
}
    \label{fig:pred_GINI_ACT}
\end{figure}

\begin{figure}[H]
    \centering
    \includegraphics[width=0.8\linewidth]{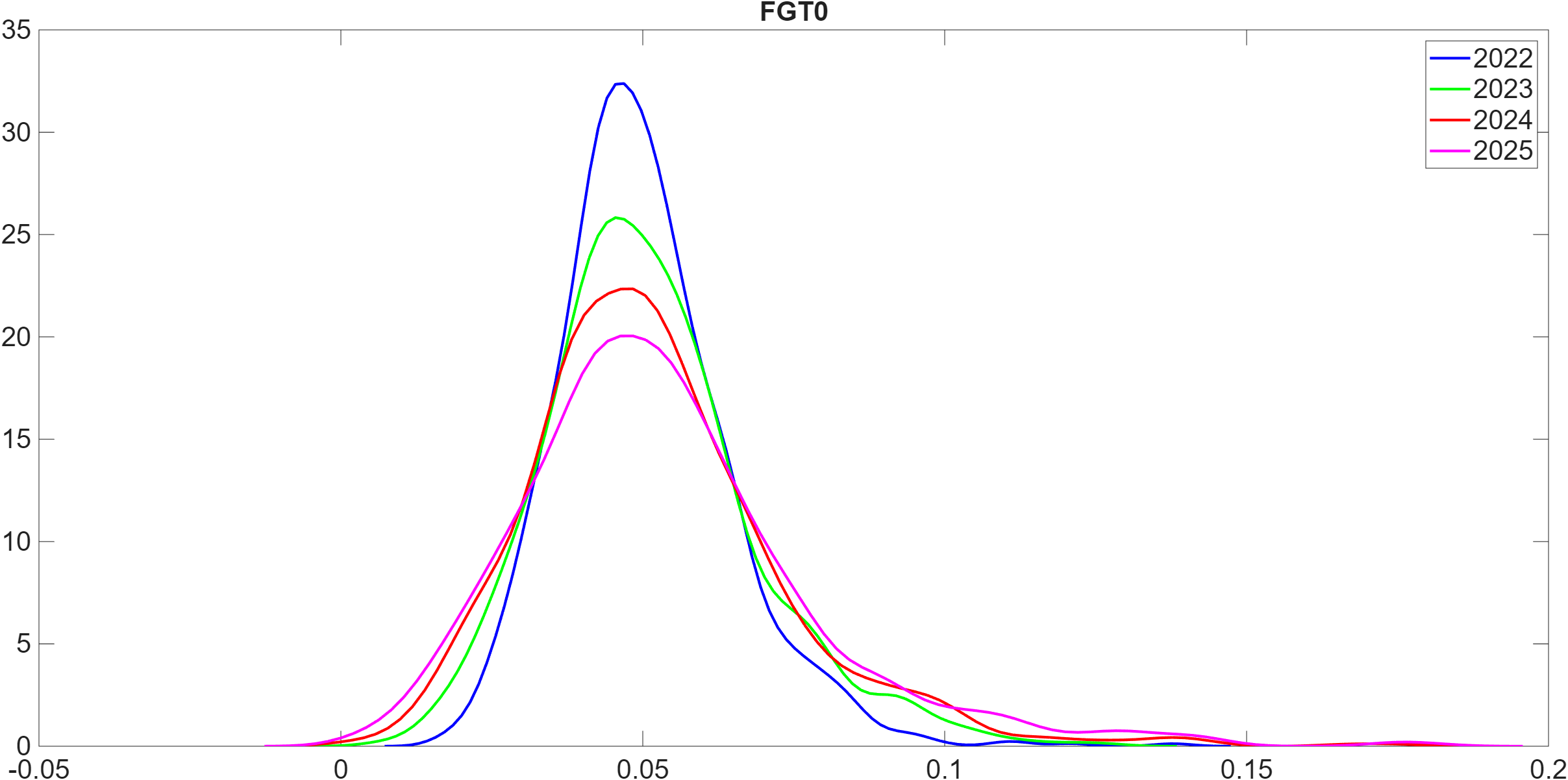}
    \caption{Kernel density estimates of the predicted FGT0 indices for the ACT population subgroup from 2022 to 2025, obtained using the random walk GB2 income model with horseshoe priors (RW-HS).  
}
    \label{fig:pred_FGT0_ACT}
\end{figure}

\begin{figure}[H]
    \centering
    \includegraphics[width=0.8\linewidth]{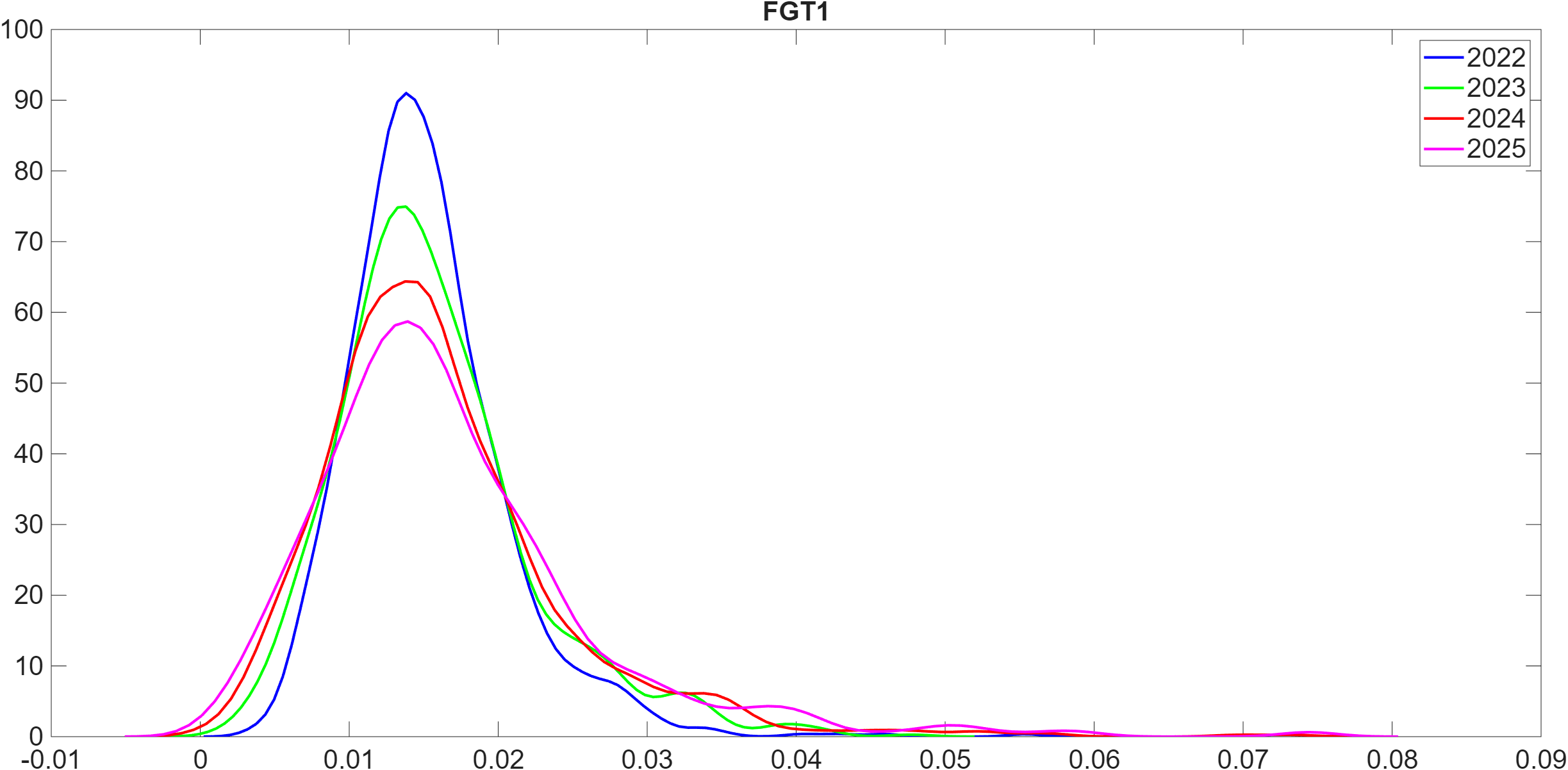}
    \caption{Kernel density estimates of the predicted FGT1 indices for the ACT population subgroup from 2022 to 2025, obtained using the random walk GB2 income model with horseshoe priors (RW-HS).  
}
    \label{fig:pred_FGT1_ACT}
\end{figure}

Figures \ref{fig:pred_means_ACT}--\ref{fig:pred_FGT1_ACT} summarise the predictive distributions of several key welfare measures for the ACT subgroup over the out-of-sample period 2022--2025 under the RW-HS GB2 model fitted to the 2001--2021 data. Figure \ref{fig:pred_means_ACT} shows that the predicted mean income remains centred at relatively high values, although the density becomes progressively flatter and more dispersed from 2022 to 2025, indicating that uncertainty about the future mean increases as the forecast horizon lengthens. A similar pattern appears in Figure \ref{fig:pred_GINI_ACT}: the predictive densities for the Gini coefficient remain concentrated around broadly similar values, with some suggestion of a modest decline in relative inequality, but the later-year densities are clearly wider, so this apparent improvement should be interpreted with greater caution. The same accumulation of uncertainty is even more evident in the poverty measures. In Figures \ref{fig:pred_FGT0_ACT} and \ref{fig:pred_FGT1_ACT}, the predictive densities for the headcount ratio and poverty gap become increasingly spread out, with longer right tails in the later years, especially by 2025. 


\bibliographystyle{apalike}
\bibliography{references_v2}

\end{document}